 \documentclass[twocolumn]{aastex631}

\shorttitle{Intermediate-mass black holes}
\shortauthors{Graham et al.}

\begin{document}

\title{Central X-ray point-sources found to be abundant in low-mass, late-type
  galaxies predicted to contain an intermediate-mass black hole}

\author[0000-0002-6496-9414]{Alister W.\ Graham}
\affiliation{Centre for Astrophysics and Supercomputing, Swinburne University of
  Technology, Victoria 3122, Australia}
\email{AGraham@swin.edu.au}

\author[0000-0002-4622-796X]{Roberto Soria}
\affiliation{College of Astronomy and Space Sciences, University of the Chinese
  Academy of Sciences, Beijing 100049, China} 
\affiliation{Sydney Institute for Astronomy, School of Physics A28, The University
  of Sydney, Sydney, NSW 2006, Australia}

\author[0000-0002-4306-5950]{Benjamin L.\ Davis}
\affiliation{Centre for Astrophysics and Supercomputing, Swinburne University of
  Technology, Victoria 3122, Australia}
\affiliation{Center for Astro, Particle, and Planetary Physics (CAP$^3$), New York
  University Abu Dhabi}

\author{Mari Kolehmainen}       
\affiliation{Observatoire Astronomique, Universit\'e de Strasbourg, CNRS, UMR 7550,
  11 Rue de l'Universit\'e, F-67000 Strasbourg, France}

\author{Thomas Maccarone}      
\affiliation{Department of Physics and Astronomy, Texas Tech University, Box 41051,
  Lubbock, TX 79409-1051, USA}

\author{James Miller-Jones}    
\affiliation{International Centre for Radio Astronomy Research, Curtin University,
  GPO Box U1 987, Perth, WA 6845, Australia}

\author{Christian Motch}       
\affiliation{Observatoire Astronomique, Universit\'e de Strasbourg, CNRS, UMR 7550,
11 Rue de l'Universit\'e, F-67000 Strasbourg, France}

\author{Douglas A.\ Swartz}    
\affiliation{Astrophysics Office, NASA Marshall Space Flight Center, ZP12,
  Huntsville, AL 35812, USA}

\begin{abstract}

Building upon three late-type galaxies in the Virgo cluster with both a
predicted black hole mass of less than $\sim$10$^5$~M$_{\odot}$ and a
centrally-located X-ray point-source, we reveal 11 more such galaxies, more
than tripling the number of active intermediate-mass black hole candidates
among this population.  Moreover, this amounts to a $\sim$36$\pm$8\% X-ray
detection rate (despite the sometimes high, X-ray-absorbing, H\,{\footnotesize
  I} column densities), compared to just 10$\pm$5\% for (the largely
H\,{\footnotesize I}-free) dwarf early-type galaxies in the Virgo cluster.
The expected contribution of X-ray binaries from the galaxies' inner field
stars is negligible.  Moreover, given that both the spiral and dwarf galaxies
contain nuclear star clusters, the above inequality appears to disfavor X-ray
binaries in nuclear star clusters.  The higher occupation, or rather
detection, fraction among the spiral galaxies may instead reflect an enhanced
cool gas/fuel supply and Eddington ratio.  Indeed, four of the 11 new
X-ray detections are associated with known LINERs or LINER/H\,{\footnotesize
  II} composites.  
For all (four) of the new detections for which the X-ray flux was strong enough to establish the
spectral energy distribution in the {\it Chandra} band, it is consistent with power-law
spectra.  Furthermore, the X-ray emission from the source with the highest
flux (NGC~4197: $L_X\approx10^{40}$~erg~s$^{-1}$) suggests a non-stellar-mass
black hole if the X-ray spectrum corresponds to the `low/hard state'.
Follow-up observations to further probe the black hole masses, and prospects
for spatially resolving the gravitational spheres-of-influence around
intermediate-mass black holes, are reviewed in some detail.

\end{abstract}

\section{Introduction}

While galaxies suspected of harboring a supermassive black hole (SMBH) with a
mass of around $10^5$--$10^6\,{\rm M}_{\odot}$  have long been
identified \citep[e.g.,][]{1985ApJS...57..503F, 1995ApJS...98..477H,
  2007ApJ...670...92G, 2013ApJ...775..116R, 2014ApJ...782...55Y, 2015ApJ...798...54G,
  2016MNRAS.455.3148S, 2018ApJS..235...40L} --- including POX~52
\citep{2004ApJ...607...90B, 2008ApJ...686..892T}, NGC~4395
\citep{2015MNRAS.449.1526L, 2015ApJ...809..101D, 2019MNRAS.486..691B} and
NGC~404 \citep{2020MNRAS.496.4061D} --- there is an observational-dearth of
centrally-located black holes with masses that are intermediate between
stellar-mass black holes ($\lesssim10^2\,{\rm M}_{\odot}$) and SMBHs
($\gtrsim10^5\,{\rm M}_{\odot}$).  
Several late-type galaxies in the Virgo cluster are known to possess an 
active galactic nucleus (AGN), including Seyferts (e.g., NGC: 4388; 4569; 4579; 4639; 
and 4698) and LINERs (e.g., NGC: 4438; 4450; and 4548).  
However, if the low Eddington ratios 
associated with the SMBHs of these AGN 
are representative of the mean, they warn that the X-ray luminosities
of potential intermediate-mass black holes (IMBHs) 
in lower-mass spiral galaxies will be challenging to observe.  

Traditional optical searches for low-mass AGN using 
spectroscopic line ratios in the Baldwin-Phillips-Terlevich
\citep[BPT:][]{1981PASP...93....5B} 
diagram, or the 
W$_{{\rm H}\alpha}$ versus [N\,{\footnotesize II}]/H$\alpha$ \citep[WHAN:][]{2011MNRAS.413.1687C}
diagram, 
are limited given that the faint AGN signal is expected to be dwarfed by the
stellar signal.  Moreover, the hardening of the AGN spectral energy
distribution (SED) 
in low-mass AGN can change the ionization structure, rendering the familiar line
diagnostics less effective 
\citep{2019ApJ...870L...2C} and further contributing to why less than 1\% of low-mass
galaxies have been reported to have evidence of an AGN 
\citep{2013ApJ...775..116R}.  In addition, studies at radio wavelengths are
not ideal given that half of the X-ray luminous AGN are radio-quiet \citep{radcliffe2021radio}.
However, members of the elusive population of 
intermediate-mass black holes (IMBHs: 10$^2$-$10^5$ M$_{\odot}$) are 
starting to be found, with 
exciting discoveries in 
SDSS J160531.84+174826.1 \citep{2007ApJ...657..700D}, 
HLX-1 in\footnote{This is an off-center target.} ESO~243-49 \citep{2009Natur.460...73F}, 
IRAS 01072+4954 \citep{2012A&A...544A.129V}, 
LEDA~87300 \citep{2015ApJ...809L..14B, 2016ApJ...818..172G}, 
NGC~205 \citep{2018ApJ...858..118N, 2019ApJ...872..104N}, 
NGC 3319 \citep{2018ApJ...869...49J, 2021PASA...38...30D}, 
IC~750 \citep{2020ApJ...897..111Z} 
and the host galaxies of
GW170817A \citep{2019arXiv191009528Z}, GW190521 \citep{2020arXiv200901075T}, 
and 3XMM J215022.4-055108 \citep{2020ApJ...892L..25L}. 

Indeed, the flood gates may be about to open. 
Recently, \citet{2018ApJ...863....1C} used the width
and luminosity of the H$\alpha$ emission line to identify 
IMBH candidates at the centers of 305 galaxies: ten of which have
X-ray data that reveal a coincident point-source and suspected 
AGN.  Four of these ten (which includes LEDA~87300) have a black
hole mass estimate less than $\sim$$10^5\,{\rm M}_{\odot}$.  In addition,
\citet{2014AJ....148..136M} has reported on 28 nearby ($<$80 Mpc) dwarf
galaxies with narrow emission line (Type 2) AGN, while \citet[][see their
  Figure~1]{2018MNRAS.478.2576M} report on X-ray emission coming from 40
predominantly star-forming dwarf galaxies with Type 2 AGN out to a redshift of
$\sim$1.3, with three galaxies stretching the sample out to $z=2.39$.
\citet[][see their Figure~8]{2018MNRAS.478.2576M} applied a roughly linear
$M_{\rm bh}$--$M_{\rm *,gal}$ relation to the galaxies' stellar masses to predict that 7
of their 40 galaxies have black hole masses less than $10^5\,{\rm M}_{\odot}$.

Closer to home, \citet{2019MNRAS.484..794G} and \citet[][hereafter
  GSD19]{2019MNRAS.484..814G} have identified 63 Virgo cluster galaxies
expected to house a central IMBH according to one or more black hole mass
scaling relations, including the newer, morphology-dependent $M_{\rm
  bh}$--$M_{\rm *,gal}$ relations \citep[][and references
  therein]{2019ApJ...876..155S}.  Reanalyzing the archival X-ray data for the
30 early-type galaxies in this set, \citet{2019MNRAS.484..794G} found that
just three of them (IC: 3442; 3492; and 3292)\footnote{X-ray point-source
  discovery in IC~3442 and IC~3492 was by \citet{2010ApJ...714...25G}.}
display a central X-ray point-source.\footnote{None of the 30 have a compact
  radio source at 8.4~GHz, with a flux limit of ~0.1 mJy
  \citep{2009AJ....138.1990C}.}  In contrast, GSD19 reported that among the
set of 33 late-type galaxies, three (NGC: 4470; 4713; and 4178)\footnote{X-ray
  point-source discovery in NGC~4713 was by \citet{2015ApJ...814...11T}, and
  in NGC~4178 by \citet{2012ApJ...753...38S}.}  of the seven with archival
X-ray data possessed a central X-ray point-source.\footnote{While GSD19 noted
  that NGC~4178's X-ray point-source may be due to a stellar-mass black hole, they
  also noted that 
  \citet{2009ApJ...704..439S} had reported a strong 
[Ne\,{\footnotesize V}] 14.32 $\mu$m emission line, with an ionization potential of 97.1 eV, 
indicative of an AGN.}

The higher activity ratio in the late-type spiral galaxies (3/7) --- when
compared to the dwarf early-type galaxies (3/30) --- may have been an
insightful clue. We have now obtained and 
analyzed {\it Chandra X-ray Observatory} ({\it CXO}) images for the
remaining (33$-$7$=$) 26 previously unobserved late-type spiral galaxies. 
As will be reported here, 9 of these 26 (or 12 of the original
33) late-type galaxies, contain a centrally-located X-ray point-source.  This
amounts to 36\% of the late-type galaxy sample.  The
addition here of NGC~4212 and NGC~4492, both of which have centrally located
X-ray point-sources and are expected to house an IMBH, 
takes the sample of IMBH candidates to 14 (see section~\ref{Sec-sample}).  
The probability that all 12 of these 33 (or 14 of 35) 
are stellar-mass X-ray binaries (XRBs) 
accreting near or above the Eddington ratio seems 
small\footnote{Arguably, if there are no IMBHs, and the  
early-type galaxy sample implies that 
the probability of an XRB residing at the center of a galaxy
  is 3-from-30, or 1-in-10, then there is roughly a $4.7\times10^{-5}$ probability
  (1-in-21000) of having a sample with as many as 12-from-33
  galaxies with a central XRB. Taken from 
$\sum_{i=0}^{(33-12)} 0.1^{33-i}\,0.9^i /(33-i)!\,i!$.
The existence of XRBs with either a low- or high-mass donor
star complicates this calculation, and is addressed later in 
section~\ref{Sec-XRB}.\label{footProb}}, and Occam's
razor favors that we are instead observing low-(Eddington ratio) AGN, possibly due to the 
higher gas fractions keeping on the AGN pilot light.  
Indeed, star-forming galaxies are known to have AGN with 
a higher Eddington ratio than quiescent, i.e., non-(star-forming), 
galaxies \citep{2009MNRAS.397..135K}. 
Relatively inactive IMBHs in dwarf 
early-type galaxies may, however, still be common \citep{2017ApJ...839L..13S,
  2018MNRAS.476..979P, 2020MNRAS.492.2268B}.  

In Section~\ref{Sec-sample}, we
introduce a subsample of low-mass Virgo cluster spiral galaxies predicted to
have an IMBH and 
reported here for the first time to have a centrally-located X-ray
point-source.  Results for the spiral galaxies with
expected black hole masses greater than $10^5$--$10^6$ M$_{\odot}$ will be
presented in a subsequent paper exploring AGN occupation fractions, Eddington
ratios, and trends with the host galaxy mass.  In Section~\ref{Sec-Data}, we
report on the X-ray data for the subsample of active IMBH candidates, while
Section~\ref{Sec-mass} discusses the prospects for estimating 
the black hole mass using the X-ray data alone (Section~\ref{subsec-alone})
or when combined with radio data (Section~\ref{subsec-radio}).
Section~\ref{Sec-XRB} addresses the issue of potential XRB contamination in more detail.  
In Section~\ref{Sec-Disc}, we discuss 
expectations for spatially resolving the gravitational sphere-of-influence around
IMBHs, and provide some direction for future observations of our, and other, IMBH candidates.
Finally, Section~\ref{Sec_dual} discusses dual and off-center AGN before a summary
of our key findings is provided in Section~\ref{Sec_summary}.

\begin{deluxetable*}{lcccccc}
\tablecaption{IMBH mass predictions based on the host galaxy properties\label{Tab-IMBH}}
\tablewidth{0pt}
\tablehead{
\colhead{Galaxy} &  \colhead{D [Mpc]}  &  \colhead{$\log M_{\rm bh}$ ($M_{\rm
*,total}$)}  &  \colhead{$\log M_{\rm bh}$ ($\phi$)}  &  \colhead{$\log M_{\rm
 bh}$ ($\sigma$)}  &   \colhead{$\log M_{\rm bh}$ ($M_{\rm nc}$)}
 &  \colhead{$\overline{\log M_{\rm bh}}$} 
}
\startdata
\multicolumn{7}{c}{Archival X-ray data presented in \citet{2019MNRAS.484..814G}} \\
NGC4178, VCC66   & 13.20$\pm$3.00 &  3.9$\pm$0.9     &   4.2$\pm$1.6  &  3.1$\pm$0.9     &  2.6$\pm$1.6   &  3.5$\pm$0.6  \\
NGC4713, ...     & 14.80$\pm$3.55 &  3.8$\pm$0.9     &   3.5$\pm$1.9  & 2.8$\pm$1.3      &  4.6$\pm$1.7   &  3.6$\pm$0.6  \\
NGC4470$^a$, VCC1205 & 16.40$\pm$6.60 &  3.4$\pm$1.2     &   4.6$\pm$2.6  & 5.1$\pm$0.8$^b$  &  4.5$\pm$1.7   &  4.6$\pm$0.6  \\ 
\multicolumn{7}{c}{New X-ray data} \\
NGC4197, VCC120  & 26.40$\pm$3.92 &  4.3$\pm$0.8     &   5.4$\pm$0.9  &  ...             &      ...       &  4.8$\pm$0.6  \\
NGC4212$^c$, VCC157  & 17.05$\pm$2.62 &  6.0$\pm$0.9     &   5.9$\pm$0.4  &  5.1$\pm$0.8     &  6.2$\pm$1.6   &  5.8$\pm$0.3  \\
NGC4298, VCC483  & 15.80$\pm$2.54 &  5.3$\pm$0.8     &   5.6$\pm$0.8  &  4.2$\pm$0.8     &  5.3$\pm$1.7   &  5.1$\pm$0.4  \\
NGC4313, VCC570  & 14.15$\pm$2.92 &  4.9$\pm$0.9     &    ...         &  5.2$\pm$0.8     &  6.7$\pm$1.6   &  5.3$\pm$0.6  \\
NGC4330, VCC630  & 19.30$\pm$1.56 &  4.4$\pm$0.8     &    ...         &      ...         &      ...       &  4.4$\pm$0.8  \\
NGC4405, VCC874  & 17.85$\pm$3.32 &  4.4$\pm$0.9     &    ...         &      ...         &      ...       &  4.4$\pm$0.9  \\
NGC4413$^d$, VCC912  & 16.05$\pm$1.40 &  3.7$\pm$0.8     &   4.5$\pm$0.6  &      ...         &      ...       &  4.2$\pm$0.5  \\
NGC4492$^e$, VCC1330 & 19.30$\pm$3.54 & 4.9$\pm$0.9  &    ...         &      ...         &       ...      &  4.9$\pm$0.9  \\
NGC4498, VCC1379 & 14.55$\pm$3.62 &  3.7$\pm$0.9     &   5.8$\pm$1.8  &  ...             &  3.8$\pm$1.5   &  4.0$\pm$0.7  \\
NGC4519, VCC1508 & 19.60$\pm$8.48 &  4.2$\pm$1.2     &   5.5$\pm$2.3  &  ...             &      ...       &  4.5$\pm$1.1  \\ 
NGC4607, VCC1868 & 19.70$\pm$6.55 &  4.5$\pm$1.1     &    ...         &      ...         &      ...       &  4.5$\pm$1.1  \\
\enddata
\tablecomments{Column 2 displays the median redshift-independent distances
  from NED.  Predicted black hole masses are in units of solar mass, derived from one to four
independent observables (see Section~\ref{Sec-pred-mass}) depending on their
availability: 
total stellar mass, $M_{\rm *,total}$;\
spiral-arm pitch angle,  $\phi$;
central stellar velocity dispersion,  $\sigma$; and 
nuclear star cluster mass, $M_{\rm nc}$.  
The final column provides the error-weighted, mean black hole mass,
$\overline{\log M_{\rm bh}}$.  
$^a$ Also known as NGC~4610. 
$^b$ Revised down from $10^6\,{\rm M}_{\odot}$ in GSD19 due
to the velocity dispersion dropping from 90 to $\sim$60 km s$^{-1}$ (see
Section~\ref{Sec-N4470}).  
$^c$ Also known as NGC~4208.
$^d$ Also known as NGC~4407. 
$^e$ NGC~4492 has archival {\it CXO} data, but is
new in the sense that we did not report on the X-ray data in GSD19.}
\end{deluxetable*}

\newpage

\section{The sample and their expected black hole masses}\label{Sec-sample}

\subsection{The sample}

The abundance of SMBHs at the centers of galaxies has led to many 
black hole mass scaling relations, some of which  were recently used by us to
estimate the masses of the black holes at the centers of 100 early-type
galaxies \citep{2019MNRAS.484..794G} and 74 late-type galaxies
(GSD19) in the Virgo galaxy cluster.  The early-type
galaxy sample was compiled by \citet{2004ApJS..153..223C} and the subsequent
{\it CXO} images from the Large Project titled `The Duty Cycle of Supermassive
Black Holes: X-raying Virgo' (PI: T.Treu, Proposal ID: 08900784) were used to
identify which galaxies had AGN \citep{2010ApJ...714...25G}.  We have established a
complementary {\it CXO} Large Project titled `Spiral galaxies of the Virgo
Cluster' (PI: R.Soria, Proposal ID: 18620568) which has imaged 52 galaxies and
utilized an additional (22+1=)\footnote{We have discovered a central X-ray 
point-source in archival, Cycle~8, {\it CXO} data for NGC~4492, an additional
Virgo spiral galaxy which is expected to harbor an IMBH, and which has taken
our parent sample from 74 to 75 spiral galaxies.\label{foot4492}} 
23 spiral galaxies for which suitable archival X-ray data existed.  

The combined sample of 75 spiral galaxies is described in \citet{Soria2021}, 
with an emphasis on the observations and data reduction, 
 star formation rate measurements, and the identification of some 80
 off-center ultraluminous X-ray sources 
(ULX: $L_{0.3-10\,{\rm keV}} \approx  10^{39}$--$10^{41}$ erg s$^{-1}$). 
Here, we are following-up on
GSD19, who determined that 33 of the original 74 spiral galaxies are expected to
harbor a central IMBH. 
More precisely, this paper started with the (33-7=) 26
spiral galaxies predicted to have an IMBH but which did not previously have
archival X-ray data.\footnote{The seven spiral galaxies which did have
  archival data were reported on in GSD19.} 
Of these 26 spiral galaxies, we focus on those found
here to have a centrally-located X-ray point-source.  
We have discovered such sources in 9 of these 26 galaxies, giving a hit-rate of 
12-from-33 when including the archival data. In addition, we have 
included NGC~4492 (not in the original sample of 74, 
see footnote~\ref{foot4492}) and NGC~4212 (already in the
original sample of 74, but not counted in the subsample of 33), 
thereby taking the tally to 14-from-35. 
NGC~4212 has a potentially exciting dual central 
X-ray point-source, and a predicted central black hole mass of
$10^5$--($2\times10^6)\,{\rm M}_{\odot}$ (GSD19), which is why it was not
counted in the initial sample of 33 because this range is above $10^5$
M$_\odot$. 

For convenience of reference, all of these 14 galaxies with central X-ray point-sources, 
and their predicted black hole masses, are presented in Table~\ref{Tab-IMBH}. 
The entries are explained in the following subsection. 
In a follow-up paper presenting the full sample of 75 galaxies, we will present
the occurrence of a central X-ray point-source with both the existence, or not, of a galaxy
bar and also as a function of the disk inclination.  Here, we simply report
that among the 14 galaxies noted above, there is no preference for those with
an X-ray point-source to have a bar (6 of 14 do) nor a particularly face-on
orientation (9 of 14 have an axis ratio greater than 0.5). 

In passing, it is relevant to again note that using powerful data mining
techniques, \citet{2018ApJ...863....1C} searched the {\it CXO} 
archives and identified a sample of 305 galaxies with both a Type I AGN, as
determined from their optical spectra, and a suspected IMBH in the range
$3\times10^4 < M_{\rm bh}/{\rm M}_{\odot} < 2\times10^5$.  Ten of these
galaxies were reported to have nuclear X-ray emission, and 4 of these 10 had
a black hole mass estimate less than $10^5\,{\rm M}_{\odot}$.  Of these four
galaxies, and of relevance here, is that the one with the smallest black hole
mass estimate is the Virgo Cluster Catalog dwarf galaxy VCC~1019 (SDSS
J122732.18+075747.7) imaged by {\it XMM-Newton}.  We downloaded and
reprocessed the {\it CXO} data for VCC~1019 --- which is a background
spiral galaxy at 150~Mpc \citep[e.g.,][]{1998ApJS..119..277G} --- and found no
X-ray emission, whereas \citet{2018ApJ...863....1C} reported a ``very faint''
source.  The nine (of ten) other galaxies with X-ray emission are not in the
Virgo cluster.

\subsection{The predicted black hole masses}\label{Sec-pred-mass}

We proceed under the hypothesis that (some of) the X-ray point-sources are
associated with massive black holes.  
There are now many approaches to predict a galaxy's central black hole mass
which do not rely upon the assumption of stable (virialized) gas clouds
orbiting the black hole in some universal geometrical configuration.  That 
assumption employs a sample mean virial
factor, $\langle f \rangle$, obtained by linking (reverberation mapping)-derived virial
products \citep[e.g.,][]{2000ApJ...540L..13P} to either the $M_{\rm
  bh}$--$\sigma$ or $M_{\rm bh}$--$M_{\rm *,gal}$ relation defined by galaxies
with directly measured black hole masses $\gtrsim$10$^6$ M$_{\odot}$
\citep[e.g.,][]{2004ApJ...615..645O, 2011MNRAS.412.2211G, 2009ApJ...694L.166B}.  In GSD19, we instead
predicted the central black hole masses of our Virgo sample of spiral galaxies
directly from the $M_{\rm bh}$--$\sigma$ and $M_{\rm bh}$--$M_{\rm *,gal}$
relations for spiral galaxies, which have a total root mean square (and intrinsic) scatter of 0.63
(0.51$\pm$0.04) and 0.79 (0.69) dex, respectively, in the $\log M_{\rm
  bh}$-direction.  We additionally predicted the black hole mass using the
host galaxy's spiral arm pitch angle, $\phi$, via the $M_{\rm bh}$--$\phi$
relation which has a scatter of just 0.43 (0.30$\pm$0.08) dex \citep{2017MNRAS.471.2187D}.  We
then highlighted galaxies for which multiple methods, from independent
observations ($\sigma$, $M_{\rm *,gal}$, $\phi$), consistently yielded an
expectation of an IMBH.  Given the absence of
bulges in some late-type spiral galaxies with massive black holes, and the
somewhat comparable levels of scatter about the $M_{\rm bh}$--$M_{\rm *,gal}$
\citep[$\Delta_{\rm rms,total}=0.79$ dex][]{2018ApJ...869..113D} and $M_{\rm
  bh}$--$M_{\rm *,bulge}$ 
\citep[$\Delta_{\rm rms,total}=0.64$--0.66 dex][]{2019ApJ...873...85D} 
relations for spiral galaxies, we have not used 
the $M_{\rm bh}$--$M_{\rm *,bulge}$ relation. 

In Table~\ref{Tab-IMBH}, the predicted black hole masses based on $\phi$ and
$\sigma$ are taken from GSD19.  Due to our use of the
median, rather than the mean, redshift-independent distance in the NASA/IPAC
Extragalactic Database (NED)\footnote{\url{http://nedwww.ipac.caltech.edu}}, 
we have revised the predicted black hole masses from GSD19 that were
based on the $M_{\rm bh}$--$M_{\rm *,gal}$ relation and the mean
redshift-independent distances.
For each galaxy, we inspected the histogram of redshift-independent distances, and for some we
removed outliers and rederived the median value, which is listed in
Table~\ref{Tab-IMBH}.  The revised distances impact upon the absolute
magnitudes and in turn the stellar mass of each galaxy, and thus the predicted
black hole masses.  

For reference, the nucleated S\'ersic galaxy NGC~205 has the lowest directly
measured black hole mass, at $(7.1_{-5.3}^{+10.7})\times10^3\,{\rm M}_{\odot}$
\citep[][their Table~6]{2019ApJ...872..104N}.  With a stellar velocity
dispersion of 33 km s$^{-1}$ from
HyperLeda\footnote{\url{http://leda.univ-lyon1.fr}}
\citep{2003A&A...412...45P}, NGC~205 agrees well with, and thus extends, the
$M_{\rm bh}$--$\sigma$ relation for S\'ersic, and thus spiral, galaxies into
the $10^3$--$10^4$ M$_{\odot}$ range \citep[][see their Figures~3 and
  11]{2019ApJ...887...10S}.  The dwarf S0 galaxy NGC~404, with a reported
black hole mass equal to $(7^{+1.5}_{-2.0})\times10^4$ M$_{\odot}$
\citep{2017ApJ...836..237N}, also follows the $M_{\rm bh}$--$\sigma$ relation
for S\'ersic galaxies \citep[][their Figures~2 and 3]{2019ApJ...887...10S}.
Having a well-resolved nuclear star cluster, with a mass of
$(1.8\pm0.8)\times10^6$ M$_{\odot}$ \citep{2009MNRAS.397.2148G,
  2018ApJ...858..118N}, NGC~205 also agrees well with and extends the $M_{\rm
  bh}$--$M_{\rm nc}$ relation into this lower mass range
\citep{2020MNRAS.492.3263G}. 

\subsubsection{Insight from nuclear star clusters}

For the nucleated galaxies, i.e., those with nuclear star
clusters, we also include the estimate of the central black hole mass derived
from the nuclear star cluster mass. 
As with massive black holes, the masses of nuclear
star clusters have previously been discovered to correlate with their host spheroid's mass
\citep{2003ApJ...582L..79B, 2003AJ....125.2936G}.  Furthermore, the
coexistence of black holes and nuclear star clusters
\citep[][and references therein]{2007ApJ...655...77G, 2008AJ....135..747G, 2008ApJ...678..116S,
  2009MNRAS.397.2148G, 2016IAUS..312..269G} implies the existence of a relation between black
hole mass and nuclear star cluster mass, which is given by 
\begin{equation}
\log \left( \frac{M_{\rm bh}}{{\rm M}_{\odot}}\right) = (2.62\pm0.42)
\log \left( \frac{M_{\rm nc}}{10^{7.83}\,{\rm M}_{\odot}}\right) + (8.22\pm0.20)
\label{Eq-new-sigma}
\end{equation}
\citep{2016IAUS..312..269G, 2020MNRAS.492.3263G}.  This relation holds in the
absence of a galaxy bulge, making it a useful tool for late-type spiral
galaxies.

A nuclear star cluster is known to reside in a couple of our spiral galaxies
with both (i) a central X-ray point-source and (ii) a suspected IMBH.  The
reported nuclear star cluster mass for NGC~4178 \citep[$5\times10^5\,{\rm
    M}_{\odot}$:][]{2009ApJ...704..439S} is assumed to have an accuracy of a
factor of 2.  Although \citet{2009ApJ...704..439S} report that NGC~4713 also
contains a nuclear star cluster, or at least a point-like source (possibly
contaminated by AGN light), they refrain from providing a mass measurement.
In NGC~4498, the Johnson/Cousins $V$-band apparent magnitude of the nuclear
star cluster has been reported as $21.53\pm0.02$~mag
\citep{2014MNRAS.441.3570G}, and the stellar-mass for the nuclear star cluster
has been taken as $(1.4\pm0.4)\times10^6\,{\rm M}_{\odot}$ from \citet[][their
  Table~A1]{2016MNRAS.457.2122G}.  The expected black hole masses, based upon
these nuclear star cluster masses, are calculated here using
equation~\ref{Eq-new-sigma}, taken from \citet[][their
  equation~7]{2020MNRAS.492.3263G}, which is more accurate than the previous
estimates obtained from the inverse of equation~12 in GSD19.  This $M_{\rm
  bh}$--$M_{\rm nc}$ relation is applicable\footnote{The upper mass cut (of
  $M_{\rm nc} \lesssim 5\times10^7 {\rm M}_\odot$) excludes systems with
  half-light radii greater than $\sim$20~pc, which may be regarded as nuclear
  disks rather than ellipsoidal-shaped star clusters.}  for $10^5 \lesssim
M_{\rm nc}/M_{\odot} \lesssim 5\times10^7$, and has an uncertainty calculated
using the expression
\begin{eqnarray}
[\delta \log (M_{\rm bh}/{\rm M}_{\odot})]^2  = 
  \left[ \log\left( \frac{M_{\rm nc}}{10^{7.83}\,{\rm M}_{\odot}}\right)
    \right]^2(0.42)^2 \nonumber \\[3pt]
 \left(\frac{2.62}{\ln10}\right)^2\left(\frac{\delta M_{\rm nc}}{M_{\rm
       nc}}\right)^2  + (0.20)^2 + (\delta_{\rm int})^2, 
\label{Eq-new-err} 
\end{eqnarray}
where the intrinsic scatter within the $M_{\rm bh}$--$M_{\rm nc}$ relation,
$\delta_{\rm int}$, has been taken to be 1.31 dex in the $M_{\rm
  bh}$-direction  \citep{2020MNRAS.492.3263G}.  
The results are given in Table~\ref{Tab-IMBH}.

\subsubsection{Error-weighted mean black hole masses}

A novel approach employed by \citet{2021PASA...38...30D} in the case of
NGC~3319 was to determine the error-weighted black hole mass from
many independent estimates, $\log(M_{{\rm bh},i})$.
Accounting for each estimate's associated uncertainty, $\delta \log(M_{{\rm
    bh},i})$, the combined probability distribution function (PDF) yields the
statistically most likely value (and its 1$\sigma$ uncertainty range) for the
black hole mass.  When many such independent estimates are
brought to bear on this derivation, as was the case for NGC~3319, one has a
rather well-defined (Gaussian-like) PDF from which one can readily establish
the probability of having detected an IMBH with $M_{\rm bh} < 10^5\,{\rm
  M}_{\odot}$.  Here, with fewer black hole estimates per galaxy than for NGC~3319, we proceed along a
simpler path by determining the error-weighted mean of the logarithm of the black
hole masses, such that
\begin{equation}
\overline{\log M_{\rm bh}} = \frac{ \sum_{i=1}^N w_i \log M_{{\rm bh},i}
}{\sum_{i=1}^N w_i}, 
\end{equation}
where we have used inverse-variance weighting\footnote{This weighting gives
  the `maximum likelihood estimate' for the mean of the probability
  distributions under the assumption that they are independent and normally
  distributed with the same mean.}  and thus $w_i = 1/(\delta \log M_{{\rm
    bh},i})^2$.  The 1$\sigma$ standard error bar attached to this 
mean is calculated as
\begin{equation}
\delta \, \overline{\log M_{\rm bh}} = \sqrt{ \frac{1}{\sum_{i=1}^N w_i } }.
\end{equation}   
Table~\ref{Tab-IMBH} provides the mean black hole mass estimates for
our sample of late-type galaxies possessing a centrally-located X-ray
point-source.  With the exception of NGC~4212, with dual X-ray
point-source, the estimates are typically less than $\sim$10$^5$ M$_{\odot}$.

\section{The X-ray Data \& Analysis}\label{Sec-Data} 

Readers not interested in the details of how the data was reduced (Sections
3.1 and 3.2) nor the individual results for each galaxy, may like to skip to
Section 4 which describes the prospects for obtaining black hole masses from
the X-ray data

\subsection{Nuclear point-like source detection}

{\it CXO} Advanced CCD Imaging Spectrometer (ACIS) data were obtained
under the `Spiral galaxies of the Virgo Cluster' Large Project (Proposal ID:
18620568). 
In addition, we used archival observations for some of the  galaxies.
We analyzed the data in a consistent manner with GSD19, employing
the Chandra Interactive Analysis of Observations ({\sc ciao}) Version 4.12
software package \citep{2006SPIE.6270E..1VF},  and Calibration Database
Version 4.9.1.  We reprocessed the event files of every observation with the
{\sc ciao} task {\it chandra\_{repro}}. For galaxies with multiple observations,
we created merged event files with {\it reproject\_{obs}}.  In those cases, we
used the stacked images to improve the signal-to-noise ratio in our search for
possible nuclear sources; however, the fluxes from the nuclear candidates were
then estimated from the individual exposures.

We used the coordinate position of the galactic nuclei reported in NED as a
reference position for our search of nuclear X-ray sources. We looked for
significant X-ray emission within 2$\arcsec$ of the reference nuclear
location. The fact that we knew in advance the (approximate) position of the
sources we were looking for meant that we could identify significant
detections with a far lower number of counts than we would require from a
blind source-finding task (e.g., {\it wavdetect}). That, combined with the
very low background level in the ACIS images, results in 99\% significant
detections even for sources with as few as 5 counts \citep[e.g., see the
  Bayesian confidence intervals in][]{1991ApJ...374..344K}.  As a rough
estimate, 5 ACIS-S counts in a typical 10~ks exposure, correspond to a
0.5--10~kev luminosity of $\sim$2 $\times 10^{38}$ erg s$^{-1}$ at a distance
of 17~Mpc.

When significant X-ray emission was detected at the nuclear position, we estimated
whether the source was consistent with being point-like, or was instead
significantly more extended than the instrumental point-spread-function (PSF)
of the ACIS detector at that location (in most cases, close to the aimpoint of
the S3 chip). In cases where we determined that the emission was extended, we
inspected the images in the soft (0.3--1 keV), medium (1--2 keV) and hard
(2--10 keV) bands separately. This enabled us to determine whether there was a
point-like (harder) X-ray source surrounded by diffuse thermal emission,
characteristic of star-forming regions; typically, the latter does not
significantly contribute to the 2--10 keV band.

For all nuclear point-like sources, we defined source extraction regions with
a radius suitable to the size of the PSF (typically, a circle with 2$\arcsec$
radius for sources at the aimpoint of the S3 chip) and local background
regions at least 4 times larger than the source region. We visually inspected
all source and background regions to make sure they did not contain other
contaminating sources. In all cases, we ran the {\sc ciao} task {\it srcflux}
to estimate the absorbed and unabsorbed fluxes. The PSF fraction in the
extraction circle was estimated with the {\it srcflux} option
`psfmethod=arfcorr', which essentially performs a correction to infinite
aperture.  In some cases, when the count rate was high enough, we also
extracted the source spectra and modeled them using the {\sc xspec}
\citep{1996ASPC..101...17A} package version 12.9.1, as described next.

\subsection{Flux and luminosity of detected nuclear sources}

The task {\it srcflux} provides two alternative estimates of the absorbed X-ray
flux: model fluxes and model-independent fluxes. Both values can be described
as approximations to the ideal ``observed'' flux that would be measurable from
a dataset with an infinitely high signal-to-noise. For the model fluxes, we
assumed a power-law spectrum with photon index $\Gamma = 1.7$ 
\citep{2009MNRAS.399.1293M, 2011A&A...530A..42C} 
and the Galactic
line-of-sight column density of H\,{\footnotesize I} gas
\citep{2016A&A...594A.116H}, taken from the High Energy                                              
Astrophysics Science Archive Research Center
(HEASARC)\footnote{\url{https://heasarc.gsfc.nasa.gov/cgi-bin/Tools/w3nh/w3nh.pl}}. 
More realistically, even for nuclear sources
with the least amount of intrinsic absorption, we may expect a total 
$N_{\rm H}$ value which is a factor of 2 higher than the Galactic $N_{\rm H}$ value, owing
to the absorbing matter in the host Virgo spiral galaxy plus that in the Milky
Way.  This conversion factor would
depend on the size and morphological type of the host galaxy, on its metalicity and
star formation rate, and on our viewing angle. However, the difference in the
estimated unabsorbed luminosities corrected for a column density of, for
example, $\sim$4 $\times 10^{20}$ cm$^{-2}$, as opposed to $\sim$2 $\times
10^{20}$ cm$^{-2}$ is only about 4\% (well below the other observational
and systematic uncertainties), because such column densities block photons
only at the low end of the ACIS-S energy range, where the instrumental
sensitivity is already very low. Thus, we avoided those complications, because
they are largely irrelevant for the purpose of this work, and list the
unabsorbed luminosities as corrected only for Galactic $N_{\rm H}$ in all
cases when there are not enough counts for any significant estimate of $N_{\rm
  H,int}$ (as explained later). It is simple to estimate the fluxes and
luminosities of the same sources corrected for higher values of $N_{\rm H}$
(if so desired), with the Portable Interactive MultiMission
Software ({\sc pimms})\footnote{\url{http://asc.harvard.edu/toolkit/pimms.jsp}}.

Model-independent fluxes from {\it srcflux} are based on the energy of the
detected photons, convolved by the detector response. For sources with a small
number of counts, the detected photons may not uniformly sample the energy
range, especially at higher energies (lower sensitivity): thus, the
model-independent flux is not necessarily a more accurate approximation of the
``ideal'' observable flux than the model-dependent value. Moreover, we need
the model-dependent fluxes in order to estimate the unabsorbed fluxes and
luminosities, a conversion that cannot be done directly from the
model-independent fluxes.
In most cases in our sample of nuclear sources, the model-dependent and
independent fluxes agree within the error bars. However, when they differ
significantly, it is a clue that either our assumed power-law is wrong, or
that the H\,{\footnotesize I} column density is $>$10$^{20}$ cm$^{-2}$. We flagged those cases
for further analysis with {\sc xspec}.

In order to mitigate the effect of an uncertain $N_{\rm H}$ on our estimate of
the unabsorbed flux and luminosity, we computed model-dependent fluxes with
{\it srcflux} in the 1.5--7 kev band rather than in the ``broad'' 0.5--7 keV
band. This is because photo-electric absorption is negligible above 1.5 keV,
at least for the range of column densities seen in Virgo spirals (up to a few
$10^{22}$ cm$^{-2}$). Thus, the observed count rate at 1.5--7 keV provides a
more accurate normalization of the true power-law spectrum. We then compute
the 0.5-7 keV flux by extrapolating the power-law model to lower energies.

There is a second reason why it is more convenient to use the 1.5-7 keV band
rather than the full ACIS band for an estimate of the nuclear fluxes with {\it
  srcflux}. Some nuclei may have thermal plasma emission from diffuse hot gas
(for example caused by star formation in the nuclear region) in addition to
point-like emission from the potential central black hole. Spatial separation of the
diffuse and point-like components is often impossible; two-component modeling
is also not an option for low-count spectra from sources with luminosities
$\lesssim$10$^{40}$ erg s$^{-1}$ at the distance of the Virgo
Cluster. Instead, it is plausible to assume that the power-law component from
the nuclear black hole dominates above 1.5 keV, and the $\sim$0.5-keV thermal
plasma emission affects mostly the softer band.  Thus, by normalizing the power-law
model to the 1.5-7 keV flux and extrapolating it down to lower energies, we
obtain a more accurate estimate of the nuclear emission than if we fit the
power-law model over the whole 0.5--7 keV range.

In Table~\ref{TableSum}, one can find the model-independent fluxes of all the
sources, and the model-dependent fluxes and luminosities of the sources, computed with {\it 
  srcflux}. We converted the unabsorbed 0.5-7 keV fluxes to unabsorbed
luminosities in the same band assuming the distances reported in
Table~\ref{Tab-IMBH}.  Finally, we converted luminosities across different
bands using {\sc pimms}, with the assumed power-law model.  
With $\Gamma = 1.7$, one has that $L_{0.5-8\,{\rm keV}} = 1.075\,L_{0.5-7\,{\rm keV}}$, 
and $L_{0.5-10\,{\rm keV}} = 1.20\,L_{0.5-7\,{\rm keV}}$,

We carried out a full spectral analysis for those nuclear sources with a sufficient
number of counts, and for sources in which our preliminary {\it srcflux}
analysis and our inspection of the X-ray colors suggested evidence of a high
absorbing column density. We extracted spectra and associated response and
ancillary response files with the {\sc ciao} task {\it specextract}. We then
regrouped the spectra to 1 count per bin with the task {\it grppha} from the
{\sc ftools} software \citep{1995ASPC...77..367B}, and modeled them in {\sc xspec}
version 12.9.1 \citep{1996ASPC..101...17A}, 
using the \citet{1979ApJ...228..939C} statistics. The count rate is
generally too low for complex modeling; however, we can spot cases of high
$N_{\rm H, int}$ and constrain its value even for sources with as low as a
dozen counts, because those counts would all be recorded at energies $>$1
keV. The second parameter left free in our {\sc xspec} fitting is the
power-law normalization. In a few cases, we had enough counts to leave also
the photon index as a free parameter; in most other cases, we fixed it at the
canonical value of 1.7. In one case, NGC\,4178, the best-fitting power-law
model is very steep (Table~\ref{TableSum}), and the disk-blackbody model {\it diskbb}
provides a more physical (although statistically equivalent) fit. Finally, for
the sources modeled in {\sc xspec}, we determined the 90\% confidence limits
on their absorbed and unabsorbed model fluxes (and hence, on their unabsorbed
luminosities) with the convolution model {\it cflux}.

\begin{rotatetable*}
\begin{deluxetable*}{lcrccccccccc}
\tablewidth{700pt}
\tablecaption{{\it CXO} observations of galaxies with a central X-ray point-source, and which are additionally expected to house a central black hole with mass $\lesssim 10^5\,{\rm M}_{\odot}$\label{TableSum}}
\tabletypesize{\scriptsize}
\tablehead{
\colhead{Galaxy}  &  \colhead{Obs.\ Date}  &  \colhead{Exp.}  &  \colhead{$F_{0.5-7\,{\rm keV}}$}  &  \colhead{$F_{0.5-7\,{\rm keV}}$}  &  \colhead{N$_{\rm H,Galaxy}$}  &  \colhead{$L_{0.5-8\,{\rm keV}}$}  &  \colhead{N$_{\rm H,intrin}$}  &  \colhead{$\Gamma$}  &  \colhead{$kT_{\rm in}$}  &  \colhead{$L_{0.5-10\,{\rm keV}}$}  &  \colhead{$L_{2-10\,{\rm keV}}$} \\
\colhead{NGC\#}  &  \colhead{}  &  \colhead{ksec}  &  \colhead{mod-indpt} &  \colhead{mod-dept}   &  \colhead{$10^{20}$ cm$^{-2}$}  &  \colhead{$10^{38}$ erg s$^{-1}$}  &  \colhead{$10^{22}$ cm$^{-2}$}   &   \colhead{}  &   \colhead{keV}  &  \colhead{$10^{38}$ erg s$^{-1}$} &  \colhead{$10^{38}$ erg s$^{-1}$}
}
\colnumbers
\startdata
\multicolumn{12}{c}{Presented in \citet{2019MNRAS.484..814G}} \\  
\,\,\,4178 & 2011-02-19      & 36.29 &    0.50$^{+0.16}_{-0.12}$  & $0.56^{+0.25}_{-0.16}$  &    2.66        &  4.04$^{+16.86}_{-2.30}$  & 0.47$^{+0.50}_{-0.35}$ & 3.43$^{+1.66}_{-1.24}$ &      ...            &  4.06$^{+16.94}_{-2.31}$  & 0.51$^{+2.12}_{-0.29}$  \\
\,\,\,4178 & 2011-02-19      & 36.29 &    0.50$^{+0.16}_{-0.12}$  & $0.56^{+0.25}_{-0.16}$  &    2.66        &  1.52$^{+1.25}_{-0.56}$   & 0.15$^{+0.33}_{-0.15}$ &      ...            & 0.56$^{+0.35}_{-0.19}$ &  1.52$^{+1.25}_{-0.56}$   & 0.35$^{+0.30}_{-0.13}$  \\
4713           & 2003-01-28  &  4.90 &    1.17$^{+0.74}_{-0.52}$  & $1.52^{+0.91}_{-0.66}$ &     1.87        &  4.40$^{+2.62}_{-1.92}$ &      ...   &      1.7            &      ...            &  4.94$^{+2.94}_{-2.15}$   & 3.19$^{+1.90}_{-1.39}$   \\
4470$^a$       & 2010-11-20  & 19.78 &    0.42$^{+0.47}_{-0.33}$  & $0.35^{+0.28}_{-0.20}$  &    1.60        &  1.31$^{+1.09}_{-0.74}$  &      ...            &      1.7            &      ...          &  1.47$^{+1.23}_{-0.83}$   & 0.95$^{+0.79}_{-0.64}$   \\
\multicolumn{12}{c}{New X-ray data} \\
4197           & 2018-07-27  &  7.96 &   11.20$^{+2.40}_{-2.36}$  & $10.60^{+3.20}_{-1.40}$ &    1.52        &   119$^{+55}_{-31}$  & 0.35$^{+0.72}_{-0.35}$ & 1.24$^{+0.84}_{-0.69}$ &    ...          &  144$^{+66}_{-38}$  & 113$^{+53}_{-29}$ \\
4212$^a$       & 2017-02-14  & 14.86 &    0.43$^{+0.36}_{-0.24}$  & $0.49^{+0.41}_{-0.27}$ &     2.67        &  1.90$^{+1.58}_{-1.03}$  &      ...            &     1.7            &       ...          &  2.13$^{+1.78}_{-1.15}$  & 1.38$^{+1.14}_{-0.75}$ \\
4298           & 2018-04-09  &  7.81 &    2.68$^{+1.54}_{-1.13}$  & $1.84^{+1.04}_{-0.76}$  &    2.62        &  6.10$^{+3.46}_{-2.51}$ &     ...             &      1.7            &      ...           &  6.85$^{+3.89}_{-2.82}$  & 4.42$^{+2.51}_{-1.82}$  \\
4313$^a$       & 2018-04-14  &  7.96 &    0.55$^{+0.51}_{-0.32}$  & $0.87^{+0.77}_{-0.48}$  &    2.40        &  2.32$^{+2.03}_{-1.29}$  &     ...             &     1.7            &      ...           &  2.61$^{+2.28}_{-1.45}$   & 1.68$^{+1.47}_{-0.93}$  \\
4330           & 2018-04-16  &  7.96 &    6.30$^{+2.35}_{-1.90}$  & $6.55^{+2.70}_{-2.08}$  &    2.07        &  80.7$^{+48.6}_{-31.7}$ & 4.33$^{+2.86}_{-1.96}$ &   1.7            &        ...         &  90.6$^{+54.6}_{-35.6}$  & 58.5$^{+35.2}_{-23.0}$  \\
4405           & 2018-04-09  &  7.96 &    0.62$^{+0.57}_{-0.36}$  & $0.84^{+0.76}_{-0.48}$  &    2.17        &  3.54$^{+3.22}_{-2.03}$  &       ...           &     1.7            &       ...          &  3.98$^{+3.61}_{-2.28}$    & 2.57$^{+2.33}_{-1.48}$   \\
4413           & 2018-05-10  &  7.80 &    0.36$^{+0.33}_{-0.22}$  & $1.00^{+0.84}_{-0.54}$ &     2.32        &  3.41$^{+2.85}_{-1.84}$  &       ...           &     1.7            &       ...          &  3.83$^{+3.76}_{-2.07}$  & 2.47$^{+2.07}_{-1.33}$   \\
\,\,\,4492$^a$ & 2007-02-22  &  4.89 &    0.69$^{+0.63}_{-0.40}$  & $1.31^{+1.01}_{-0.66}$  &    1.43        &  6.77$^{+6.58}_{-3.58}$ & 0.05$^{+0.25}_{-0.05}$ &   1.7            &       ...          &  7.60$^{+7.39}_{-4.02}$  & 4.91$^{+4.77}_{-2.60}$   \\
\,\,\,4492$^a$ & 2014-04-25  & 29.68 &    0.54$^{+0.30}_{-0.22}$  & $0.85^{+0.46}_{-0.34}$  &    1.43        &  4.40$^{+2.81}_{-1.89}$ & 0.05$^{+0.25}_{-0.05}$ &   1.7            &       ...          &  4.94$^{+3.16}_{-2.12}$  & 3.19$^{+2.04}_{-1.37}$   \\
4498           & 2018-04-07  &  8.09 &    0.60$^{+0.44}_{-0.30}$  & $1.17^{+0.85}_{-0.57}$  &    2.25        &  3.29$^{+2.37}_{-1.62}$  &      ...            &     1.7            &       ...          &  3.69$^{+2.67}_{-1.81}$  & 2.39$^{+1.71}_{-1.18}$   \\
4519           & 2018-05-05  &  8.45 &    1.10$^{+0.69}_{-0.49}$  & $1.39^{+0.89}_{-0.61}$  &    1.36        &  7.02$^{+4.44}_{-3.12}$ &      ...            &      1.7            &       ...          &  7.88$^{+4.99}_{-3.50}$  & 5.09$^{+3.22}_{-2.26}$   \\
4607           & 2018-05-09  &  7.96 &    4.93$^{+2.22}_{-1.92}$  & $3.84^{+2.20}_{-1.59}$  &    2.53        &  51.0$^{+49.3}_{-26.3}$ & 4.68$^{+5.86}_{-2.64}$ &   1.7            &       ...          &  57.3$^{+55.3}_{-29.6}$  & 37.9$^{+34.8}_{-20.0}$   \\
\enddata
\tablecomments{Column~1: Galaxy name. 
$^a$ For galaxies with a dual X-ray point-source, we are reporting the flux from the more central source. 
Column~4:  The observed, i.e., partially-absorbed, model-independent photon flux is for the
centrally-located X-ray point-source in the {\it CXO}/ACIS-S 0.5--7 keV bands.  
The units are $10^{-14}$ erg cm$^{-2}$ s$^{-1}$, and the associated uncertainties show the 90\% confidence range. 
Column~5: Absorbed model-dependent flux in units of $10^{-14}$ erg cm$^{-2}$ s$^{-1}$. 
Column~6:  Galactic column density of neutral atomic hydrogen, H\,{\footnotesize I}. 
Column~7:  The X-ray luminosity 
$L_{0.5-8\,{\rm keV}}$ represents the unabsorbed 0.5--8 keV 
luminosity of each point-source, derived using the distances provided 
in Table~\ref{Tab-IMBH} and corrected for our Galaxy's obscuring
H\,{\footnotesize I} plus (when indicated in column~8) the obscuring line-of-sight H\,{\footnotesize I} intrinsic to the external galaxy. 
Column~9:  The measured or adopted X-ray SED power-law slope, $\Gamma$.  
Column~10 gives the blackbody temperature of the model's inner disk  ({\it diskbb}). 
Column~11: Unabsorbed X-ray luminosity $L_{0.5-10\,{\rm keV}}$ from 0.5--10
keV based on the extrapolated power-law SED.
Column~12: Similar to column~11 but from 2--10 keV.
}
\end{deluxetable*}
\end{rotatetable*}

\subsection{Galaxies with archived X-ray data reported in GSD19}

Three galaxies (NGC~4178, NGC~4713, and NGC~4470) 
from GDS19 had both archival X-ray data revealing a central
X-ray point-source and at least two estimates for $M_{\rm bh} < 10^5\,{\rm
  M}_{\odot}$.  NGC~4178 (GSD19, their 
Figure~11) represents a somewhat edge-on counterpart to NGC~4713
(GSD19, their Figure~13), 
for which some additional comments are provided next.  In the case of
NGC~4470, it was not the primary target of the past {\it CXO} observations,
and as such it was always located several arcminutes from the aimpoint,
resulting in a broadened Point Spread Function (PSF) at this galaxy's center.
Below, we present the X-ray contours for NGC~4470, not previously shown in GSD19.

\subsubsection{NGC 4178: blackbody versus power-law}

As noted, NGC~4178 was presented in GSD19. Attempts to fit a power-law model
to the X-ray SED, with its rapid drop off from the soft to the hard energy
band, resulted in an unrealistically steep slope (see Table~\ref{TableSum}).
The X-ray SED was instead quite well fit with a blackbody disk model having an
intrinsic temperature T$_{\rm intrin}$ 0.56$^{+0.35}_{-0.19}$ keV.  The inner-disk
radius $R_{\rm {in}}$ associated with the {\it diskbb} fit is such that
$R_{\rm {in}} (\cos \theta )^{1/2} \approx 1.19 N^{1/2}_{\rm {dbb}} \,d_{10}$.
With $N_{\rm {dbb}}$ the normalization of the {\it diskbb} model in {\tt
  xspec}, $d_{10}$ the distance to the source in units of 10 kpc, and $\theta$
our viewing angle ($\theta = 0$ is face-on), we obtain $R_{\rm {in}} (\cos
\theta )^{1/2} \approx$ 104 km (53--613 km, 90\% confidence).

We compared the power-law and disk-blackbody models using the Anderson-Darling
(AD) test statistics \citep[e.g.,][]{2011hxra.book.....A}. Specifically, we
performed Monte Carlo calculations of the goodness-of-fit in {\sc xspec}, with
the command {\it goodness}, and compared the percentage of simulations with
the test statistic less than that for the data. For the power-law model, 
75\% of the realizations were better than the AD test value for the
data ($\log$ AD $= -3.67$).  For the disk-blackbody model, 43\% of
realizations were better than the AD value ($\log$ AD $= -3.88$). Thus, the
disk-blackbody model is only weakly preferred.
Moreover, 
\citet{2009ApJ...704..439S} have reported clear 
[Ne\,{\footnotesize V}] emission, indicative of an ionizing AGN, in NGC~4178.

\begin{figure}
 \includegraphics[angle=0, width=1.0\columnwidth]{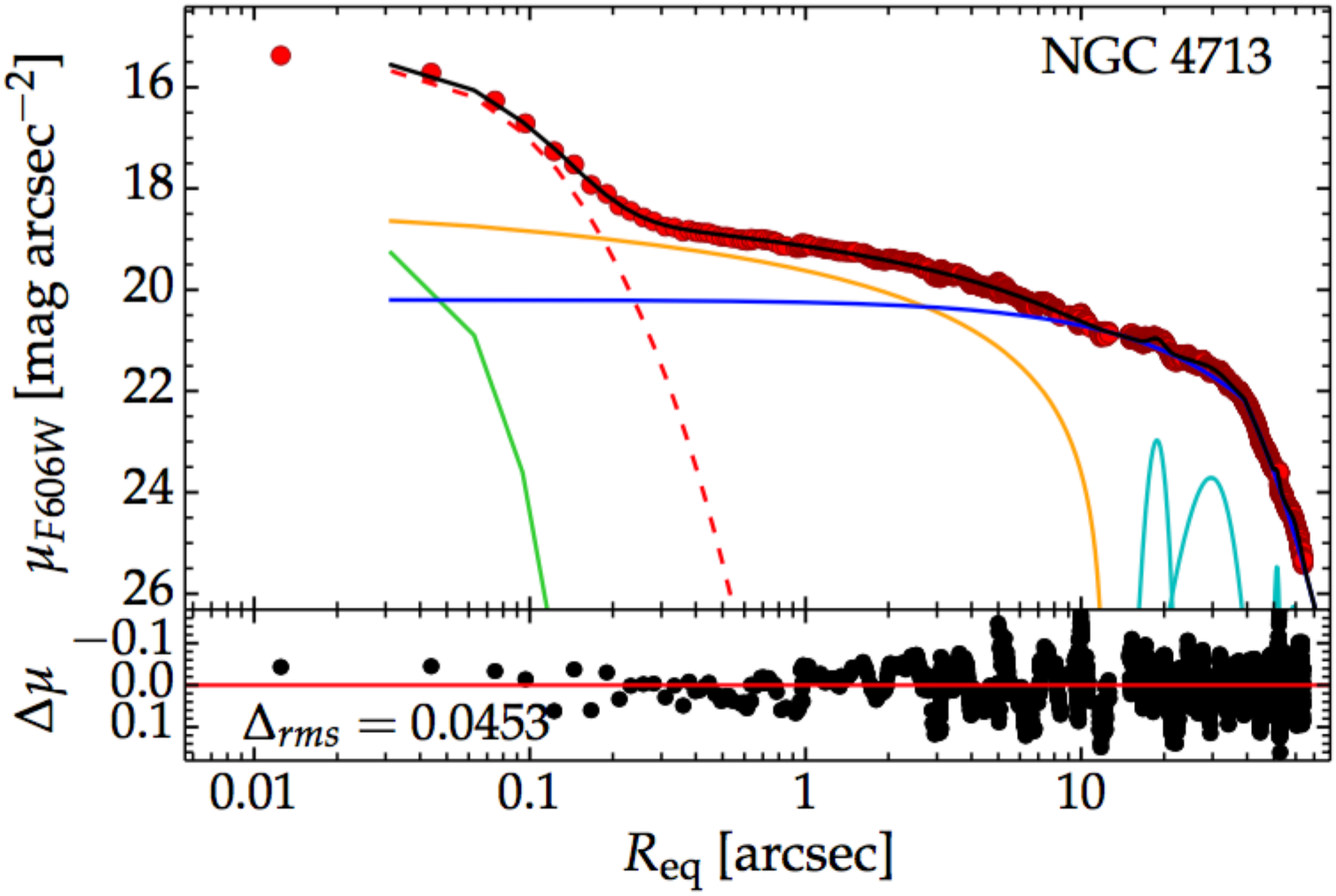} 
 \caption{
Geometric-mean axis, aka 
equivalent axis, light profile for the bulgeless galaxy NGC~4713, fit with a truncated
   exponential disk (dark blue) plus some faint spiral arm-crossings (light
   blue), a bar (orange), a nuclear star cluster (dashed red) and a very
   faint point-source (green).}
\label{Fig-4713-prof}
\end{figure}

\subsubsection{NGC 4713: a LEDA~87300 analog} 

Both the image of NGC~4713 (GSD19, their Figure~13) and its light profile
(Figure~\ref{Fig-4713-prof}), resembles LEDA~87300 \citep[][see their Figures~2 and
  5]{2015ApJ...809L..14B, 2016ApJ...818..172G}.  From {\it Hubble Space
  Telescope} ({\it HST}) images, both galaxies can be seen to contain a central
point-source and a bar with spiral arms emanating from the ends
\citep{2017ApJ...850..196B}.  The better spatial resolution provided by {\it
  HST} has removed the uncertainty between the bar plus bulge
components --- collectively referred to as the `barge'
\citep{2016ApJ...818..172G} --- that was previously affecting the ground-based
images.  Both galaxies now appear to be bulgeless.  The
centrally-located point-source in the optical image of LEDA~87300 may be
partly due to its active galactic nucleus (AGN), which was bright enough to
enable \citet{1997ApJS..112..315H} to flag this galaxy as having a `transition
nucleus' with a luminosity-weighted [O\,{\footnotesize I}] strength intermediate between
H\,{\footnotesize II} nuclei and LINERS (low-ionization nuclear emission-line
regions).  NGC~4713 was subsequently flagged by \citet{2007MNRAS.381..136D} as
having a LINER/H\,{\footnotesize II} nucleus.

\begin{figure}
 \includegraphics[angle=0, width=1.0\columnwidth]{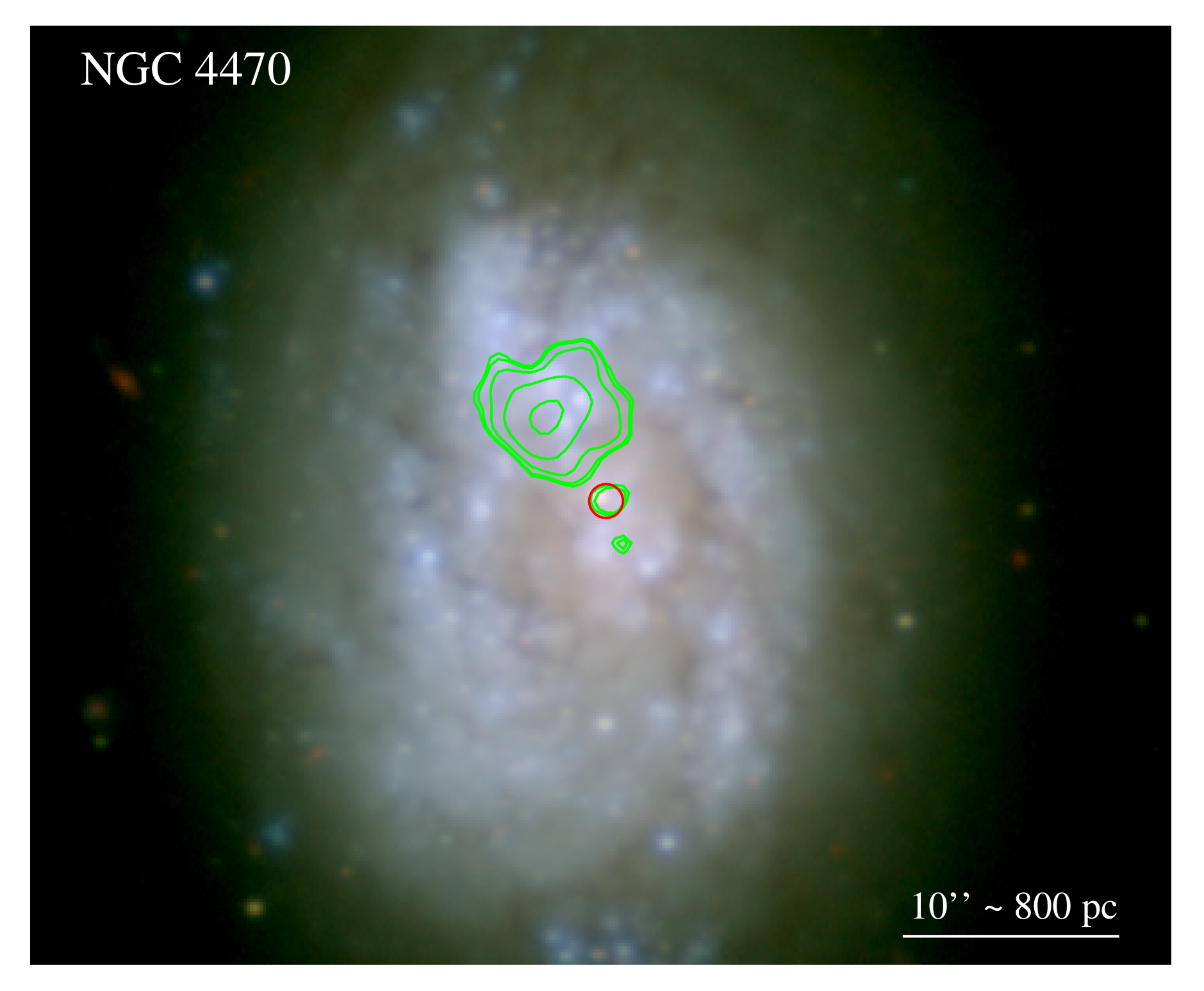}
 \caption{Next Generation Virgo Cluster Survey
   \citep[NGVS:][]{2012ApJS..200....4F} image of NGC~4470, aka NGC~4610, (red = $i$ filter;
   green = $g$; blue = $u^*$), with {\it Chandra}/ACIS-S contours (0.5--7.0
   keV band) overlaid in green.  The contours are just a visual device to show
   the location of the X-ray source (accurate to $\approx$0$\farcs$6).
   North is up, east is to the left.  The red circle shows the NED-provided
   position for the galaxy's optical nucleus, and it has a radius of 1$\arcsec$ in
   this and subsequent figures, roughly reflecting the associated
   uncertainty/range coming from different galaxy isophotes.} 
\label{Fig-4470}
\end{figure}

We have modeled the distribution of HST/WFC/ACS/F606W  light in 
NGC~4713 following the process described in 
\citet{2019ApJ...873...85D}.  We used the {\it Isofit} task 
\citep{2015ApJ...810..120C}, run within the Image Reduction and Analysis
Facility ({\sc IRAF})\footnote{\url{http://ast.noao.edu/data/software}}
package {\sc ellipse} to capture the galaxy light.  We then modeled this
using the {\sc Profiler} software \citep{2016PASA...33...62C}
and an empirical PSF (with airy rings) measured from a star.  The result is
shown in Figure~\ref{Fig-4713-prof}. 

We have found that NGC~4713 contains a slightly resolved nuclear star
cluster with an equivalent-axis half light radius equal to 0$\farcs$07 
(5~pc), a S\'ersic index $n=1.23$, and an {\it F606W} apparent magnitude of
19.57$\pm$0.18 mag (AB mag). 
Correcting for 0.06~mag of Galactic extinction, and using a distance modulus
of 30.85, one has an absolute magnitude of $-$11.34 mag (AB). 
Using an absolute magnitude for the Sun of $\mathfrak{M}_{\odot,F606W} = 4.72$
mag, and a stellar
mass-to-light ratio\footnote{Without a color for the nuclear star cluster, we
  note that the galaxy has a (Galactic absorption)-corrected color of 
$B_T - V_T = (12.19-0.101) - (11.72-0.077) = 0.446$, and
  $B_{F435W}-V_{F606W} = 0.446$ corresponds to $M/L_{F606W}=0.98$ 
  \citep[][their Eq.~2]{2013MNRAS.431..430W}.}  of 1.0$\pm$0.5, 
we obtain 
$\log(M_{\rm nc}/M_{\odot}) = 6.43 \pm 0.31$ and a predicted black hole mass
$\log(M_{\rm bh}/M_{\odot}) = 4.56 \pm 1.66$ 
from Eq.~\ref{Eq-new-sigma}. 

A point-source for AGN light was additionally included in the modeling, but
we were not able to obtain a useful constraint.  Removal of the point-source
shown in Figure~\ref{Fig-4713-prof} brightens the star cluster by 0.01~mag.
If the AGN point-source is 
brighter than that shown, then the nuclear star 
cluster will be fainter and the predicted black hole mass will be smaller.
This may explain why this prediction for the black hole mass is an order of magnitude 
higher than the predictions from the other methods (see Table~\ref{Tab-IMBH}),
although the error bars are large. 
Finally, we note that $-$11.34 mag (AB) corresponds\footnote{AB$_{\nu}
  \, [{\rm mag}] = -2.5
  \log(f_{\nu}$ [erg cm$^{-2}$ s$^{-1}$ Hz$^{-1}$])$ - 48.60$.} 
to an $F606W$ luminosity of $7.4\times10^{39}$ erg s$^{-1}$, which is $\sim$17
times brighter than the 0.5--8 keV X-ray luminosity.  Should the IMBH
candidate have $L_{\rm opt} \approx L_X$, then the nuclear star cluster light will
dominate the 606~nm continuum, as observed. 

LEDA~87300 is of interest because of its AGN, 
evidenced by its nuclear X-ray
point-source, broad H$\alpha$ emission, and narrow emission line ratios 
\citep{2015ApJ...809L..14B}.  Using a virial $f$-factor of $2.3^{+0.9}_{-0.6}$
from \citet{2011MNRAS.412.2211G} gives a virial black hole mass of
$2.9^{+6.7}_{-2.3}\times10^4 f_{2.3}$ M$_{\odot}$ in LEDA 87300 \citep{2016ApJ...818..172G}. 
The heightened uncertainty on the black hole mass, with its 1$\sigma$ error
range from $0.6\times10^4$ to $10^5$, is because the $f$-factor is the mean
value derived from $\sim$30 AGN with reverberation mappings, and when using this
value to predict 
the virial black hole mass for an individual galaxy like LEDA~87300, 
in addition to the observational measurement errors, one needs
to fold in the intrinsic scatter between the individual AGN, which is roughly a
factor of 3, coming from the scatter in the $M_{\rm bh}$--$\sigma$ diagram.

While LEDA~87300 has a stellar mass of $2.4\times10^9\,{\rm M}_{\odot}$
(accurate to a factor of 2) and an estimated stellar velocity dispersion of
$40\pm11$ km s$^{-1}$ \citep[][see their section~3.2]{2016ApJ...818..172G},
NGC~4713 has a stellar mass of $4\times10^9\,{\rm M}_{\odot}$ (GSD19) and a
measured velocity dispersion of $23.2\pm8.9$ km s$^{-1}$
\citep{2009ApJS..183....1H}.  We will endeavor to obtain optical spectra of
NGC~4713 to detect a broad H$\alpha$ line.  From this, we would be able to derive a
virial mass for the black hole in NGC~4713 associated with the nuclear X-ray
point-source and LINER/H\,{\footnotesize II} nucleus. As noted in GSD19, the
X-ray photons from the central point-source in the archived {\it CXO}/ACIS-S
image of NGC~4713 were detected in all three standard bands (soft, 0.3-1 keV;
medium, 1-2 keV; hard, 2-7 keV), consistent with a power-law spectrum
rather than purely a blackbody spectrum.

\subsubsection{NGC 4470: dual X-ray point sources 170~pc apart}\label{Sec-N4470}

NGC~4470 (aka NGC~4610) is a face-on spiral galaxy (Figure~\ref{Fig-4470}).  The Reference
Catalog of galaxy Spectral Energy Distributions
\citep[RCSED:][]{2017ApJS..228...14C}\footnote{\url{http://rcsed.sai.msu.ru/catalog}}
places NGC~4470 in the H\,{\footnotesize II} region of the narrow-line 
[O\,{\footnotesize III}]/H$\beta$ versus [N\,{\footnotesize II}]/H$\alpha$ diagnostic diagram.
However, it is becoming increasingly apparent that faint or `hidden' AGN can
be missed when using BPT \citep{1981PASP...93....5B} diagnostic diagrams
\citep[e.g.,][]{2005ApJ...627..711Z, 2015MNRAS.454.3722S, 2017MNRAS.467..540L,
  2019ApJ...870L...2C, cann2021relics}.  This is perhaps not surprising in low-mass galaxies
because, unless the Eddington ratio is high, the AGN signal in the central
aperture/fibre/spaxel will be swamped by the galaxy's starlight
in these systems with low black hole masses \citep{2020ApJ...898L..30M}.  Although, by concentrating on a
nearby ($D\le80$~Mpc) sample of dwarf galaxies, \citet{2014AJ....148..136M}
did find 28 galaxies dominated by narrow emission line (Type 2) AGN, and
assuming an [O\,{\footnotesize III}]-to-bolometric luminosity correction factor
of 1000 they reported minimum black hole masses of $10^3$--$10^6$ M$_{\odot}$
for their sample.

\begin{figure}
 \includegraphics[angle=0, trim=0.0cm 0.0cm 0.cm 0cm, width=1.0\columnwidth]{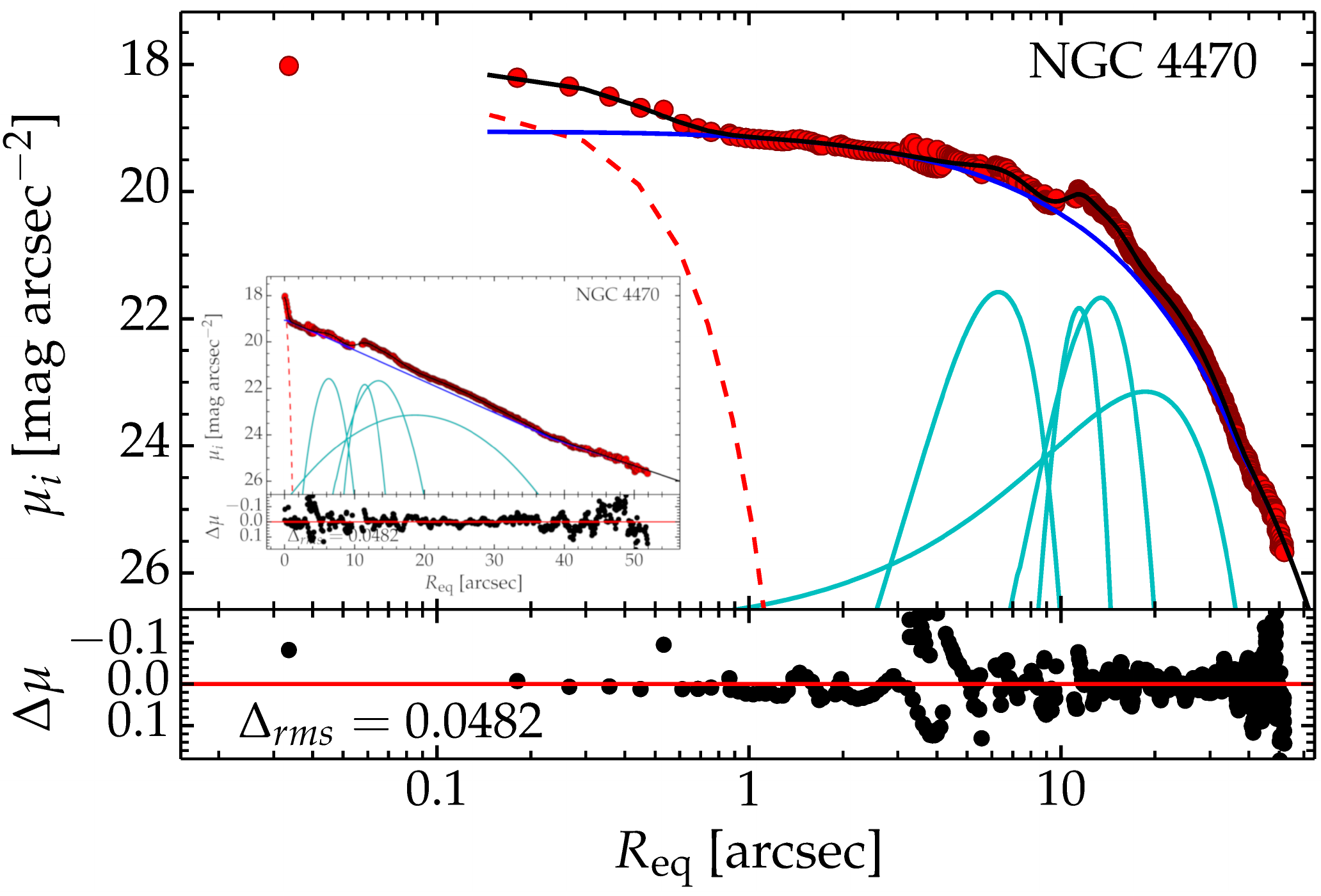}
 \caption{Similar to Figure~\ref{Fig-4713-prof}, but for NGC~4470 (aka NGC~4610).}
\label{Fig-4470-prof}
\end{figure}

The RCSED reports a velocity dispersion of $61\pm6$ km s$^{-1}$ for
NGC~4470 (SDSS J122937.77+074927.1).  This is lower than the value of
$90\pm13.5$ km s$^{-1}$ that was used in GSD19, and results in a lower
($M_{\rm bh}$--$\sigma$)-derived black hole mass of $\log M_{\rm bh} =
5.1\pm0.8$.  This mass is now consistent (overlapping uncertainties) with the
($M_{\rm bh}$--$M_{*,gal}$)-derived value of $\log M_{\rm bh} = 4.1\pm1.0$
(GSD19).

Given that the pre-existing {\it CXO} data for NGC~4470 was
(intentionally)\footnote{The primary {\it CXO} target was NGC~4472 and its halo.}
offset from the aimpoint of the telescope, it required a careful reanalysis.
Six of the past seven ACIS observations (spanning 2010 to 2019) only captured
the nuclear region of NGC~4470 on the external chips, where the PSF was
unfortunately too broad and distorted to obtain a reliable flux measurement.
However, a $\sim$20~ks exposure from 2010 ({\it CXO} Obs.\ ID 12978), directed
4 arcminutes away at the globular cluster RZ~2109 around NGC~4472, proved
fruitful, and we have re-analyzed these data to report on NGC~4470's central
X-ray point-source (Table~\ref{TableSum}).  The galaxy's optical center, as
given by NED, coincides with a red feature which we cannot resolve in the
NGVS image with 0$\farcs$7 seeing. Figure~\ref{Fig-4470} 
displays the overlapping X-ray point-source at this central location.  Both
this X-ray source and the brighter source to the south proved too faint to
acquire a spectrum.

Fitting a point-source to the CFHT/NGVS/$i$-band data, see
Figure~\ref{Fig-4470-prof}, yields a luminosity for the star cluster of 
$\log L_i/L_{\odot,i}=6.57\pm0.35$, based on $\mathfrak{M}_{\odot,i} = 4.58$. 
With an $i$-band mass-to-light ratio of 
0.70$\pm$0.04, based on a galaxy $g-i$ color equal to 0.69$\pm$0.03 and using
the color-dependent stellar mass-to-light ratios from
\citet{2015MNRAS.452.3209R}, the corresponding stellar-mass of the nuclear
star cluster is $\log M_{\rm nc}/M_{\odot}=6.42\pm0.35$, and the predicted
black hole mass is $\log M/M_{\odot}=4.53\pm1.72$
(Equation~\ref{Eq-new-sigma}).  However, this may be an upper limit due to
contamination by AGN light increasing the light that we have assigned to the
star cluster. That is, we have effectively erred on the side of caution and
are not under-predicting the black hole mass in an attempt to predict/find IMBHs.
We also modeled the galaxy components in both the $g$- and $i$-band NGVS
images, and we measured a $g-i$ color equal to 0.57 for the nuclear component.
This resulted in a 23\% smaller stellar-mass estimate for the star cluster,
and a 50\% smaller estimate for the black hole mass. Although, this color for
the nuclear component may be influenced by AGN light and as such we have
erred on the side of caution and adopted the former measurement. 

There is another equally bright X-ray source 2$\farcs$1 (170~pc) to the south, and
a more extended source ($\sim$10$^{39}$ erg s$^{-1}$) located 6$\arcsec$
north-east of the nuclear position and associated with an excess of blue stars and
ongoing star formation.  Based on the location of the second X-ray
point-source, at 170~pc from the nucleus, it may be a stellar-mass ULX, 
but is potentially more interesting than that if it represents one 
half of a dual IMBH system.  A longer {\it CXO} exposure with the aimpoint on
NGC~4470 would enable this to be answered.

\begin{figure*}
$
\begin{array}{cc}
 \includegraphics[angle=0, trim=0cm 0.0cm 0cm 0cm, width=1.0\columnwidth]{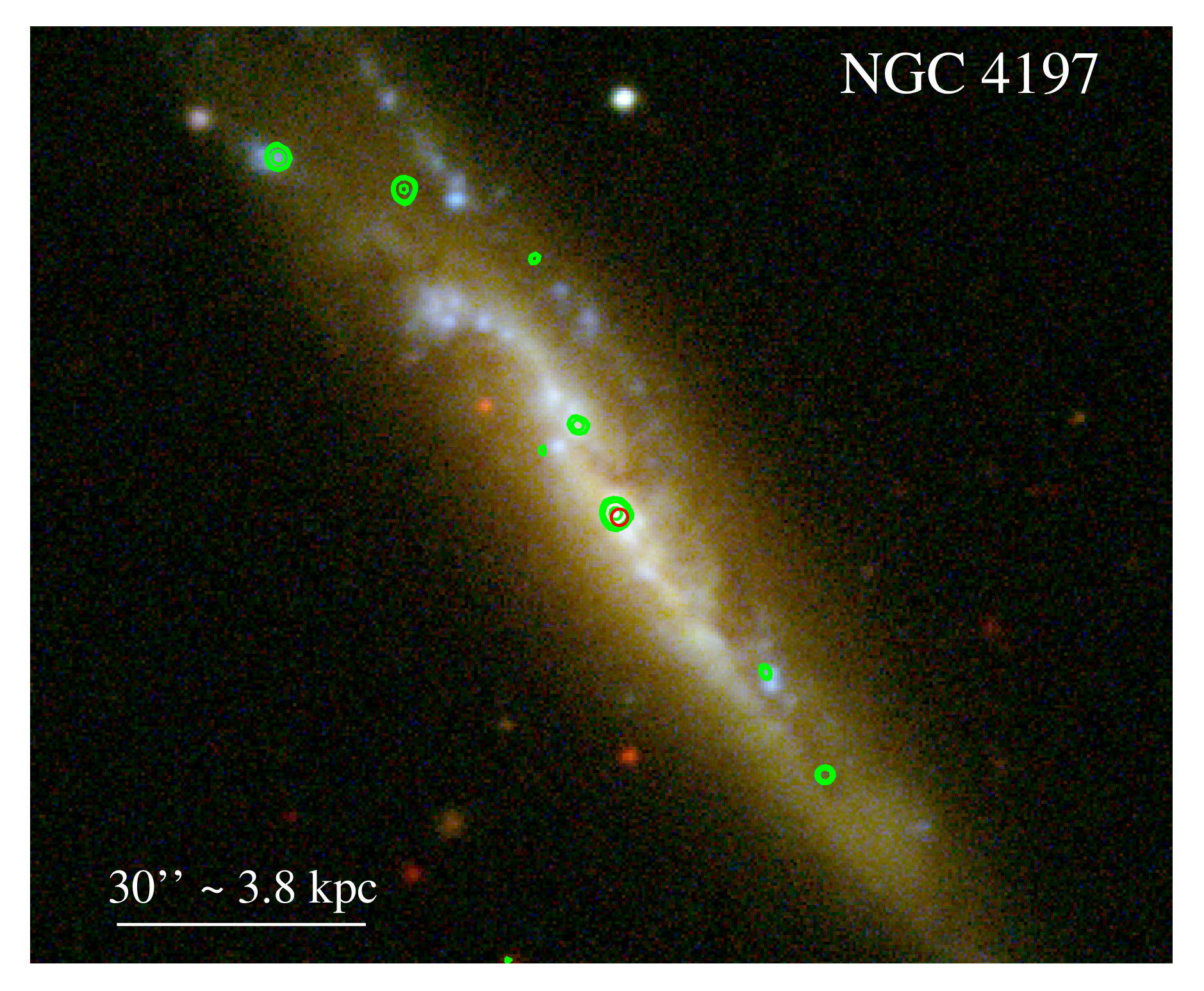} & 
 \includegraphics[angle=0, trim=0cm 0.0cm 0cm 0cm, width=1.0\columnwidth]{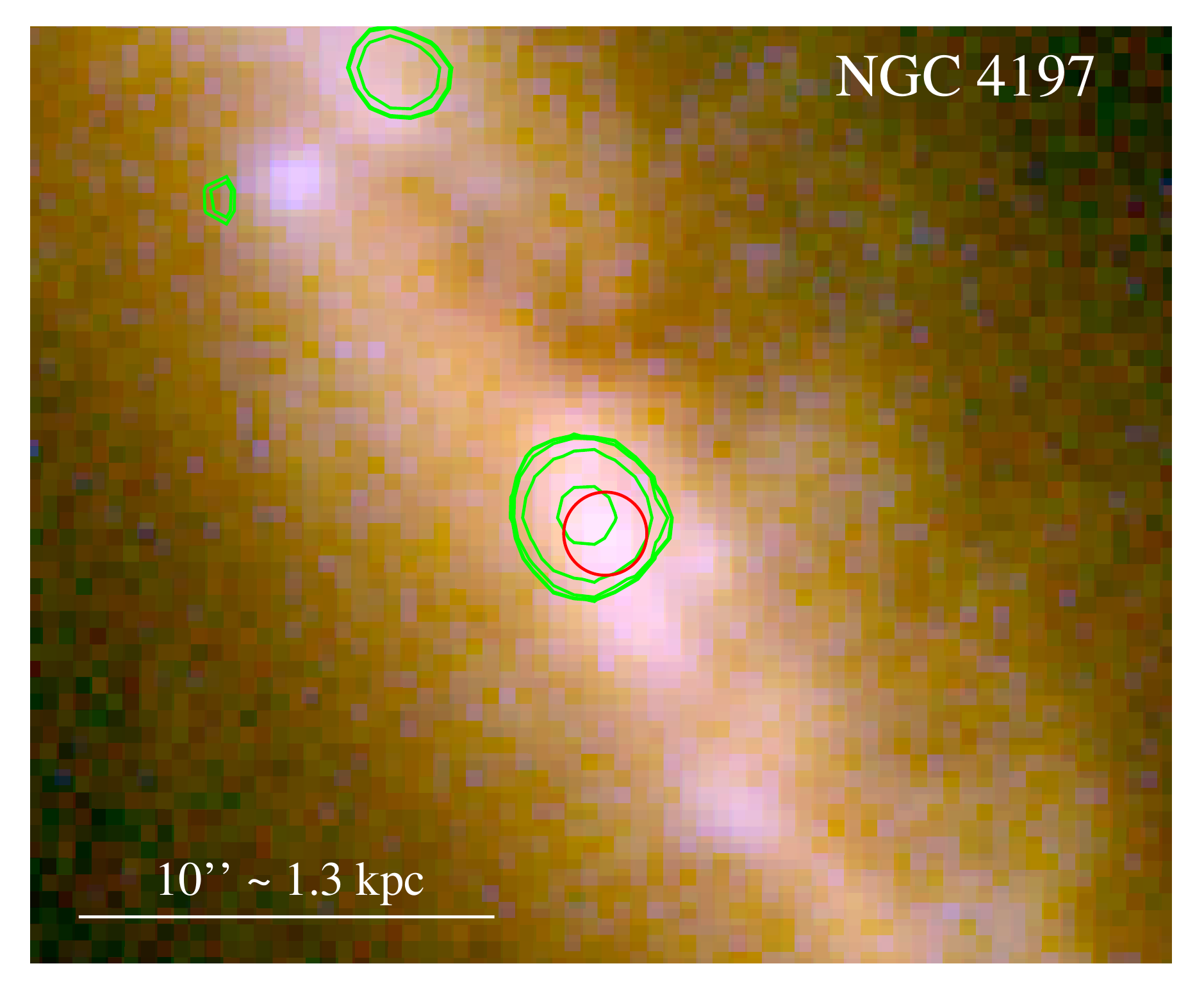} \\
\end{array} 
$
 \caption{Left: Similar to Figure~\ref{Fig-4470}, but 
displaying a Sloan Digital Sky Survey \citep[SDSS][]{2015ApJS..219...12A} image of NGC~4197
(red = $i^{\prime}$ filter; green = $g^{\prime}$; blue = $u^{\prime}$). 
 Right: Zoom-in on the inner region.} 
\label{Fig-4197}
\end{figure*}

\begin{figure}
 \includegraphics[angle=270, trim=2cm 1cm 0cm 4cm, width=1.0\columnwidth]{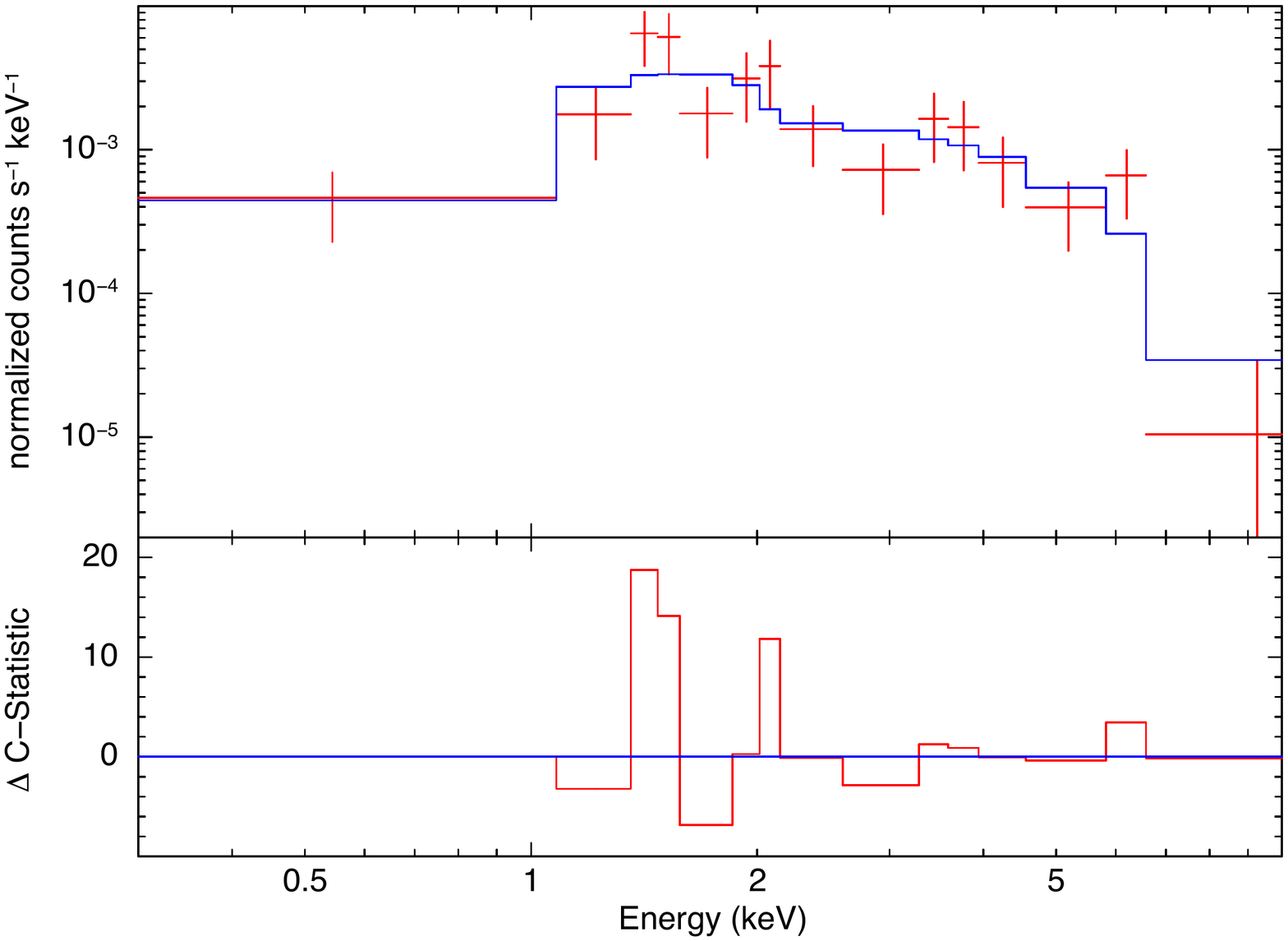}
 \caption{{\it CXO}/ACIS-S spectrum of the nuclear source in NGC~4197,
   fit with a power-law model. The datapoints have been grouped to a
   signal-to-noise $>$1.8 for plotting purposes only. The fit was done on the
   individual counts, using Cash statistics. See Section~\ref{Sec-4197} for the fit
   parameters.}
\label{Fig-ngc4197-spec}
\end{figure}

\subsection{Galaxies with new X-ray data}

\subsubsection{NGC~4197: a likely bright IMBH}\label{Sec-4197}

As with NGC~4178 above, NGC~4197 appears in the flat galaxy catalog of
\citet{1993AN....314...97K} due to the somewhat edge-on (inclination = 79
degrees) orientation of its disk relative to our line-of-sight.
\citet{1985ApJS...57..643D} have reported weak H$\alpha$ emission coming
from the nucleus of this galaxy.
 
As a part of our Virgo
cluster X-ray survey, NGC~4197 was observed by {\it CXO} for $\approx$8 ks, on 2017
July 26 \citep[for more details, see][]{Soria2021}.  
We find a strong, point-like X-ray source (Figure~\ref{Fig-4197})
located at RA $= 12^{\mathrm{h}}$ 14$^{\mathrm{m}}$ 38$^{\mathrm{s}}$.59, Dec
$= +05^{\circ}$ 48$^{\prime}$ 21$^{\prime\prime}$.2 [J2000.0]. 
Considering the scatter in the positions reported by NED, 
this is consistent with the position of the optical nucleus: 
it is $\approx$0$\farcs$7 ($\approx$90~pc) away from the $r$-band SDSS position.

\begin{figure*}
$
\begin{array}{cc}
 \includegraphics[angle=0, trim=0cm 0.0cm 0cm 0cm, width=1.0\columnwidth]{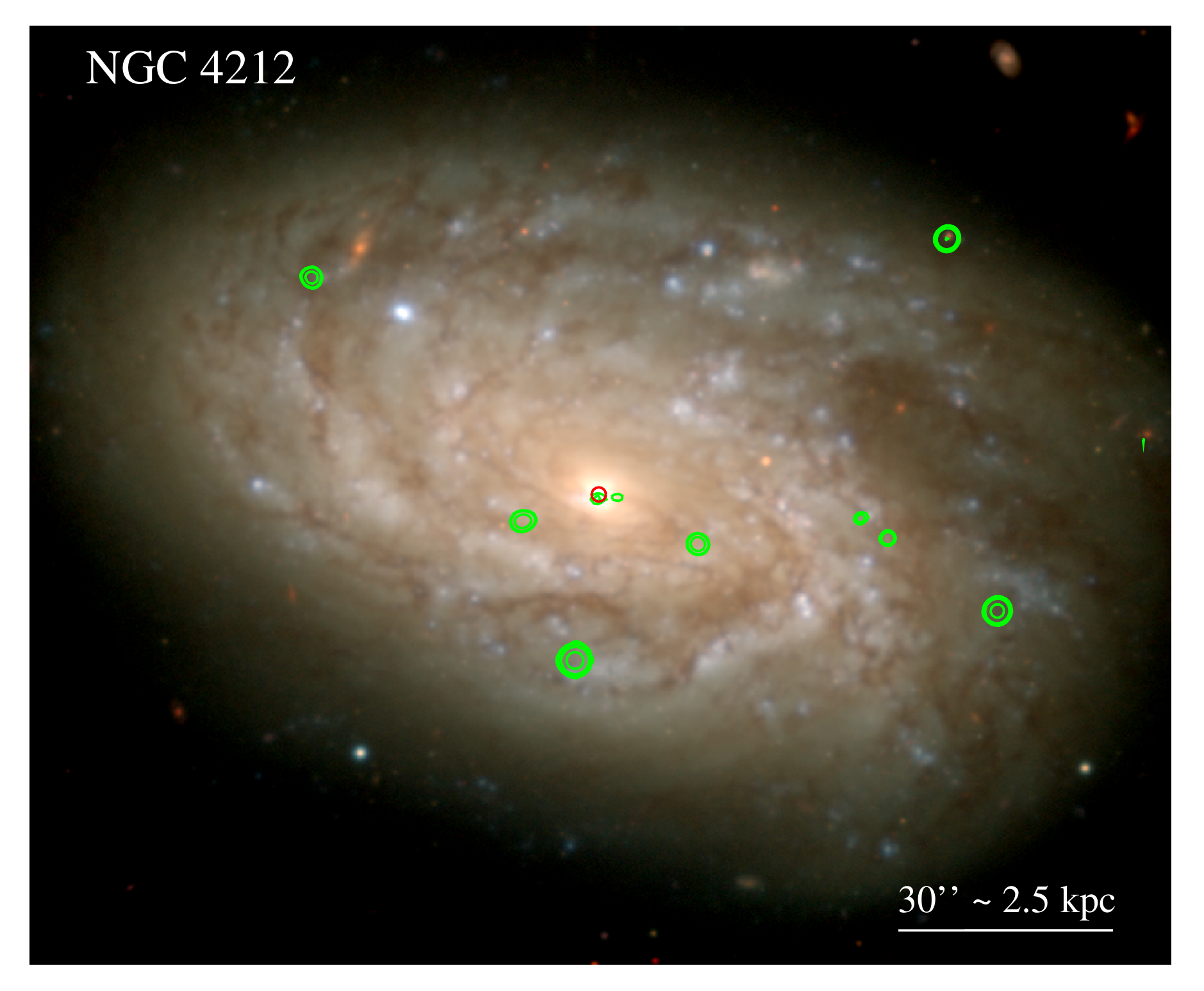} & 
 \includegraphics[angle=0, trim=0cm 0.0cm 0cm 0cm, width=1.0\columnwidth]{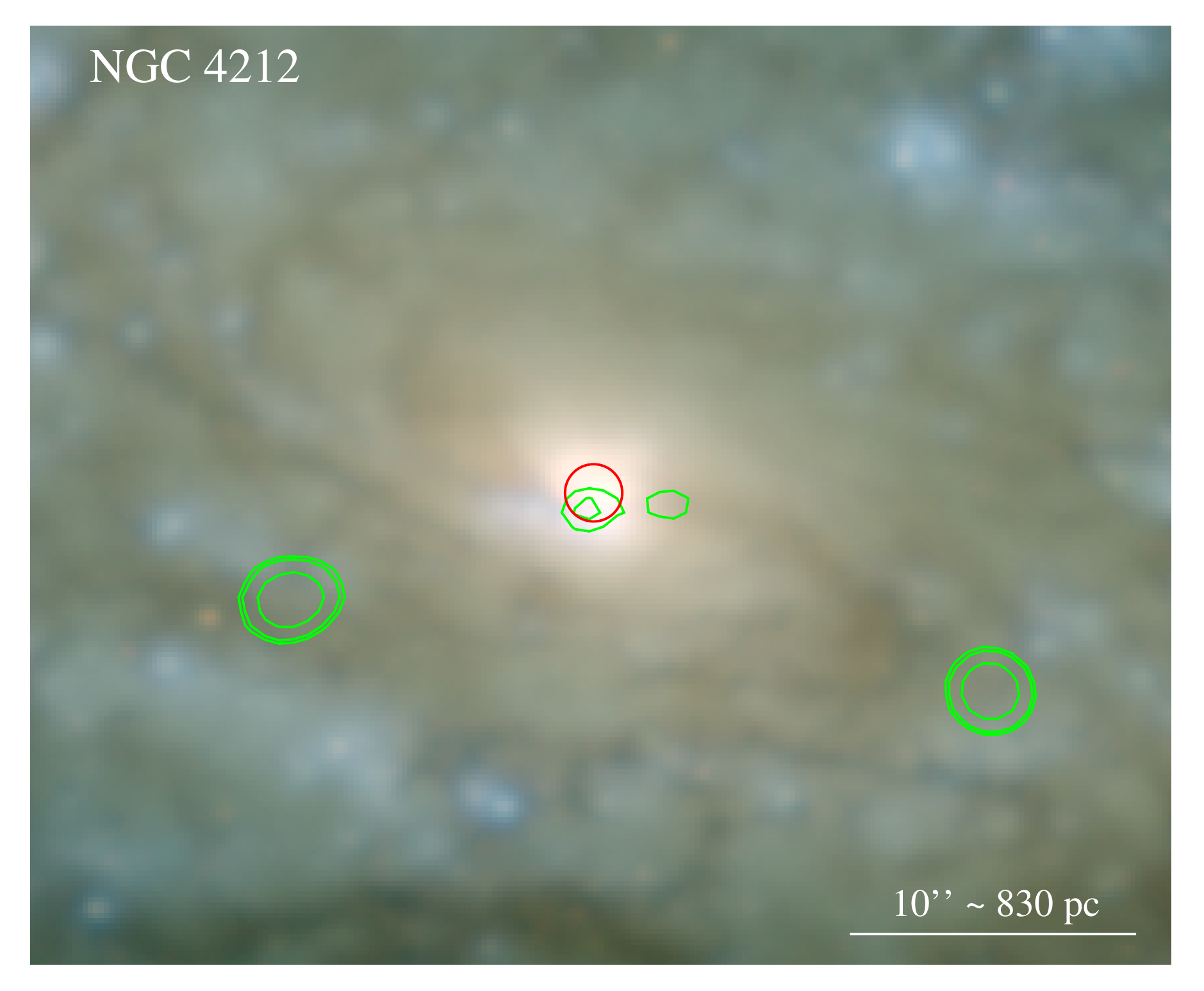} \\
\end{array} 
$
 \caption{Similar to Figure~\ref{Fig-4197}, but displaying an NGVS image of 
   NGC~4212 (aka NGC~4208).} 
\label{Fig-4212}
\end{figure*}

\begin{figure}
 \includegraphics[angle=0, width=1.0\columnwidth]{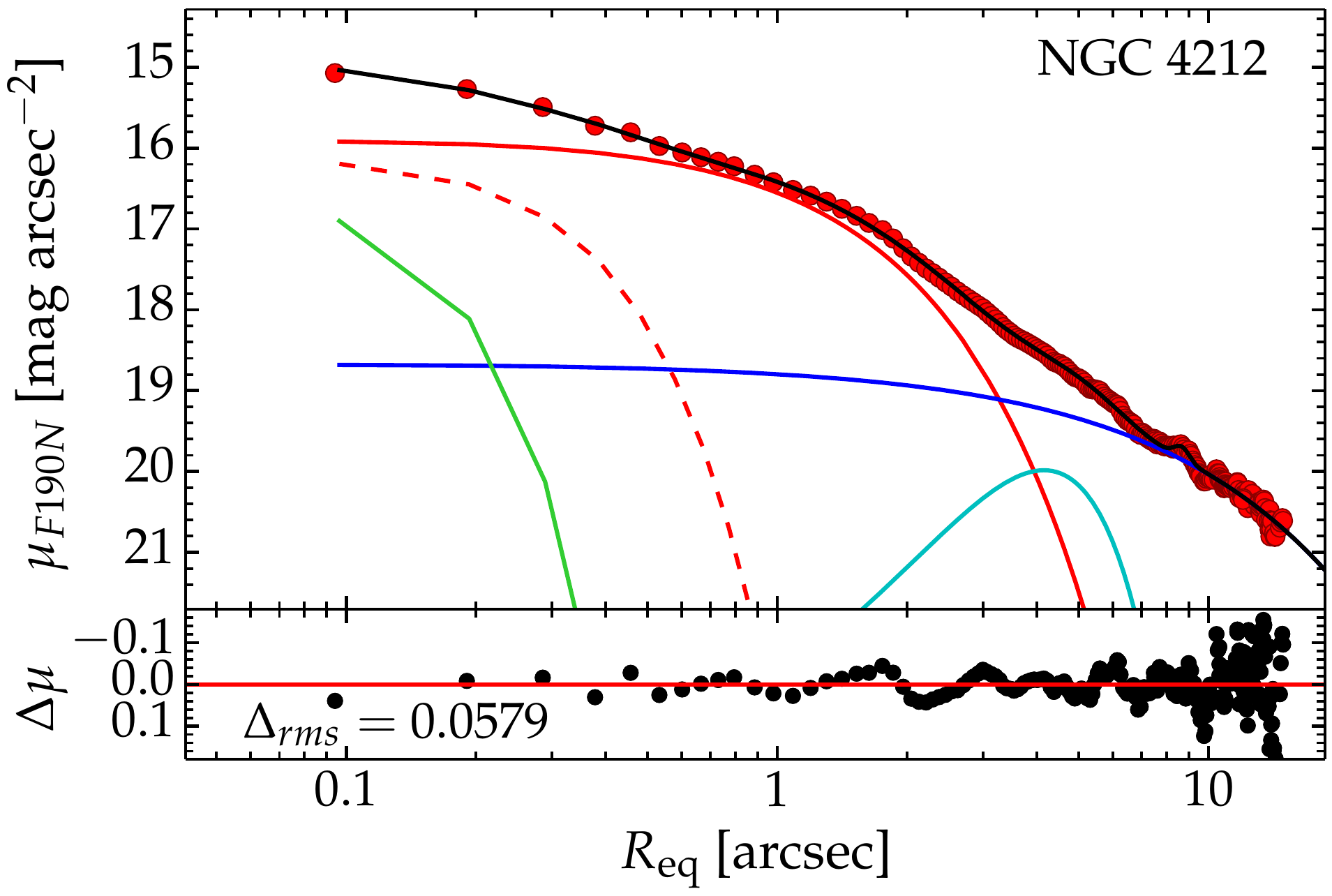}
 \caption{Similar to Figure~\ref{Fig-4713-prof}, but for NGC~4212.}
\label{Fig-4212-prof}
\end{figure}

We extracted a spectrum within a 2$\arcsec$ circular source region (see
Figure~\ref{Fig-ngc4197-spec}), with the local background extracted from the
annulus between radii of 3$\arcsec$ and 9$\arcsec$.  We then fit the spectrum
in {\sc xspec}, using the Cash statistics. We find that the spectrum
(Figure~\ref{Fig-ngc4197-spec}) is well described (C-statistic of 62.7 for 50
degrees of freedom) by a power-law with photon index $\Gamma =
1.24^{+0.84}_{-0.69}$ and an intrinsic\footnote{By `intrinsic', we are
  referring to the intervening number density beyond our Galaxy, primarily
  within the external galaxy.} column density $N_{\rm H} = 3.5^{+7.2}_{-3.5}
\times 10^{21}$ cm$^{-2}$.  The unabsorbed 0.5--7 keV flux is F$_{0.5-7\,{\rm
    keV}} = 1.1^{+0.3}_{-0.2} \times 10^{-13}$ erg cm$^{-2}$ s$^{-1}$.  After
correcting for absorption according to our best-fitting power-law model, we
derive a luminosity $L_{0.5-10\,{\rm keV}} = 1.4^{+0.7}_{-0.3} \times 10^{40}$
erg s$^{-1}$ at the assumed distance of 26.4 Mpc for this galaxy.  If the
X-ray spectrum corresponds to the low/hard state of an IMBH, the black hole
mass would be $\ga$ a few $10^3$ solar masses.

We also tried to fit the spectrum with a disk-blackbody model. We rule out a
peak disk temperature $T_{\rm in} \lesssim 1.6$ keV at the 90\% confidence
levels. Models with disk temperatures higher than that are acceptable (and,
for $T_{\rm in} \gtrsim 3$ keV, essentially identical to the power-law model)
because the peak emission moves close to or beyond the {\it Chandra} band, and
we are only seeing the (power-law)-like section of the disk-blackbody below its
peak. Disk temperatures of up to $\sim$2 keV are sometimes seen in
stellar-mass ULXs with a super-critical disk (slim disk). Thus, we cannot rule
out that the source is a $\approx$10$^{40}$ erg s$^{-1}$ stellar-mass ULX
\citep[e.g.,][]{2014Natur.514..202B} 
located exactly at the nuclear position, but the simplest explanation
consistent with the data is that it is the nuclear BH of this galaxy.

The Eddington luminosity can be expressed as $L_{\rm Edd} \approx
1.26\times10^{38}\, (M_{\rm bh}/M_{\odot})\, (\sigma/\sigma_{\rm T})^{-1}$ erg
s$^{-1}$.  Given that NGC~4197 has $L_{0.5-10\,{\rm keV}} = 144\times10^{38}$
erg s$^{-1}$, and assuming there is a hydrogen plasma with $\sigma =
\sigma_{\rm T}$ (the Thomson scattering cross-section), this luminosity
equates to a $\sim$10$^2$ M$_{\odot}$ black hole accreting at the Eddington
limit, or a $\sim$10 M$_{\odot}$ black hole accreting at ten times the
Eddington limit.  Alternatively, given that we have predicted $\overline{\log
  M_{\rm bh}}=4.8\pm0.6$ for NGC~4197 (Table~\ref{Tab-IMBH}), the Eddington
luminosity for such a black hole is $8.2^{+24.5}_{-6.1}\times10^{42}$ erg
s$^{-1}$.  Expressing the Eddington ratio as $L_X/L_{\rm Edd}$, with $L_X
\equiv L_{0.5-10\,{\rm keV}}$, implies an Eddington ratio of 0.0018, or
0.18\%.

\subsubsection{NGC 4212: dual X-ray point sources 240~pc apart}

\citet{2007MNRAS.381..136D} report that NGC~4212 (aka NGC~4208) is a
LINER/H\,{\footnotesize II} 
composite galaxy.  \citet[][see their Table~1]{2002ApJS..142..223F} searched
for, but did not detect, a radio point-source in this galaxy which
\citet{1973PASP...85..103S} noted had a peculiar amorphous nucleus, likely due
to dust.  \citet{2004AJ....128.1124S} report dust absorption almost down to
the center of the {\it HST/STIS} $R$-band image, but they show a noticeable
brightening within the core which is also evident in the NICMOS/F190N image
from {\it HST} observing program 11080 (P.I.: D.\ Calzetti).

NGC~4212 is the only galaxy in our list of 14 spiral galaxies to have a
predicted black hole mass greater than $\approx$10$^5$ M$_{\odot}$, weighing
in at $6^{+7}_{-3}\times10^5$ M$_{\odot}$.  However, it is particularly
interesting and worthy of inclusion because we have discovered that there are
{\em two} faint {\it CXO} sources near the nucleus, with one of them displaced by a
little less than 1$\arcsec$ from the optical nucleus.  Considering the
positional uncertainty at such faint levels, and the presence of a dust lane
likely shifting the optical center northward, this X-ray point-source may be consistent with the
optical nucleus.\footnote{The X-ray point-source is too faint to establish whether or not
  it is moderately absorbed.}
The second, nearby, X-ray point-source is 2$\farcs$9
($\approx$240~pc) away.  Their separation is resolvable with {\it CXO} (see
Figure~\ref{Fig-4212}).  As with NGC~4470, this off-centre  X-ray point source
could be an XRB. 

It is tempting to investigate the archived {\it HST} image of this galaxy in
order to get at the galaxy's nuclear star cluster magnitude and mass.
However, like the LINER/H\,{\footnotesize II} galaxy NGC~4713, we need to be mindful that
this is also a LINER/H\,{\footnotesize II} galaxy, and as such some of the excess
nuclear light will be optical emission emanating from the
unresolved, non-thermal AGN, as is the case in, for example, the LINER galaxy
NGC~4486 \citep{2006ApJS..164..334F} and the Seyfert~1.5 galaxy NGC~4151
\citep[][see their Figure~4]{2014ApJ...791...37O}.
Modeling the HST/NICMOS/F190N image, we find the galaxy is well fit with a nuclear star
cluster having a magnitude of 17.64$\pm$0.75 (AB mag) and a half-light radius of
0$\farcs$23 (19~pc) (see Figure~\ref{Fig-4212-prof}).  For
$\mathfrak{M}_{\odot,F190N} = 4.85$ and 
$M/L_{F190N}=0.5\pm0.1$, this translates to a mass of 
$\log(M_{\rm nc}/M_{\odot}) = 7.05\pm0.34$, from which one would predict a black
hole mass of $\log(M_{\rm bh}/M_{\odot}) = 6.2\pm1.6$, supportive of the
expectation from the galaxy's stellar mass and spiral arm pitch angle (see
Table~\ref{Tab-IMBH}).  As with the AGN in NGC~4713, we were unable to provide
a useful constraint.

\subsubsection{NGC~4298}

\begin{figure}
 \includegraphics[angle=0, width=1.0\columnwidth]{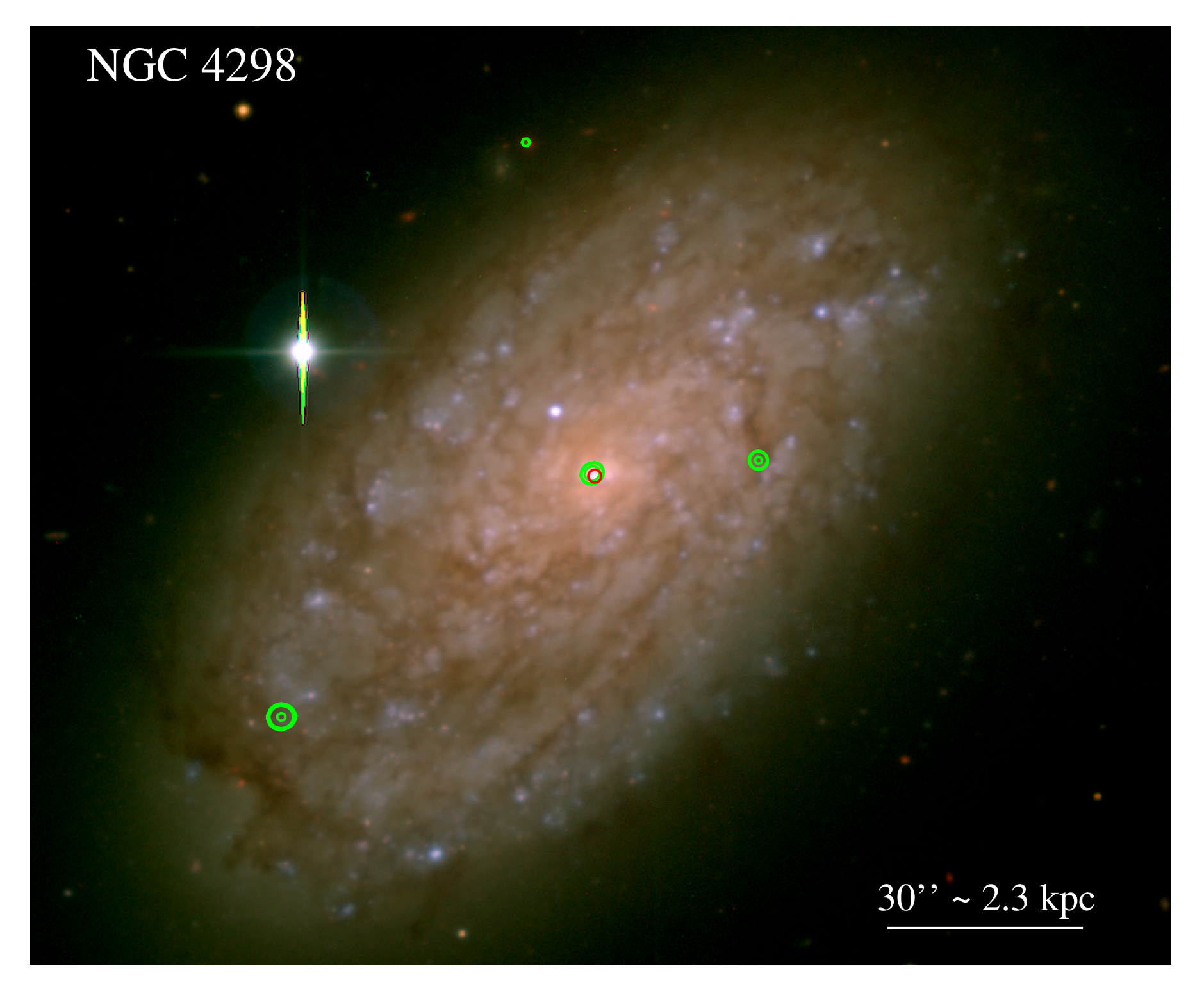}
 \caption{Similar to Figure~\ref{Fig-4470}, but  displaying an NGVS image of
   NGC~4298.}  
\label{Fig-4298}
\end{figure}

\begin{figure}
 \includegraphics[angle=0, width=1.0\columnwidth]{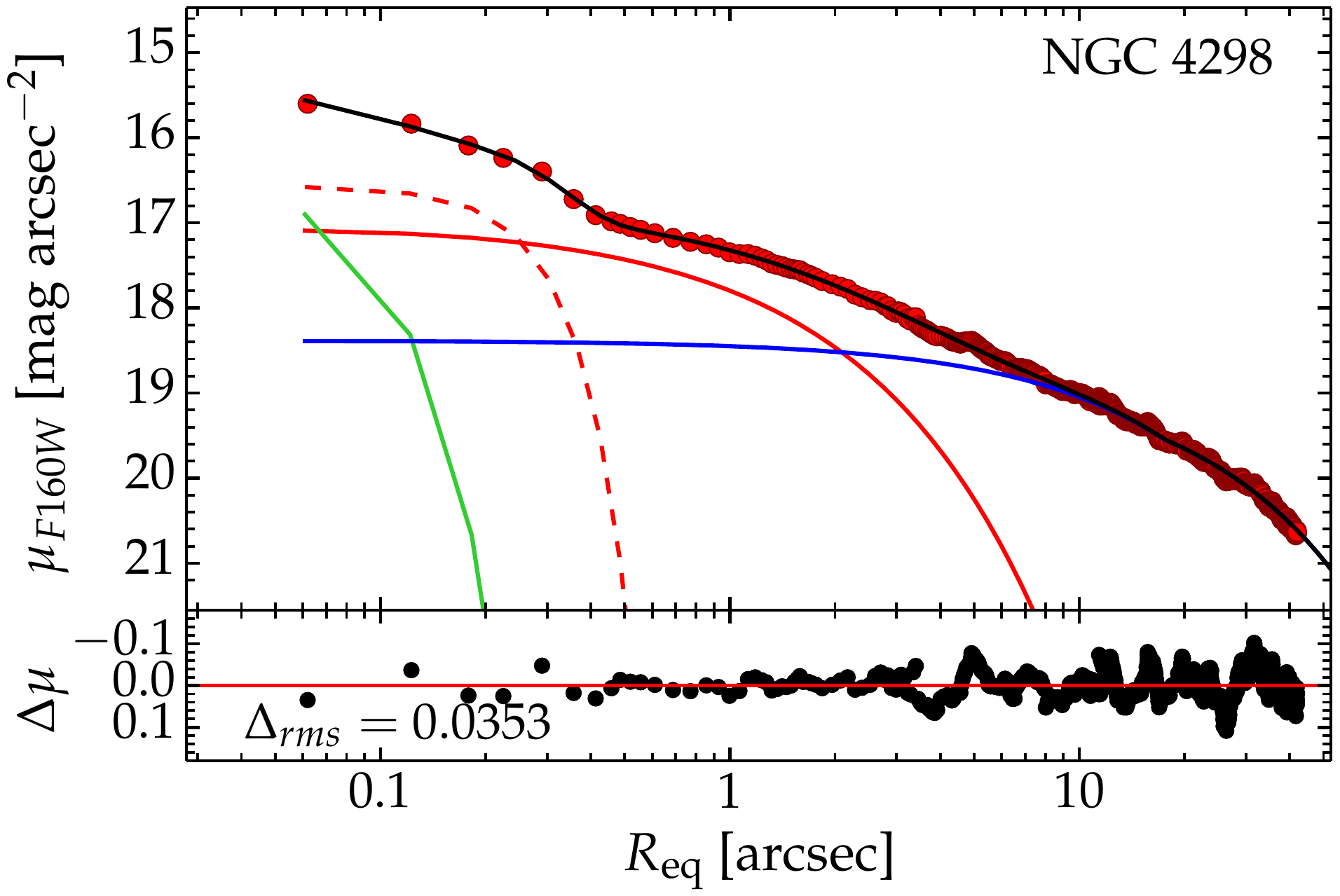}
 \caption{Similar to Figure~\ref{Fig-4713-prof}, but for NGC~4298.} 
\label{Fig-4298-prof}
\end{figure}

Figure~\ref{Fig-4298} presents the optical and X-ray image for NGC~4298, while 
Figure~\ref{Fig-4298-prof} provides a decomposition of the galaxy light as
seen in the HST/WFC3/IR F160W image from {\it HST} 
observing program 14913 (P.I.: M.\ Mutchler). 

Optical/near-IR nuclei in {\it HST} images
may be active BHs and/or star clusters. \citet{2006ApJS..165...57C} showed that
the nuclear star clusters in the Virgo cluster galaxies are slightly resolved
with {\it HST/ACS}, enabling one to differentiate between point-sources and the
spatially-extended star clusters.  While {\it HST's} spatial resolution is
better in the UV and optical than it is in the near-infrared --- simply because of
how the diffraction limit 
scales linearly with wavelength --- NGC~4298 is too dusty to see the nucleus
at UV/optical wavelengths.  However, NGC~4298 is clearly nucleated at 1.6
$\mu$m (and
also 1.9 $\mu$m), and Figure~\ref{Fig-4298-prof} reveals that, using Profiler 
\citep{2016PASA...33...62C}, 
its nucleus can be well approximated by a S\'ersic 
function (convolved with the {\it HST's} PSF) plus a tentative detection of a
faint point-source.  
The S\'ersic nucleus has a half-light radius equal to 0$\farcs$20 (15 pc)
and an apparent (absolute) AB magnitude of 18.1$\pm$0.3
mag ($-$12.9$\pm$0.5 mag) in the F160W 
band\footnote{Performing the decomposition without the point-source yields an apparent magnitude
for the star cluster of 17.9$\pm$0.2 mag.}. 
The tentative point-source, representing the putative AGN, 
has an apparent (absolute) magnitude of 20.4$\pm$0.3 mag ($-$10.6$\pm$0.5 mag). 

Using an absolute magnitude for the Sun of $\mathfrak{M}_{\odot,F160W} = 4.60$
mag (AB) \citep{2018ApJS..236...47W}, and $M/L_{F160W} = 0.5$, gives a nuclear
star cluster mass $\log(M_{\rm nc}/M_{\odot}) = 6.7\pm0.4$ and thus a
predicted black hole mass of $\log(M_{\rm bh}/M_{\odot}) = 5.3 \pm 1.7$ using
Equations~\ref{Eq-new-sigma} and \ref{Eq-new-err}.

\begin{figure*}
$
\begin{array}{cc}
 \includegraphics[angle=0, trim=0cm 0.0cm 0cm 0cm, width=1.0\columnwidth]{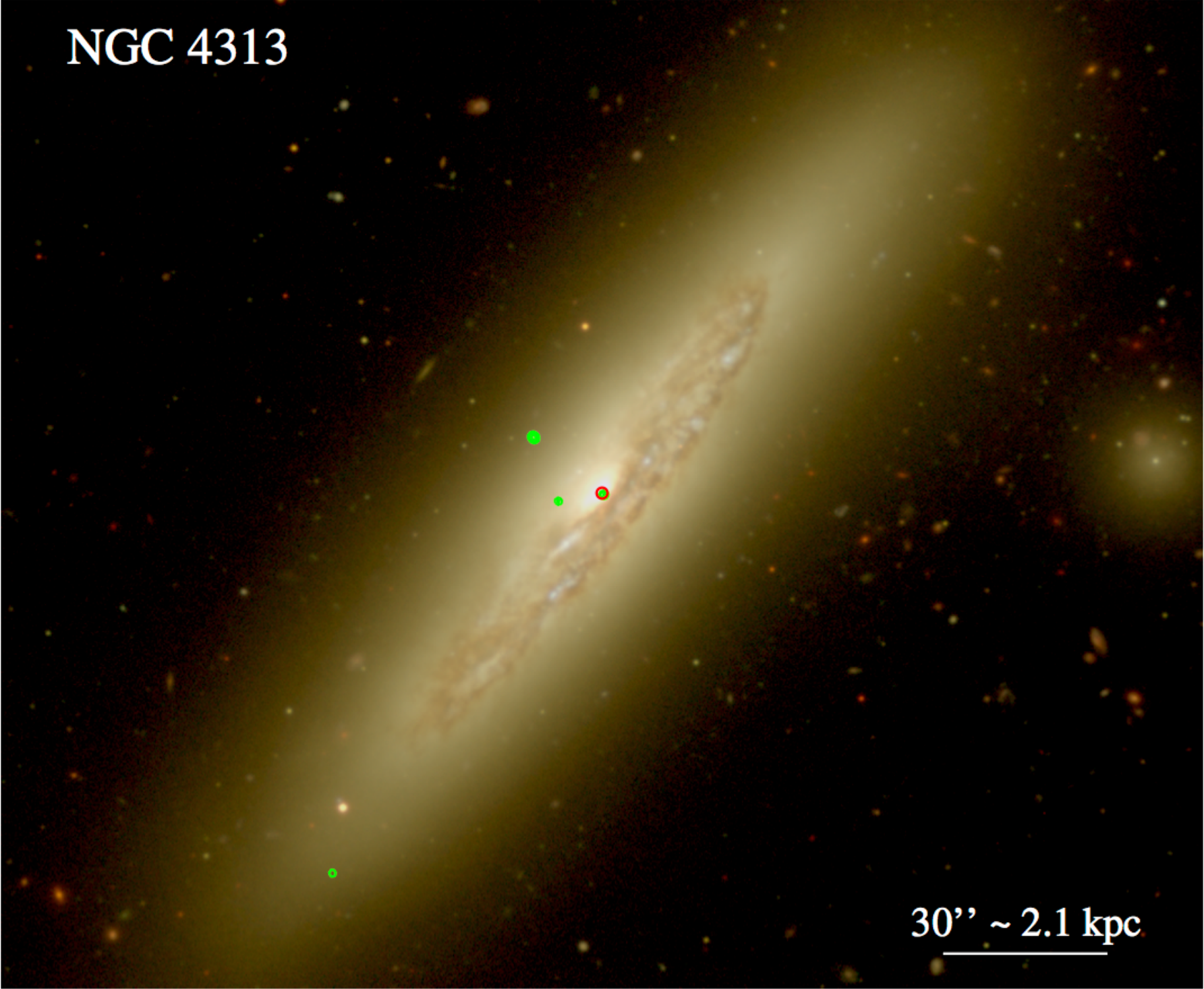} &
 \includegraphics[angle=0, trim=0cm 0.0cm 0cm 0cm, width=1.0\columnwidth]{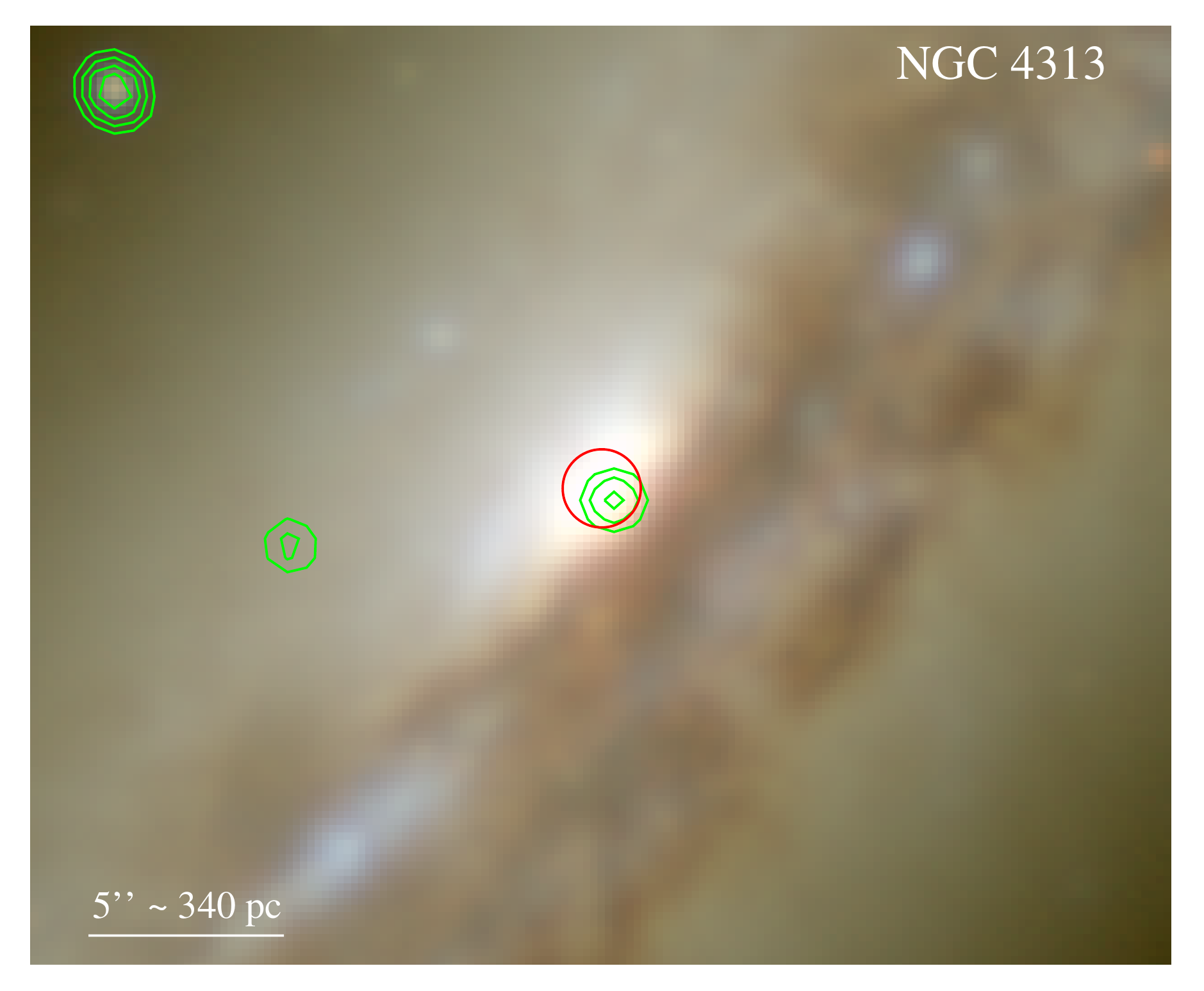} \\
\end{array}
$
\caption{Similar to Figure~\ref{Fig-4197}, but displaying an NGVS image of NGC~4313.}
\label{Fig-4313}
\end{figure*}

\begin{figure}
 \includegraphics[angle=0, trim=0.0cm 0.0cm 0.0cm 0cm, width=1.0\columnwidth]{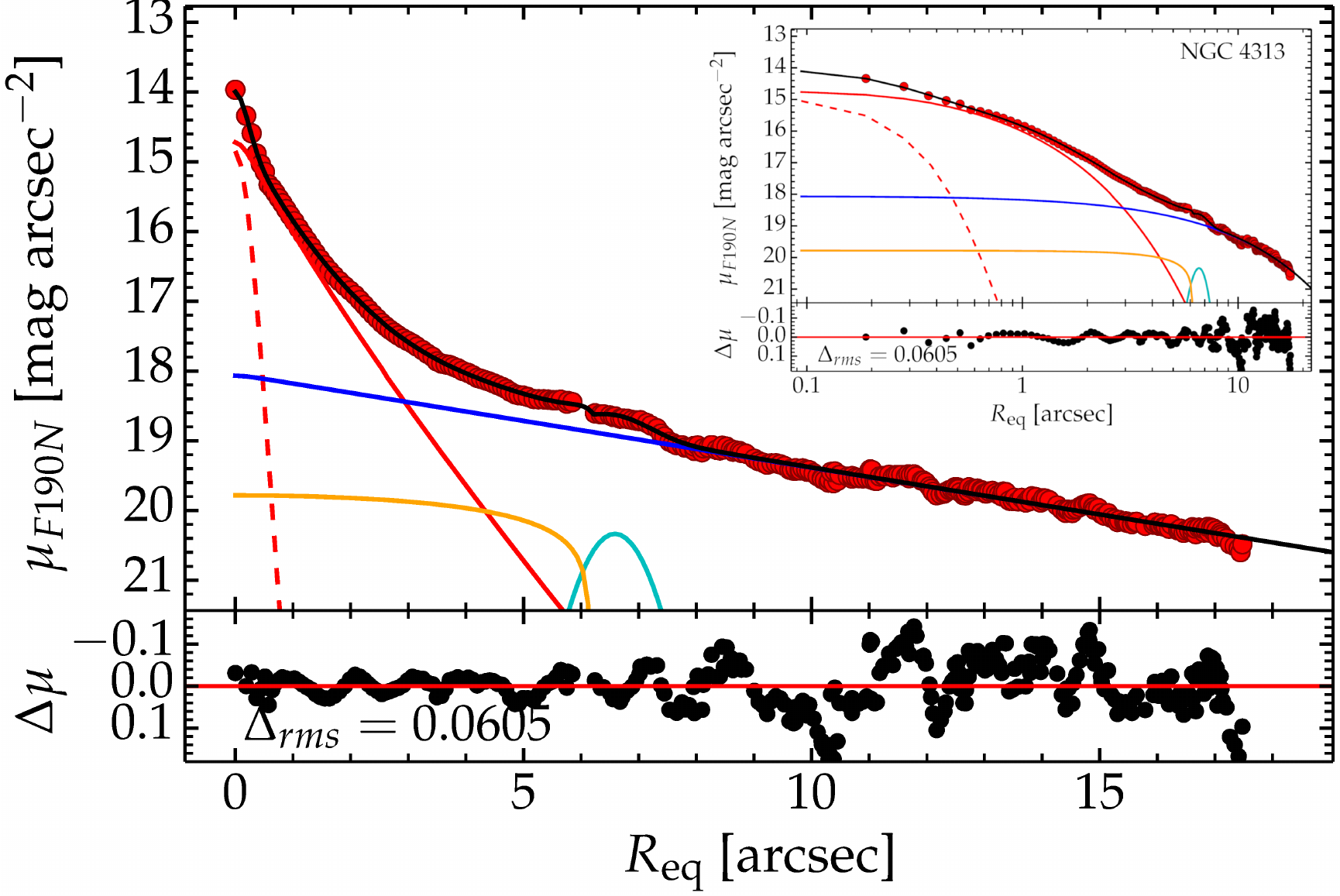}
 \caption{Similar to Figure~\ref{Fig-4713-prof}, but for NGC~4313.}
\label{Fig-4313-prof}
\end{figure}

\subsubsection{NGC~4313: dual X-ray point sources 590~pc apart}

\citet{2007MNRAS.381..136D} report that NGC~4313 is a Seyfert/LINER galaxy. 
From our {\it CXO} data, we report the discovery of an apparently faint, point-like, X-ray
source coincident\footnote{As with NGC~4212, a dust lane has likely shifted
  the optical nucleus of NGC~4313.} with the optical nucleus (Figure~\ref{Fig-4313}). 
Due to the galaxy's somewhat edge-on orientation, it may have a high intrinsic
absorption of X-ray photons. 
Just one of the six X-ray photons is in the 0.3-1 keV band, 3 are in the 1-2
keV band, and 2 are in the 2-10 keV band. 
Offset by 8$\farcs$4 (590~pc) is a second, slightly fainter, X-ray point-source. 

We have been able to inspect an HST/NICMOS/F190N image from {\it HST} observing
program: 11080 (P.I.: D.\ Calzetti) and decompose the galaxy light, which appears to
consist of a bulge, plus a large-scale disk with a weak bar and ansae, 
and a nuclear star cluster with 
$m=16.72\pm0.75$ (AB mag), S\'ersic index $n\approx 0.8$ and effective half-light
radius $R_{\rm e}\approx 10$~pc 
(Figure~\ref{Fig-4313-prof}). 
To obtain the mass of this nuclear component, 
we have used $\mathfrak{M}_{\odot,F190N} = 4.85$ mag (AB)
\citep{2018ApJS..236...47W}, corrected for 0.011 mag of Galactic extinction, 
and assumed\footnote{For reference, \citet{2009ApJ...690.1031B} found
  an $M/L_{F190N}$ value of 0.47 for the
nucleus of the late-type Sd galaxy NGC~3621.} $M/L_{F190N} = 0.5\pm0.1$. 
This yields a stellar mass for the nuclear component of $\log(M_{\rm
  nc}/M_\odot) = 7.27\pm0.36$, and this value may be an underestimate given that there will
be some internal extinction at 1.9 $\mu$m coming from within the inclined
galaxy NGC~4313. 
This leads to a higher than anticipated black hole mass prediction of 
$\log(M_{\rm nc}/M_\odot) = 6.74\pm1.64$.

\begin{figure*}
$
\begin{array}{cc}
 \includegraphics[angle=0, trim=0cm 0.0cm 0cm 0cm, width=1.0\columnwidth]{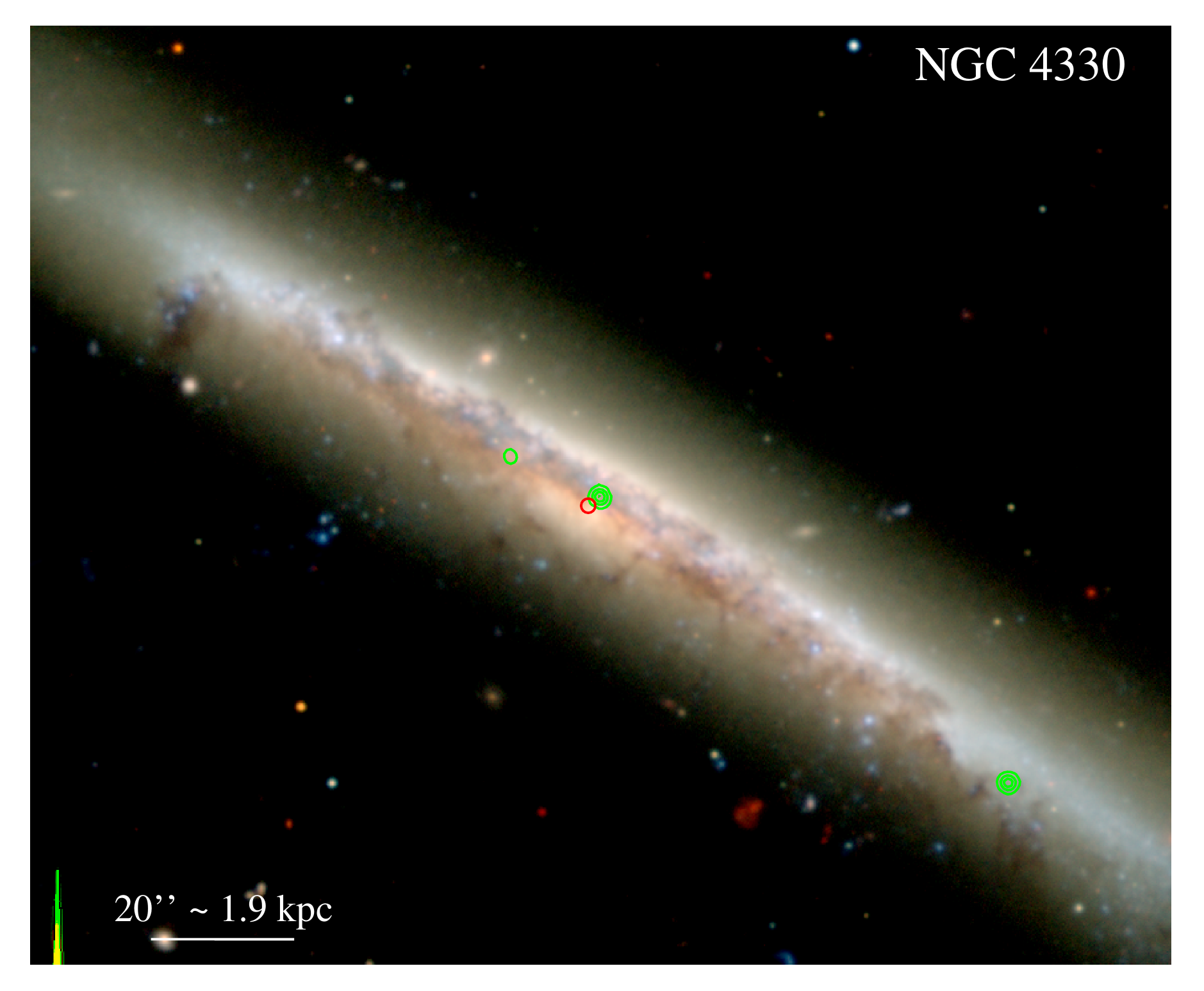} &
 \includegraphics[angle=0, trim=0cm 0.0cm 0cm 0cm, width=1.0\columnwidth]{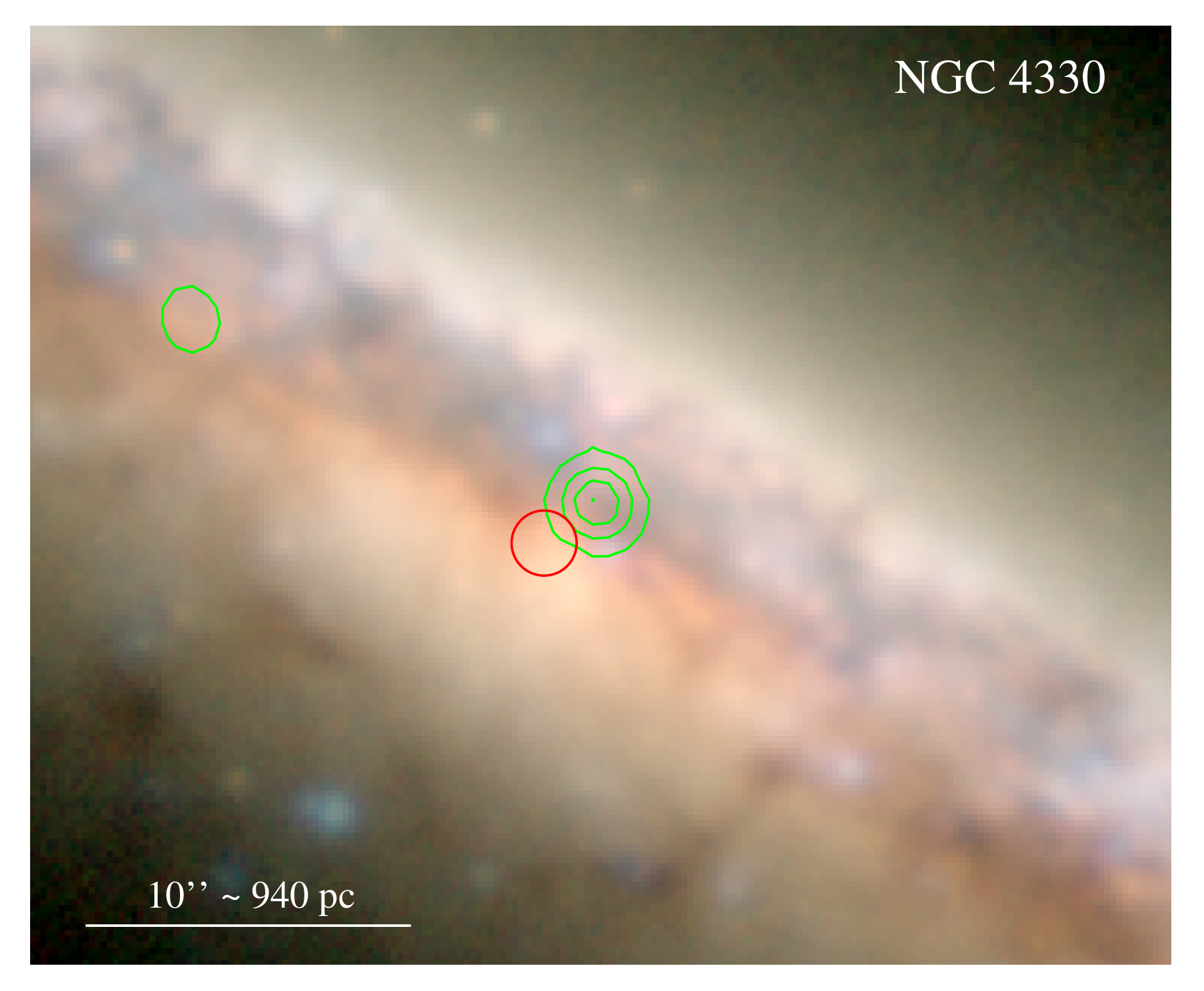} \\
\end{array}
$
 \caption{Similar to Figure~\ref{Fig-4197}, but displaying an NGVS image of NGC~4330.}
\label{Fig-4330}
\end{figure*}

\begin{figure}
 \includegraphics[angle=270, trim=2cm 1cm 0cm 4cm, width=1.0\columnwidth]{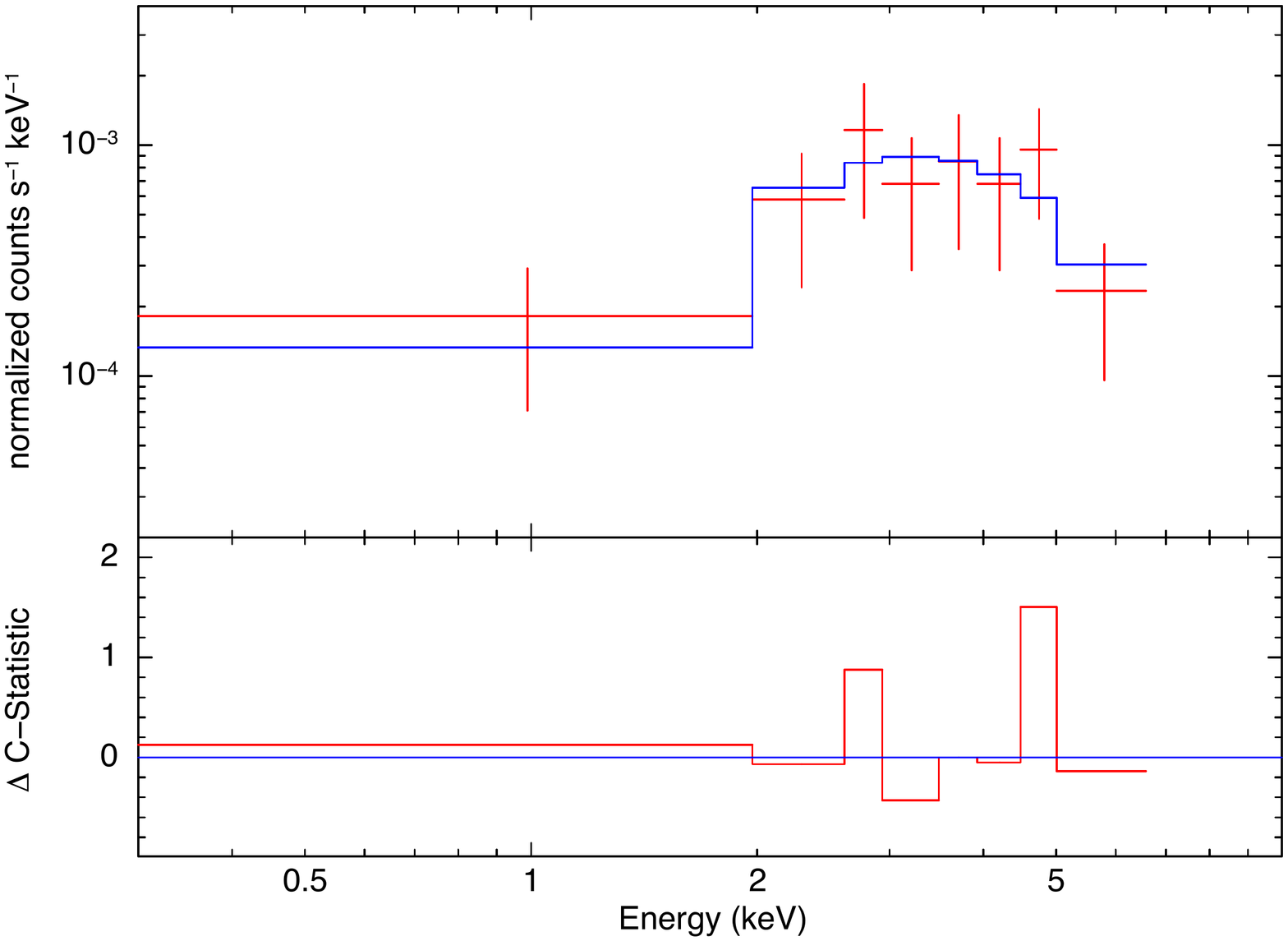}
 \caption{Similar to Figure~\ref{Fig-ngc4197-spec}, but for the edge-on galaxy
   NGC~4330, whose published optical center may be displaced due to the
   intervening dust.}
\label{Fig-ngc4330-spec}
\end{figure}

\subsubsection{NGC~4330: $L_X \approx 10^{40}$ erg s$^{-1}$}

NGC~4330 is experiencing ram pressure stripping of both its neutral HI gas
\citep{2007ApJ...659L.115C, 2011AJ....141..164A} 
and its ionized gas \citep{2012A&A...537A.143V, 2018A&A...614A..57F}. 

The edge-on orientation of its disk to our line-of-sight, coupled with the
detection of a nuclear X-ray source, suggests that it may harbor an
intrinsically bright AGN given that some X-rays have penetrated their way
through and out of 
the disk plane (see Figure~\ref{Fig-4330}).  For comparison, NGC~4197
and NGC~4313 (Figures~\ref{Fig-4197} and \ref{Fig-4313}) and NGC~4178 (GSD19)
represent other examples of spiral galaxies with {\it somewhat} edge-on disks in which we have
detected a nuclear X-ray point-source. 

In NGC~4330, the {\it CXO} source is $\approx$2$\arcsec$ from the optical center reported by NED. However, as
Figure~\ref{Fig-4330} shows, the 
location of the dust lane, coupled with the slight banana shape of the galaxy, 
may result in the optical centroid derived from the outer isophotes not
corresponding to the true nucleus of the galaxy, which might instead be 
flagged by the location of the {\it CXO} source. 
Among our sample of galaxies expected to possess a central IMBH,
NGC~4330 has the second brightest central X-ray point-source, after NGC~4197 (see
Table~\ref{TableSum}).  

As with NGC~4197, we were able to obtain a meaningful X-ray spectrum from the
central point-source in NGC~4330 (see Figure~\ref{Fig-ngc4330-spec}). 
The background-subtracted spectrum was fit in {\sc xspec}, using
the Cash statistics, and is well described 
by a power-law with a (fixed) photon index
$\Gamma = 1.7$ and a high intrinsic column density, 
$N_{\rm H} = 4.3^{+2.9}_{-2.0} \times 10^{22}$ cm$^{-2}$. 
The unabsorbed 
flux F$_{0.5-7\,{\rm keV}} = 6^{+3}_{-2} \times 10^{-14}$ erg
cm$^{-2}$ s$^{-1}$.  At a distance of 19.30 Mpc, 
this corresponds to a luminosity 
$L_{0.5-7\,{\rm keV}} = 1.727^{+0.645}_{-0.515} \times 10^{39}$ erg s$^{-1}$. 
Extrapolating the power-law, one has 
$L_{0.5-10\,{\rm keV}} = 0.9^{+0.6}_{-0.3} \times 10^{40}$ erg s$^{-1}$. 
\citet{2010ApJ...724..559L} report that nuclear X-ray luminosities 
$>10^{40}$ erg s$^{-1}$ can confidently be ascribed to AGN emission, making
NGC~4330, along with NGC~4197 ($L_{0.5-10\,{\rm keV}} = 1.4 \times 10^{40}$
erg s$^{-1}$), strong candidates for IMBHs. 

\subsubsection{NGC~4405 \& NGC~4413}

\begin{figure*}
$
\begin{array}{cc}
 \includegraphics[angle=0, trim=0cm 0.0cm 0cm 0cm, width=0.95\columnwidth]{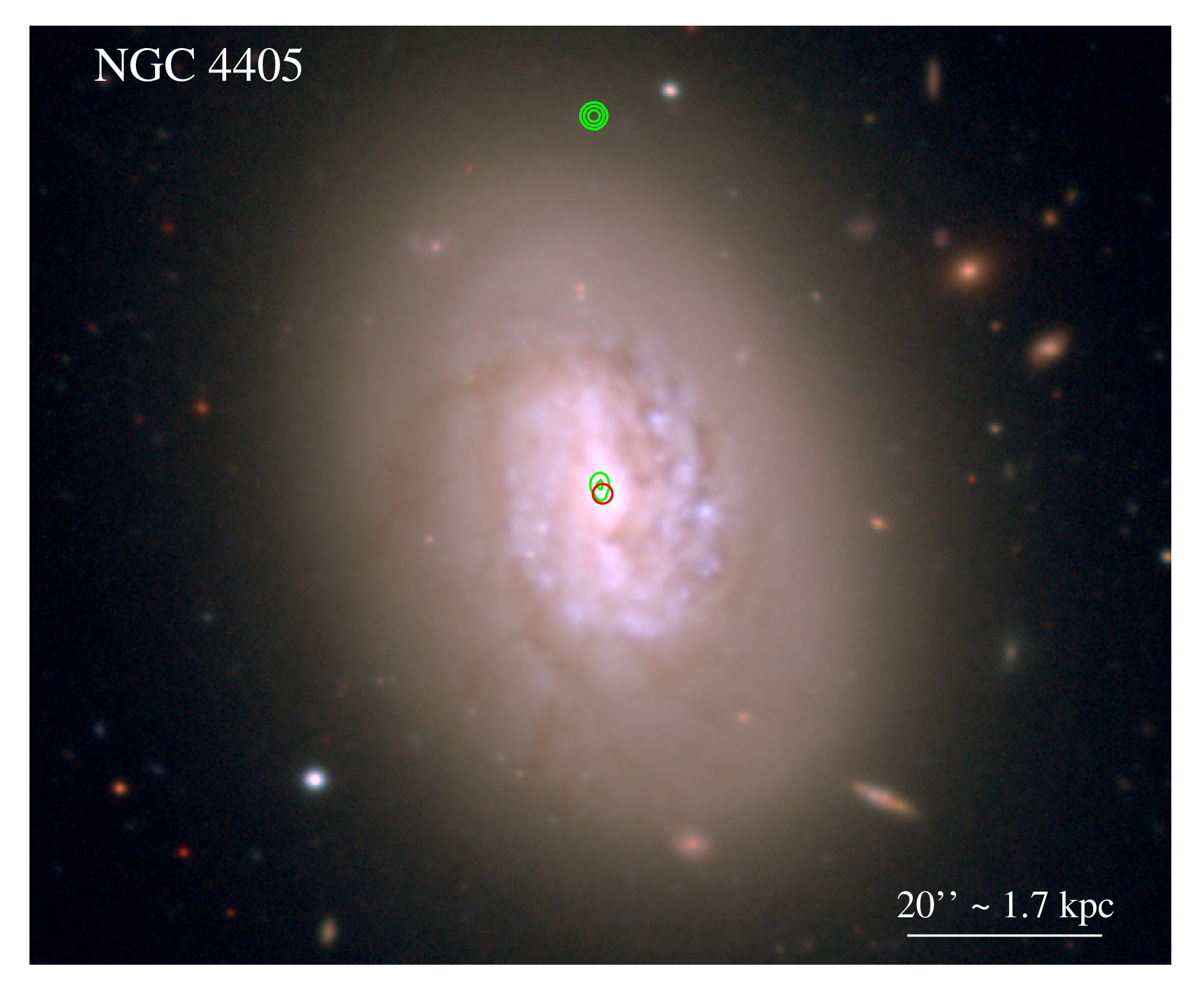} &
 \includegraphics[angle=0, trim=0cm 0.0cm 0cm 0cm, width=0.95\columnwidth]{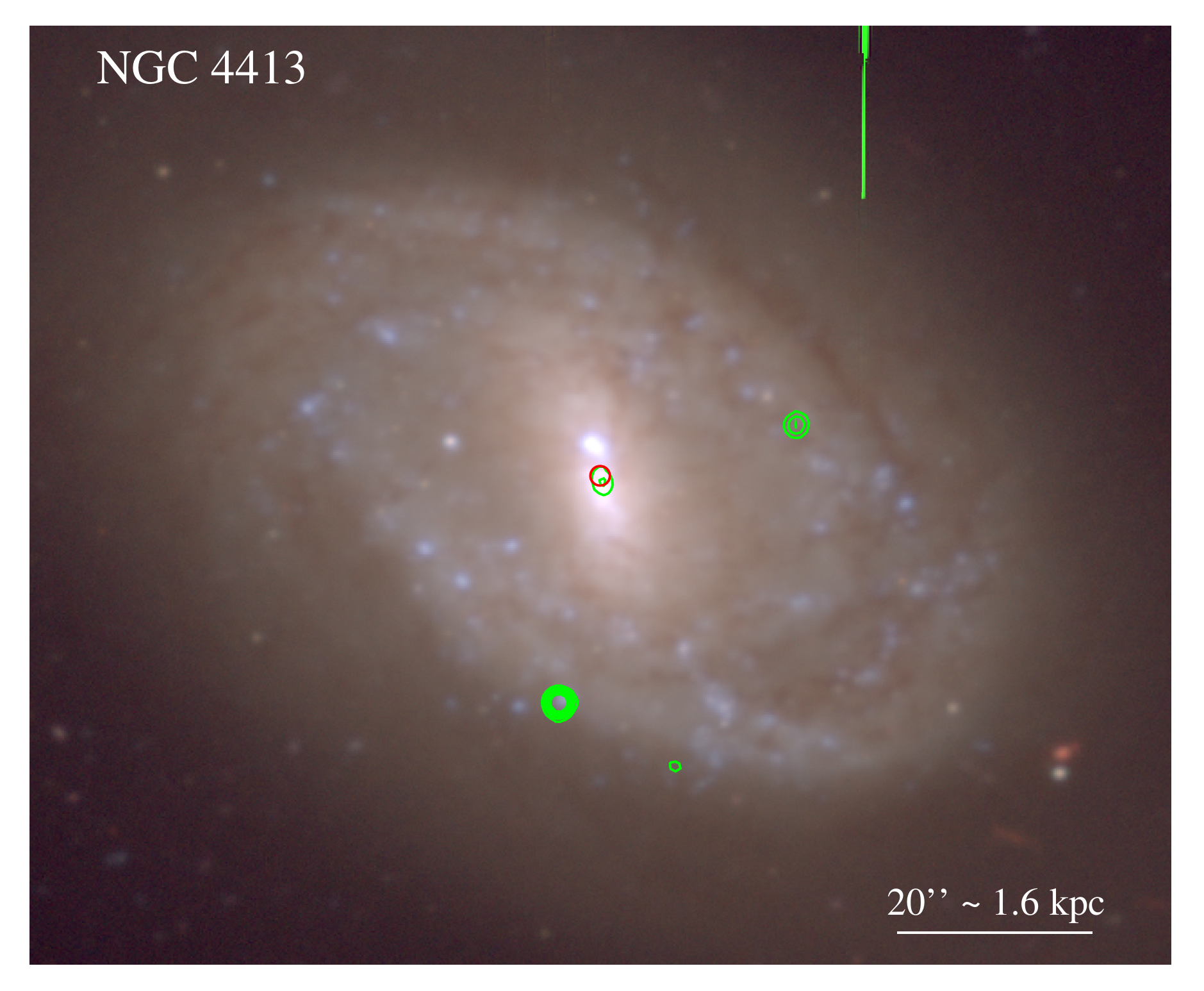} \\
\end{array}
$
 \caption{Similar to Figure~\ref{Fig-4470}, but displaying NGVS images of
  NGC~4405 and NGC~4413 (aka NGC~4407).  For these two galaxies, the central
  X-ray photon count was low, and as such we {\it tentatively} consider them to be
  X-ray point-sources.}
\label{Fig-multi-2}
\end{figure*}

NGC~4405 (Figure~\ref{Fig-multi-2}) and 
NGC~4413 (aka NGC~4407, see Figure~\ref{Fig-multi-2}) have fairly face-on
disks.  We have discovered a central X-ray point-source in both, although 
neither are particularly strong.  Their fluxes and luminosities are provided
in Table~\ref{TableSum}.

\begin{figure*}
$
\begin{array}{cccc}
 \includegraphics[angle=0, trim=0cm 0.0cm 0cm 0cm, width=0.95\columnwidth]{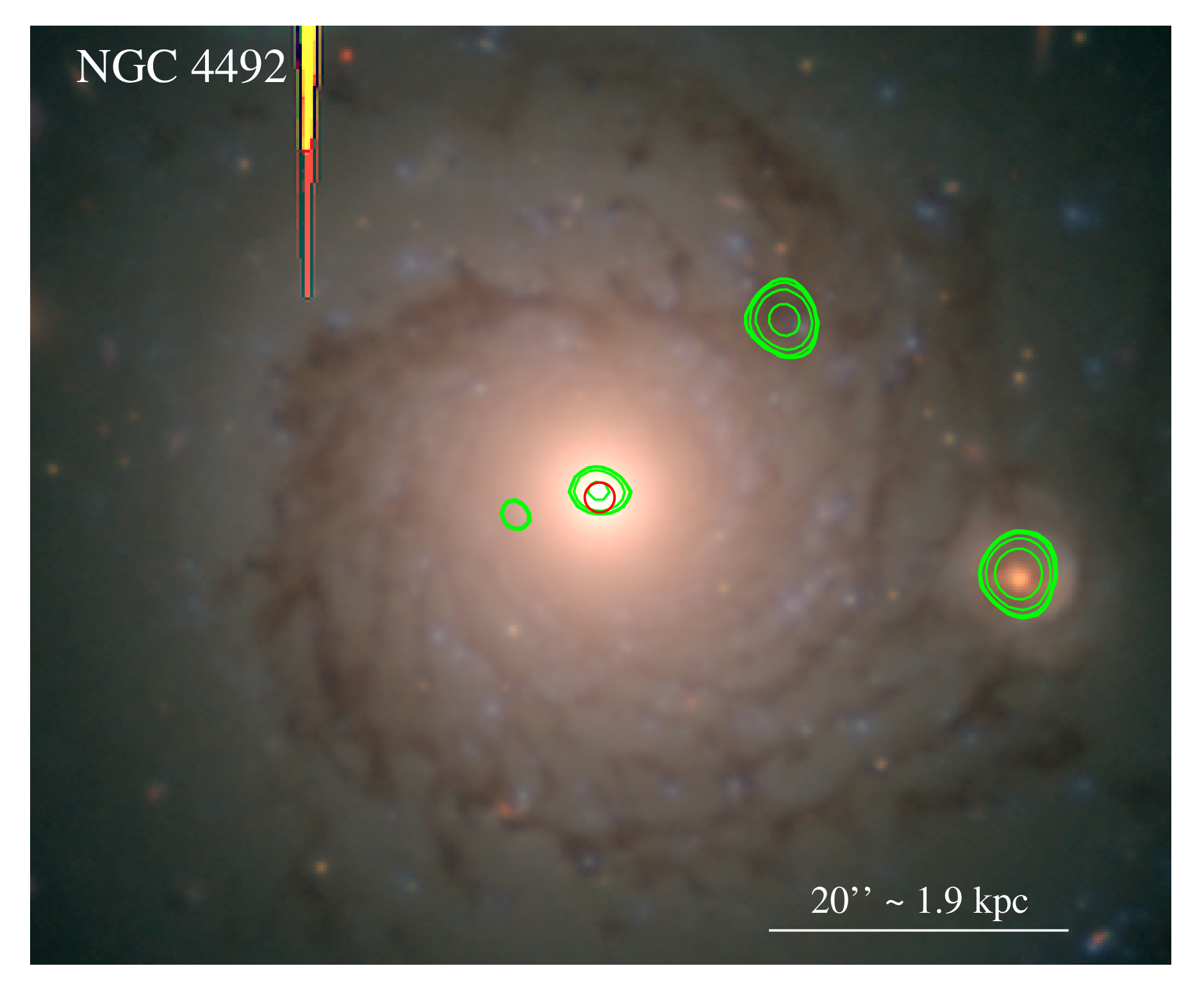} &
 \includegraphics[angle=0, trim=0cm 0.0cm 0cm 0cm, width=0.95\columnwidth]{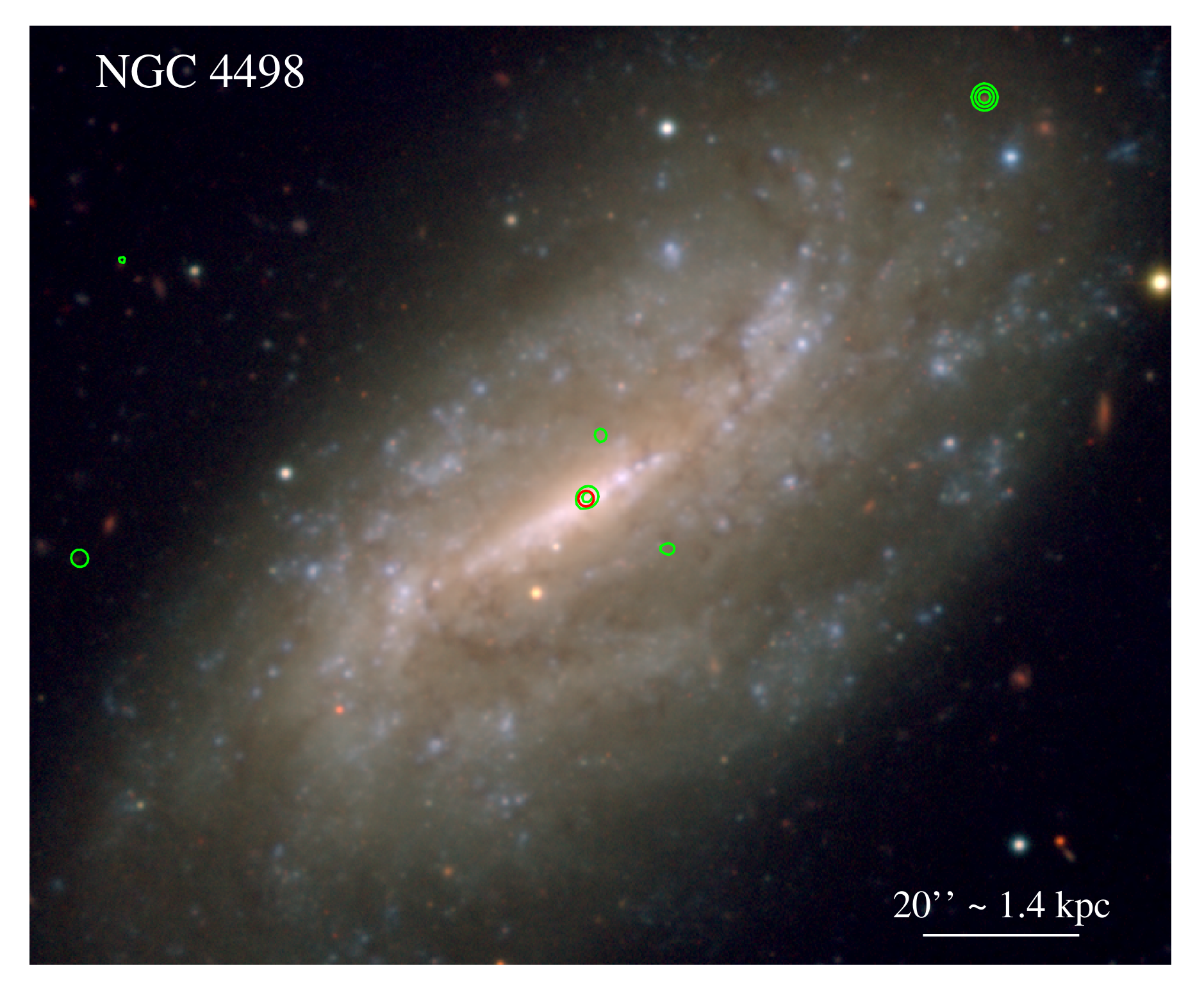} \\
 \includegraphics[angle=0, trim=0cm 0.0cm 0cm 0cm, width=0.95\columnwidth]{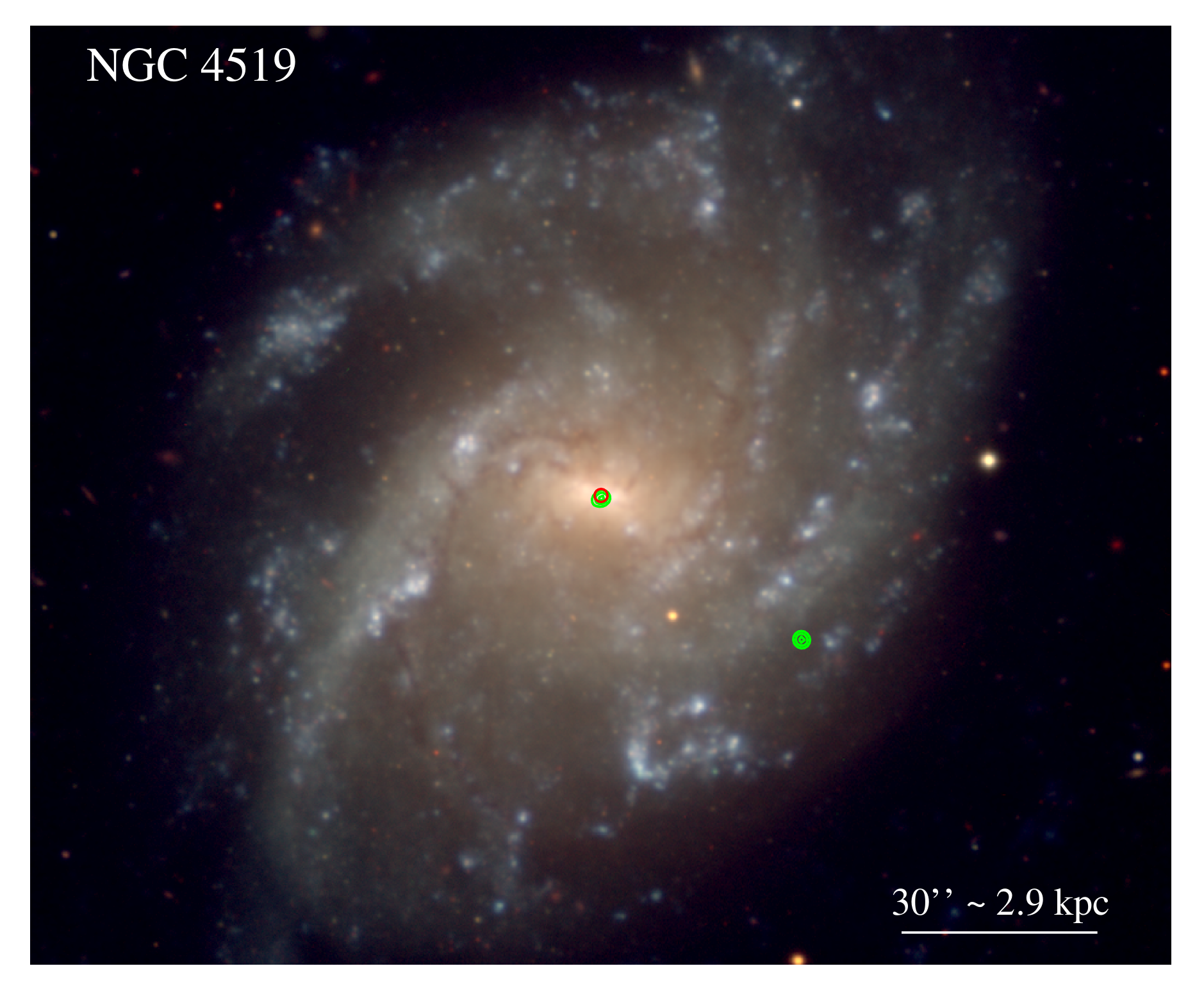} &
 \includegraphics[angle=0, trim=0cm 0.0cm 0cm 0cm, width=0.95\columnwidth]{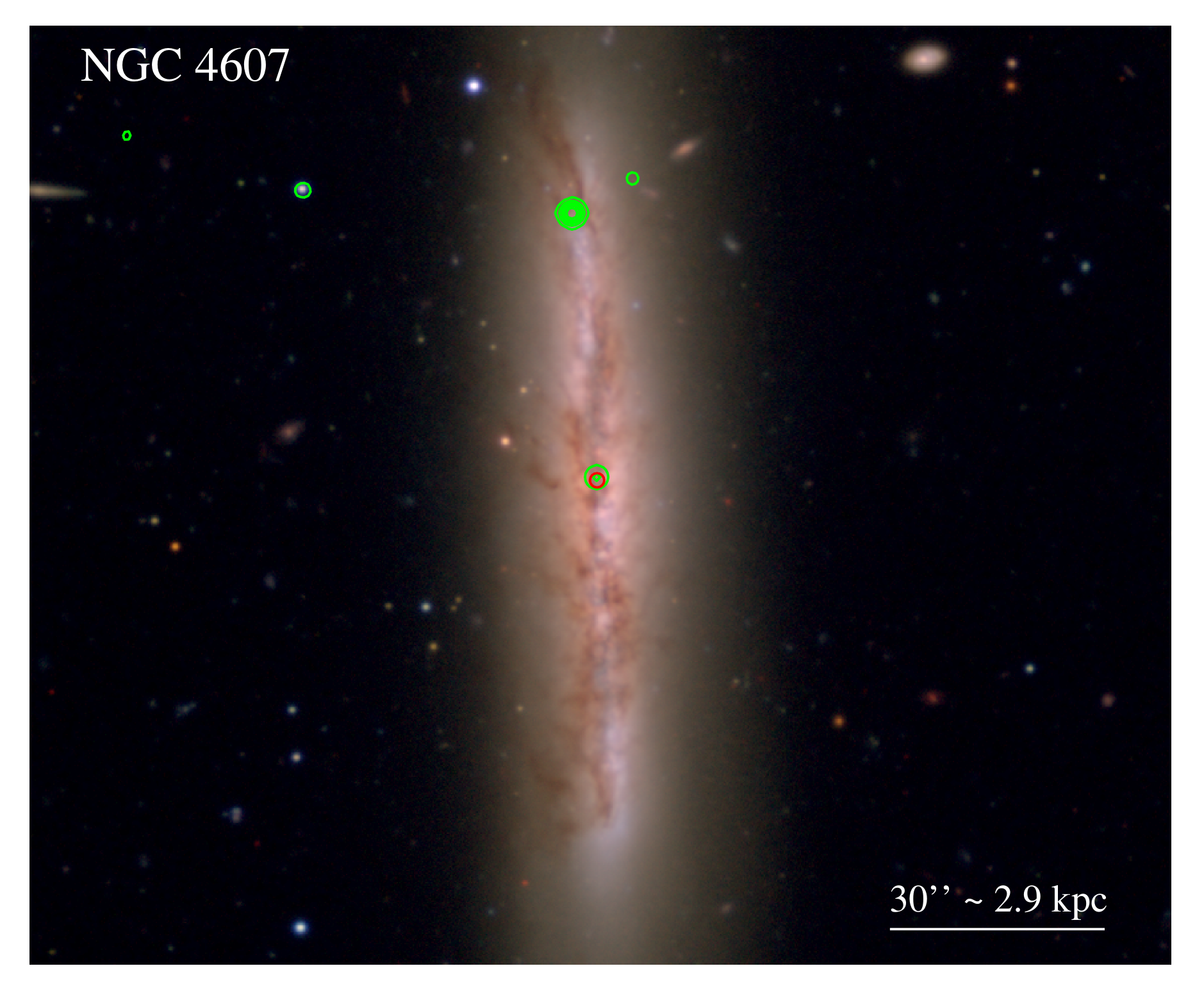} \\
\end{array}
$
 \caption{Similar to Figure~\ref{Fig-4470}, but displaying NGVS images of NGC~4498, NGC~4519,
   NGC~4492, and NGC~4607.  For NGC~4492, the bright source to the right is 
 the nucleus of the galaxy SDSS J123057.82+080434.7
 \citep{2016ApJS..224...40W}, and the red/yellow `bleed' is from a bright adjacent star. 
} 
\label{Fig-multi-4}
\end{figure*}

\begin{figure}
 \includegraphics[angle=270, trim=2cm 1cm 0cm 4cm, width=1.0\columnwidth]{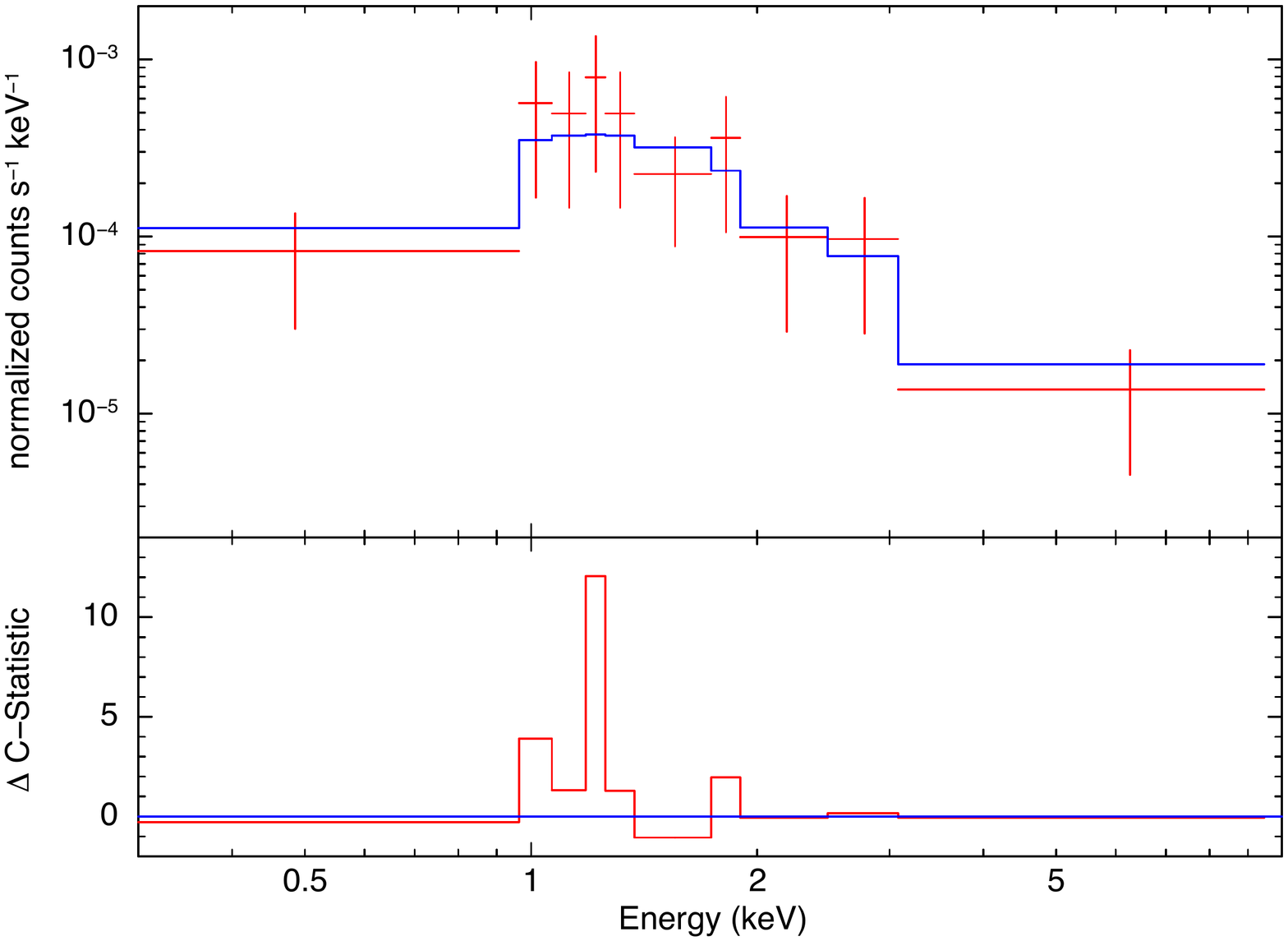}
 \caption{Similar to Figure~\ref{Fig-ngc4197-spec}, but for NGC~4492.}
\label{Fig-ngc4492-spec}
\end{figure}

\subsubsection{NGC~4492; dual X-ray point sources 550~pc apart} 

NGC~4492 is a new addition to the sample in GSD19, having useful {\it CXO}
data from Cycle~8 (4.89~ks, Proposal 08700652, P.I.: S.Mathur) and Cycle~15
(29.68~ks, Proposal 15400260, P.I.: T.Maccarone), and an expected IMBH at its
center based upon the galaxy's stellar luminosity (see Table~\ref{Tab-IMBH}).
\citet{2007MNRAS.381..136D} classified this galaxy as having no 
(H$\alpha$ nor [N\,{\footnotesize II}]) emission 
lines based upon their 2$\arcsec$-slit spectra from the Bologna
Faint Object Spectrograph (BFOSC) attached to the Loiano 1.5~m telescope.
However, we find that it possesses a central X-ray point-source, and a second
X-ray point-source 550~pc to the east of the nuclear point-source
(Figure~\ref{Fig-multi-4}). 

We have combined the above two {\it CXO} exposures (using the {\sc ciao}
task {\it {specextract}}, with the option `combine\_spectra = yes') to obtain
the spectrum of the faint nuclear source at the center of  NGC~4492.
Figure~\ref{Fig-ngc4492-spec} reveals a power-law SED, with $\Gamma=1.7$, as opposed
to the blackbody radiation curve of a hot accretion disk.  
After fitting this combined spectrum, we held $N_{\rm H}$ and $\Gamma$
constant (see Table~\ref{TableSum}),
leaving the normalization parameter free, and we fit the two individual
spectra in {\sc xspec}.  The results are given in Table~\ref{TableSum}.

\subsubsection{NGC~4498}

NGC~4498 contains a few knots of star formation along its spine,
as seen in {\it HST} observing program 5446 (P.I.: G.\ D.\ Illingworth,
WFPC2/F606W). 
As noted in Section~\ref{Sec-sample}, 
\citet{2016MNRAS.457.2122G} have reported a stellar-mass for the central
star cluster of $(136^{+46}_{-41})\times10^4\,{\rm M}_{\odot}$ using
a $V$-band mass-to-light ratio of $0.63^{+0.21}_{-0.19}$. 
Associated with this, we report the discovery of an X-ray point-source 
(Figure~\ref{Fig-multi-4}). 

\subsubsection{NGC~4519}

Globally, like NGC~4713, NGC~4519 (Figure~\ref{Fig-multi-4}) somewhat
resembles the barred, bulgeless spiral galaxy LEDA~87300
\citep{2017ApJ...850..196B}.  Centrally, NGC~4519 contains knots of star
formation near its nucleus, and defining a single nuclear star cluster may
prove problematic as there are multiple candidates seen in the {\it HST}
images from observing program 9042 (P.I.: S.J.Smartt, F814W) and observing
program 10829 (P.I: P.Martini, F606W).

\subsubsection{NGC~4607:  $L_X \approx 0.6\times10^{40}$ {\rm erg s}$^{-1}$}

\begin{figure}
 \includegraphics[angle=270, trim=2cm 1cm 0cm 4cm, width=1.0\columnwidth]{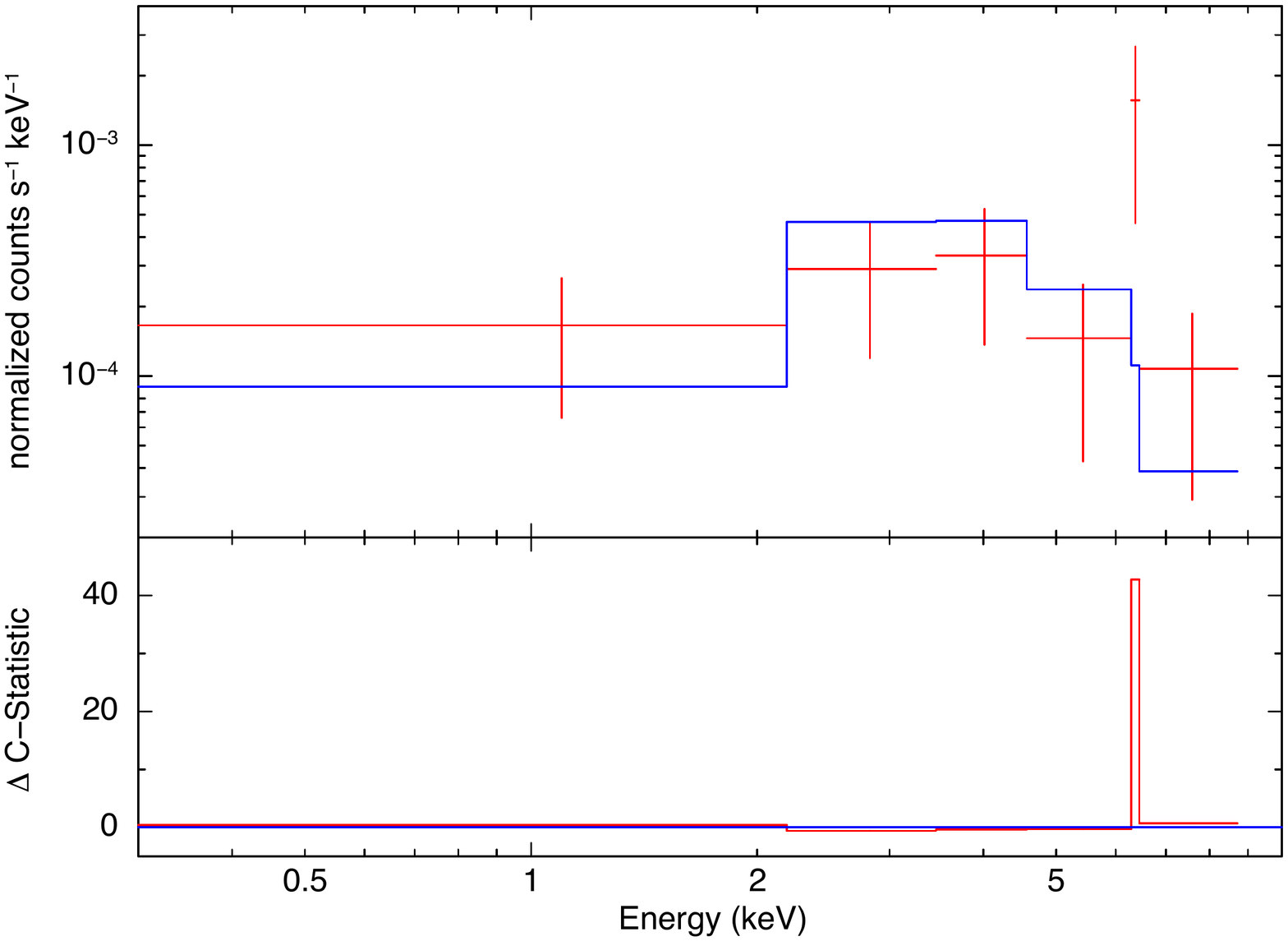}
 \caption{Similar to Figure~\ref{Fig-ngc4197-spec}, but for NGC~4607.}
\label{Fig-ngc4607-spec}
\end{figure}

As with 
the edge-on galaxy NGC~4330, NGC~4607 is aligned edge-on to our
line-of-sight.  Given the expectedly high line-of-sight column density of
neutral hydrogen through the disk of this galaxy, our discovery of a central
X-ray source (Figure~\ref{Fig-multi-4}) suggests that it must be intrinsically bright
and contain X-ray photons in the higher energy bands.  While the central X-ray
point-source has only a few counts, they all have energies above 1~keV, 
which suggests that the  N$_{\rm H,intrin}$ absorption in NGC~4607 
is quite high.  The position of the source disfavors an
XRB in the outskirts of this galaxy due to the required coincidence
that it lines-up with our 
sight-line to the center of this galaxy.  
Moreover, \citet{2007MNRAS.381..136D} have reported that NGC~4607 is a known LINER. 

We constructed the X-ray spectrum using the {\sc ciao} task {\it
  {specextract}}, before regrouping the spectra to 1 count per bin prior to
the Cash statistics analysis in {\sc xspec} using a fixed power-law slope
$\Gamma = 1.7$ and a free N$_{\rm H,intrin}$.  The spectrum is shown in
Figure~\ref{Fig-ngc4607-spec}.  Although ratty, we wish to point out the
grouping of counts around 6.4~keV.  Although it could be due to the randomness
of small number statistics, it is more likely due to a strong Fe K$_{\alpha}$
fluorescence line (and a Compton reflection bump above 10~keV) from cold,
near-neutral, material (disk, torus or clouds) irradiated by the nuclear X-ray
source \citep{1990Natur.344..132P}.  We hope to acquire a longer Chandra, or a
new {\it NuSTAR} \citep{2013ApJ...770..103H}, observation in the future, also
enabling a better constraint on the slope of the X-ray SED.

\section{Prospects for black hole masses from the X-ray data}\label{Sec-mass}

\subsection{X-ray spectra}\label{subsec-alone}

Longer {\it CXO} exposures, plus {\it NuSTAR} and {\it XMM-Newton} exposures
--- if its 
larger PSF does not encounter `crowding' issues and the source count is high
enough to overcome the larger background level --- 
would be of benefit.  This would 
enable the X-ray SED modeling of the
high-energy X-ray photons coming from hot accretion disks, 
and/or Compton scattering in a hot inflow, or inverse Compton emission from the
magnetically-powered jet base or corona
above the disk, and/or synchrotron X-ray emission related to the 
unobscured part of an inner jet 
\citep[e.g.,][]{1972A&A....21....1P, 1995ApJ...452..710N,
  2007A&A...468..129T}. 
Longer X-ray exposures 
would be valuable for establishing, through higher quality spectra,  the existence of a possible dual AGN, 
similar to the X-ray pair in NGC~6240 \citep{2003ApJ...582L..15K,
  2020arXiv200901824F}.

Although it is not yet feasible with the available X-ray data 
to definitively distinguish between a stellar-mass and an IMBH/AGN, we 
outline two possible spectral clues that can be pursued via 
deeper {\it CXO} observations and/or with the proposed next generation of X-ray
satellites including 
Athena \citep{2013arXiv1306.2307N, 2013arXiv1308.6785R}, 
AXIS \citep{2018SPIE10699E..29M}, and 
Lynx \citep{2019JATIS...5b1001G}.

The first distinction we address is between stellar-mass black holes and
IMBHs. 
Standard accretion-state models predict that an IMBH or AGN at a luminosity of
$\sim$10$^{38}$--10$^{39}$ erg s$^{-1}$ should have an unbroken power-law
spectrum in the 0.5-10 keV band (low/hard state), while a stellar-mass source
(especially, a stellar-mass black hole) should have a disk-blackbody spectrum
with a temperature $\sim$0.5--1 keV (high/soft state) and a normalization
corresponding to characteristic radii $\sim$50--100 km. For example, this was
the main argument in favor of the identification of the nuclear source in M33
($\sim$20 times closer to us than the Virgo cluster) as a stellar-mass black
hole \citep{2004A&A...416..529F}. There are, however, additional caveats in
this simple classification. 
While transient stellar-mass black holes near their Eddington luminosities can
exhibit hard power-law spectra in the hard intermediate state, these phases are
short-lived, typically lasting only a few days, and hence it is rather
unlikely that a Chandra observation would catch a source in such a state 
\citep[e.g.,][]{2005Ap&SS.300..107H, 2009MNRAS.400.1603M}. 
Moreover, neutron star X-ray binaries (X-ray
pulsars) can also reach or exceed such luminosities, and at a moderate
signal-to-noise ratio the Comptonized spectrum of those sources is also well
approximated by a hard power-law, in the relatively narrow {\it Chandra} or
{\it XMM-Newton} bands \citep{2009A&A...498..825F, 2016A&A...591A..29F,
  2017ApJ...836..113P}.  The X-ray-to-optical
flux ratio criterion is also inapplicable for nuclear sources, because in most
cases we can only measure the optical brightness of the host star cluster, not
the direct optical emission from the disk or the donor star of the X-ray
source.  As an indication of the difficulty of the task for galaxies in the Virgo cluster,
one need only read the discussion about the nuclear black hole identification in 
the much closer M83 galaxy by \citet{2020MNRAS.495..479R}. 

The second distinction, although likely an artificial one created out of
observational selection bias, is between IMBHs (10$^2$--10$^5$ M$_{\odot}$)
and SMBHs in the 10$^6$--10$^7$ M$_{\odot}$ mass range.  In the X-ray
luminosity range $\sim$10$^{38}$--10$^{39}$ erg s$^{-1}$, both an IMBH and a
`normal' SMBH, with say $M_{\rm bh} \sim 10^{6}\, M_{\odot}$, would be lumped
together in the low/hard state, according to the traditional state
classification.  However, more recent studies of accreting BHs in this
low-luminosity regime show further physical changes as a function of the
Eddington ratio $\equiv L_{\rm bol}/L_{\rm Edd}$ ($\approx L_{\rm
  X}/L_{\rm Edd}$ for stellar-mass BHs). The hardest spectra ($\Gamma \approx
1.7$) occur at an Eddington ratio of $\sim 10^{-3}$; below that threshold, the X-ray
spectrum progressively softens again, reaching an asymptotic value of $\Gamma
\approx 2.1$ at an Eddington ratio of $\sim 10^{-5}$ and below
\citep{2011MNRAS.417..280S, 2013MNRAS.428.3083A, 2013ApJ...773...59P,
  2015MNRAS.447.1692Y, 2017ApJ...834..104P}.  For example, an SMBH with $M_{\rm
  bh}$ around a few $10^6\, M_{\odot}$ and $L_{\rm X} \approx$ a few $10^{38}$
erg s$^{-1}$ will have an Eddington ratio $\sim$10$^{-5}$
\citep[assuming a bolometric correction $L_{\rm bol}/L_{\rm X} \approx
  10$:][]{2012MNRAS.425..623L}.  Instead, an IMBH at the same X-ray luminosity
may have an Eddington ratio of $\sim$10$^{-3}$. Thus, a low-state IMBH should have a
moderately harder spectrum than a more massive nuclear BH, at the same X-ray
luminosity.  We suggest that for a sufficiently high signal-to-noise ratio in
the X-ray spectra, it will be possible to discriminate between the two cases:
if not for individual sources, at least based on the statistical distribution
of fitted photon indices.

\subsection{X-ray luminosities: the fundamental plane of black hole activity}\label{subsec-radio}

The ``fundamental plane of black hole activity'' for black holes with low accretion rates 
\citep{2003MNRAS.345.1057M, 2004A&A...414..895F, fischer2020fundamental} 
encompasses stellar-mass black hole X-ray binaries and AGN, and
enables one to estimate the black hole mass based upon the nuclear radio
emission, $L_R = \nu L_{\nu}$ erg s$^{-1}$ (at 5~GHz), and the unabsorbed nuclear X-ray
luminosity, $L_X$ erg s$^{-1}$ (at 0.5-10~keV).  
Although, it is noted that some AGN are X-ray luminous but radio silent
\citep{radcliffe2021radio}. 
\citet{2012MNRAS.419..267P} report the following correlation for radio active
systems: 
\begin{eqnarray} 
\log L_X = (1.45\pm0.04)\log L_R  \nonumber \\
- (0.88\pm0.06)\log M_{\rm bh} - (6.07\pm1.10). \label{Eq_bhfp}
\end{eqnarray} 
For $M_{\rm bh}=10^5\,{\rm M}_{\odot}$ and $L_X = 10^{40}$ erg s$^{-1}$, one
obtains an expected radio luminosity $\nu\,L_{\nu}$(5~GHz) of $10^{34.8}$ erg
s$^{-1}$, while for $M_{\rm bh}=3\times 10^3\,{\rm M}_{\odot}$ and $L_X = 10^{38}$ erg
s$^{-1}$, one obtains an expected radio luminosity $\nu\,L_{\nu}$(5~GHz) of
$10^{32.5}$ erg s$^{-1}$.  
At an average distance of 17~Mpc, this corresponds
to 0.04~mJy and 0.18~$\mu$Jy, respectively.
Using the correlation from \citet{2019ApJ...871...80G}, which is based on the
2-10 keV luminosity\footnote{For $\Gamma=1.7$, $L_{2-10\,{\rm keV}} = 0.646
  L_{0.5-10\,{\rm keV}}$.}, these estimates are 0.03~mJy and
0.08~$\mu$Jy. This latter value is $\sim$7 times smaller than the estimate of
0.54~$\mu$Jy obtained using the earlier correlation from \citet{2009ApJ...706..404G}.

For reference, \citet{2012ApJ...750L..27S} searched for radio emission from
potential IMBHs in three globular clusters but found no source down to rms
noise levels of 1.5--2.1 $\mu$Jy beam$^{-1}$ with the Very Large Array (VLA).
\citet{2018ApJ...862...16T} have also reported a non-detection of IMBHs with
$M_{\rm bh} \gtrsim 10^3$ M$_{\odot}$ (3$\sigma$) in globular clusters, with a
VLA image stack of 24 GCs having an rms sensitivity of 0.65 $\mu$Jy
beam$^{-1}$, and an Australia Telescope
Compact Array (ATCA) image stack of 14 GCs having an rms sensitivity of
1.42 $\mu$Jy beam$^{-1}$.  It will, however, be interesting to explore the
Andromeda Galaxy's very low metalicity globular cluster RBC EXT8
\citep{2020arXiv201007395L} --- taken from the Revised Bologna Catalogue
\citep{2004A&A...416..917G} --- which may have formed more massive stars than
is usual, and might have formed an IMBH \citep{2021MNRAS.505..339M}, 
see also \citet{2020Natur.583..768W} in regard to the Phoenix GC and stream.

Reversing tact, one can use X-ray luminosities and radio observations to predict the 
black hole masses, or at least obtain an upper limit if only upper limits to
the radio luminosity of potential compact nuclear sources are found.  
We scanned the literature for radio data from our sample, and found one observation. 
With 0$\farcs$15 
spatial resolution, \citet{2005A&A...435..521N} reported an upper limit to the
nuclear flux in NGC~4713 of $\approx$1.10~mJy at 15~GHz (2~cm), or 
$\log(L_{\nu,15~{\rm GHz}}\,{\rm W}\,{\rm Hz}^{-1}) < 19.63$ for $D=17.9$~Mpc.  This
equates to 
$\log (\nu\,L_{\nu,15~{\rm GHz}}\,{\rm erg}\,{\rm s}^{-1}) < 36.81$.
However, we do not know the slope of the radio SED, required to obtain the
luminosity at 5~GHz. 
Moreover, given the available X-ray flux and anticipated black hole mass in
NGC~4713, one requires constraints on the radio flux which are
some three orders of magnitude tighter, at around 1~$\mu$Jy.  
Adding to the challenge, \citet{2005A&A...435..521N} report significant
inter-year variability at 15~GHz and, for a given black hole, the radio and
X-ray flux are correlated \citep{1998A&A...337..460H, 1999MNRAS.309.1063B,
  2003A&A...400.1007C}, hence Equation~\ref{Eq_bhfp}.

\citet{2009AJ....138.1990C} obtained VLA images for 63 of the 100 early-type galaxies in the
Virgo Cluster compiled by \citet{2004ApJS..153..223C}, but they detected
compact radio sources at the centers of just 12 of these (with fluxes from
0.13~mJy to 2.7~Jy), and no compact radio cores in any of the 30 lowest mass
galaxies with $M_{\rm *,gal} < 1.7 \times10^{10}\,M_{\odot}$.  In GSD19, we
reported that only 3 of the 30 galaxies (from the larger set of 100) that were
expected to have a central IMBH also had a central X-ray point-source.
However, for our sample of 75 spiral galaxies with ongoing star formation, and
thus cold gas to potentially fuel a greater level of accretion
onto a central IMBH, we have found that  
13 of the 34 galaxies expected to have 
an IMBH (or 14 from 35 when including NGC~4212) also have an X-ray point source in their center.  
As such, there is hope for detecting the brighter
sources in our sample at radio wavelengths, and the large collecting area of
radio facilities with the spatial resolution to detect compact radio sources,
such as the VLA, the next generation Very Large Array 
\citep[ngVLA,][]{2015arXiv151006438C}, 
and the upcoming Square Kilometer Array radio telescope 
\citep[SKA:][]{2009IEEEP..97.1482D}, will play a key role
in detecting the fainter sources.

\subsection{X-ray contamination: X-ray binaries}\label{Sec-XRB}

Here we consider the possibility that some of
the nuclear X-ray point-sources are XRBs, powered by an accreting neutron star or a
$\sim$10~M$_\odot$ black hole, rather than massive black hole.  
Distinguishing between the two possibilities for individual sources can be
difficult even at distances much closer than the Virgo Cluster. A classic
example of this situation is the bright ($L_{\rm X} \approx 2 \times 10^{39}$
erg s$^{-1}$) point-like X-ray source in the nucleus of the Local Group's
late-type spiral galaxy M\,33 \citep{2002MNRAS.336..901D}, which is almost
certainly a stellar-mass X-ray binary rather than an IMBH, based on its
spectral and timing properties
\citep{2004A&A...425...95D,2011MNRAS.417..464M,2015A&A...580A..71L,2018MNRAS.480.2357K}. At
the distance of the Virgo Cluster, it would be impossible to tell. Therefore,
we need to assess the XRB contamination on a statistical level.

X-ray binaries are usually divided into two classes, based on the age and mass
of their donor star.  High-mass X-ray binaries (HMXBs) have a young donor,
typically more massive than $\sim$10~M$_\odot$. Since massive stars live
for less than a few $10^7$ years, HMXBs are located in regions of current or
recent star formation. As a first approximation, their number is linearly
proportional to the star formation rate
\citep{2012MNRAS.419.2095M,2019ApJS..243....3L}. 
Low-mass X-ray binaries
(LMXBs), instead, have a low-mass donor star, and ages $\gtrsim$10$^9$
years.  Their number and spatial distribution in a galaxy approximately follows
the stellar-mass distribution \citep{2004ApJ...611..846K,2012A&A...546A..36Z}
In more detail, their number is a function of the integrated star formation
rate over the galaxy history, see \citet{2013ApJ...764...41F}. 
The LMXB population of a galaxy is itself 
composed of two sub-populations: one formed in globular clusters and the other
in the field of the galaxy.  
Due to the higher stellar density and higher rate of stellar encounters in
globular clusters, they are $\sim$1000 times more efficient than field stars at forming 
LMXBs, per unit stellar mass 
\citep{2007ApJ...660.1246S,2013ApJ...764...98K,2020ApJS..248...31L}. 
The HMXBs, on the other hand, are not known to get such a boost 
\citep{2012ApJ...755...49G, 2019ApJ...871..122J}; 
they are generally not formed from
close encounters and captures like LMXBs, but instead come from the initial
fragmentation of molecular clouds. 

Spiral galaxies usually contain a mix of HMXB and LMXB populations. The HMXBs
dominate for specific star formation rates 
$\log(\rm{sSFR}) \gtrsim -10.5$ yr$^{-1}$ \citep{2019ApJS..243....3L}. 
They also dominate the high-luminosity
end of the distribution, at $L_{\rm X} \gtrsim 10^{39}$ erg s$^{-1}$
\citep{2019ApJS..243....3L}. \citet{Soria2021} extensively reports on the
star formation rate, and off-center ULX population, among our parent sample of
75 Virgo cluster spiral galaxies. 
Our subsample in Table~\ref{Tab-IMBH} have star formation rates ranging from
0.2 to 1.1 M$_\odot$ yr$^{-1}$, with two galaxies having a rate of just 0.1
M$_\odot$ yr$^{-1}$ \citep{2015A&A...579A.102B}, and 
none of these galaxies appear to contain a nuclear starburst. 
This corresponds to specific star formation
rates $\log(\rm{sSFR}\,{\rm yr}^{-1})$ ranging from $-$9.6 to $-$10.5, with the two low
star formation rate galaxies having logarithmic specific star formation rates
of $-$11.0 and $-$11.5.  While these latter two galaxies are expected to have
more LMXBs than HMXBs brighter than $L_{0.5-8\,{\rm keV}} = 10^{38}$ erg
s$^{-1}$, this reverses for the galaxies with higher specific star-formtion rates.
For a rate of $-$10.5, one expects 3 HMXBs brighter than $10^{38}$ erg
s$^{-1}$ and 1 brighter than $6\times 10^{38}$ erg s$^{-1}$.  For a specific
star formation rate of
$-$9.6, one expects 8 HMXBs brighter than $10^{38}$ erg s$^{-1}$ and 3
brighter than $6\times 10^{38}$ erg s$^{-1}$. 
However, the star formation rates
reported above are for the whole galaxy, including their spiral arms.  
For the galaxies in our subsample,
the rates in the nuclear regions alone (within 100 parsec of the nuclear
position) are roughly 3 to 4 orders of magnitude lower. 
That is, around $\log(\rm{sSFR}\,{\rm yr}^{-1}) \approx -12.5$ or lower. 
Therefore, the probability of finding an HMXB at the nuclear position is low. 

The second possibility of contamination is from field LMXBs, in galaxies where
the stellar mass in the nuclear region is dominated by field stars rather than
a nuclear star cluster.  For a stellar mass\footnote{Our subsample of spiral
galaxies with candidate IMBHs have stellar masses (0.3--10)$\times10^{10}$
M$_\odot$ (GSD19).}  of $10^{10}$ M$_\odot$, 
we expect $\sim$5 LMXBs more luminous than $10^{38}$ erg
s$^{-1}$, and $\sim$1 LMXB more luminous than $5 \times 10^{38}$ erg
s$^{-1}$ (those numbers are a few times lower in elliptical galaxies, because
their stellar population is older). The question then becomes how much of that
stellar mass is within the inner region.  For a disk with a typical
exponential scale-length, $h$, of 3 kpc \citep[e.g.,][]{2008MNRAS.388.1708G},
the half-light radius $R_{\rm e} = 1.678 \times 3 \approx 5$ kpc.  Using the 
equations from \citet{2005PASA...22..118G}, the
fraction of light within the inner 80~pc ($\approx$1$\arcsec$) radius is
$3.5\times10^{-4}$, and within the inner 160~pc ($\approx$2$\arcsec$) 
radius it is $1.4\times10^{-3}$.  Therefore, the probability of a 
field-LMXB located at the nuclear position is also small. 
To give a specific example, one would expect $3.5\times10^{-4}\times5 \approx
0.002$ LMXBs more luminous than $10^{38}$ erg
s$^{-1}$ within the inner 80~pc ($\approx$1$\arcsec$) radius of a spiral
galaxy with a stellar mass of $10^{10}$ M$_\odot$.  That is, just 1-in-50 will be
expected to have such an LMXB.  

Finally, we consider the possibility of finding a bright LMXB inside a nuclear
star cluster. For this case, we already know that the probability is not
negligible, as the example of M33 shows.  To estimate this, we assume that a
nuclear star cluster is equivalent to a GC of similar mass, and examine the
number of LMXBs observed in the latter class of systems. From a study of GCs
around Virgo and Fornax elliptical galaxies, \cite{2013ApJ...764...98K} found
that about 5\% of red GCs, and about 1.5\% of blue GCs, contain an LMXB more
luminous than $10^{38}$ erg s$^{-1}$ at 0.3--8~keV.  This drops to 2.7\% (red) and
0.8\% (blue) for $L_{0.3-8\,{\rm keV}} \geq 2\times10^{38}$ erg s$^{-1}$. 
Considering only GCs more luminous than 
$M_z \approx -10$ mag (a better comparison for our nuclear star clusters,
which have masses from $0.5\times 10^6$--$2\times 10^7$ M$_\odot$)\footnote{These
  masses are typical for star clusters \citep{2013ApJ...763...76S}, and they are
  similar to the nuclear star cluster masses in the 30 early-type galaxies
  predicted to harbor an IMBH by \citet{2019MNRAS.484..794G}.}, 
\cite{2013ApJ...764...98K} showed that the bright LMXB occupation fraction was
higher by a factor of 3, for both red and blue GCs. This is consistent with
the results of \cite{2007ApJ...660.1246S}, who also found an occupation
fraction of $\sim$10--20\% for LMXBs above $10^{38}$ erg s$^{-1}$ in the most
massive red GCs, and a factor of 3 lower in the most massive blue GCs. In our
sample of Virgo spiral galaxies expected to host an IMBH, we found an X-ray
point-source occupation fraction of 12-from-33, or 14-from-35, which is
roughly 36\% or 40\%. 

Moreover, the above percentage of 36\% 
represents a lower limit to the true value of spiral galaxies having
central sources with $L_X > 10^{38}$ erg s$^{-1}$ because our data was usually
not deep enough to detect X-ray sources as faint as $10^{38}$ erg s$^{-1}$, 
even under the assumption of no intrinsic neutral gas absorption.
For sources not significantly affected by intrinsic absorption (e.g., at
the outskirts of a galactic disk or in regions dominated by old stellar
populations), our {\it CXO} data, with typical exposure times of $\sim$10 ks,
has a detection limit around 3--4$\times 10^{38}$ erg s$^{-1}$
\citep{Soria2021}.  For comparison, the early-type galaxy sample was observed
to a similar depth of $3.7\times10^{38}$ erg
s$^{-1}$\citep{2010ApJ...714...25G}.  Only a few large galaxies in our sample
have longer archival observations that enabled us to reach detection limits
around 10$^{38}$ erg s$^{-1}$.

Furthermore, we do know that late-type galaxies have generally more
X-ray-absorbing cold gas in their nuclear regions than early-type galaxies (in
which the gas is mostly ionized). For disk galaxies seen at high-inclination,
the absorbing column density through the disk plane also has to be added to
the intrinsic absorption in the nuclear region. For plausible H\,{\footnotesize I}
column densities of $\approx$3$\times 10^{22}$ (100 times the Galactic
line-of-sight value in the Virgo direction, corresponding to about 15 mag of
extinction in the V band), the detection limit at 17 Mpc in a typical 10-ks
observation is reduced to $\approx$10$^{39}$ erg s$^{-1}$. Sources with
luminosities of a few $10^{38}$ erg s$^{-1}$ are much more likely to be
detected in the nuclear region of early-type Virgo galaxies than disk
galaxies. Thus, we argue that if we could remove the effect of the obscuring
H\,{\footnotesize I} gas, the 36\% detection rate would rise while the 10\% figure for
the early-type galaxies would not change.

The above value of 36\% is 3.6 times higher
than the ratio of 3-from-30 (10\%) found by \citet{2019MNRAS.484..794G} among the
Virgo cluster dwarf early-type galaxies expected to have an IMBH and similarly
imaged through a {\it CXO} Large Project with long exposure times.  The 
explanation may be that the more 
gas-rich environment of the star-forming, late-type spiral galaxies is 
more favorable for igniting the central black hole and turning on an AGN
than the environment within the dwarf early-type galaxies
\citep{2009MNRAS.397..135K}.  
Collectively, the findings above may be suggesting that we are not
just detecting XRBs involving a 
donor star feeding the accretion disk around a compact stellar-mass black hole
or neutron star \citep[e.g.,][]{1992Natur.355..614C, 2003A&A...410...53S,
 2014Natur.505..378C}.  One could spend an inordinate amount of energy and
text trying to refine the negligible probabilities of detecting a field-XRB within the inner
arcsecond or two of each galaxy, including considerations of the (host
galaxy's) stellar mass, 
star formation rate and metalicity within the  region sampled by {\it
  CXO}. Ultimately, all that effort would be undermined because 
there would need to be a recognition that we do not yet 
know with certainty how many XRBs are expected for a given nuclear star
cluster (mass, metalicity and star formation rate)\footnote{The pursuit of
  nuclear star cluster metalicities, stellar ages, and star formation rates is
  well beyond the scope of the current investigation.}, and 
their XRB contribution is expected to dominate over the field population given the
findings in globular clusters.  
However, there is already a clincher against a population of solely XRBs. 
We already know that 
four\footnote{NGC~4713 and NGC~4212 are LINER/H\,{\footnotesize II} composites, NGC~4313
  is a Seyfert/LINER and NGC~4607 is a LINER.} of our late-type galaxies are LINERs, and
NGC~4178 has a high-ionization [Ne\,{\footnotesize V}] emission line
\citep{2009ApJ...704..439S}, likely flagging the existence of a massive black
hole. 
Nonetheless, it remains desirable to know what the prospects are for
constraining the masses of these suspected IMBHs.  In what follows, we provide
a concerted and substantial discussion for where and how further progress on this
front can be made through recourse to non-(X-ray) data.

\section{Prospects for black hole masses from non-(X-ray) data}\label{Sec-Disc}

In what follows, we review the prospects for spatially resolving the sphere of
gravitational influence around IMBHs.  This offers the promise of a direct
mass measurement.  We further discuss additional prospects, 
involving non-(X-ray) data, for determining the presence and mass of IMBHs. 

\subsection{Chasing the sphere-of-influence}

From our parent sample of 74+1 late-type galaxies in the Virgo cluster, only
three (NGC: 4303; 4388; and 4501)\footnote{The recent compilation of directly
  measured black hole masses given by \citet{2019ApJ...887...10S} reports two
  additional late-type galaxies with directly measured black hole masses 
  in/near the Virgo cluster. They are NGC~4151 \citep{1971ApJ...165L..43G,
    1984ApJS...56..507W} with $M_{\rm bh} \sim 5\times10^7\,{\rm M}_{\odot}$,
  and NGC~4699 \citep{1997ApJS..108..155G} which belongs to the NGC~4697 Group
  \citep{2011MNRAS.412.2498M} and has $M_{\rm bh} \sim 2\times10^8\,{\rm
    M}_{\odot}$.}  have had directly measured black hole masses reported (see
Table~A2 in GSD19).  That is, their black hole's sphere-of-influence should
have been 
spatially resolved.  While two of these are reported to have black hole masses
greater than $\sim$$10^7\,{\rm M}_{\odot}$, NGC~4303 ($D\approx 12$-13 Mpc,
$\sigma=95$ km s$^{-1}$) has the smallest reported black hole mass of the
three, at $4\times10^6\,{\rm M}_{\odot}$ \citep[][observed with {\it
    HST}/STIS]{2007A&A...469..405P}.
For reference, from the sibling sample of 100 early-type galaxies in the Virgo
cluster \citep{2004ApJS..153..223C} that were also observed with {\it CXO}
\citep{2008ApJ...680..154G}, there are 11 galaxies that have
directly measured black hole masses \citep[see Table~1
  in][]{2019MNRAS.484..794G}, and all but one of those\footnote{NGC~4486A
  (VCC~1327, $D=18.3$~Mpc, $\sigma=131\pm13$ km s$^{-1}$) has the lowest
  reported black hole mass of the 11 early-type galaxies 
  at $(1.3\pm0.8)\times10^7\,{\rm M}_{\odot}$.  It
  was observed with the integral field spectrograph SINFONI on the Very Large
  Telescope under $0\farcs1$ spatial resolution by
  \citet{2007MNRAS.379..909N}.}  have black hole masses greater than 
$\sim$2$\times10^7\,{\rm M}_{\odot}$.

Table~\ref{Tab-Calib} reveals the typical spatial resolution required to
resolve the gravitational sphere-of-influence (soi) around black holes of
different mass and located at a typical Virgo cluster distance of 17 Mpc.
These estimates are based upon the expression $r_{\rm soi} = G M_{\rm
  bh}/\sigma^2$ \citep{1972ApJ...178..371P, 1976MNRAS.176..633F} --- which is
informatively reviewed in \citet{2001ASPC..249..335M} and
\citet{2013degn.book.....M} --- and the spiral galaxy $M_{\rm bh}$--$\sigma$
relation from \citet{2017MNRAS.471.2187D}.  Obviously for local galaxies, if they are half
this assumed distance then their apparent $r_{\rm soi}$ (in arcsec, not in
parsec) will double.

\begin{deluxetable}{ccl}
\tablecaption{Black hole calibration points\label{Tab-Calib}}
\tablehead{
\colhead{$M_{\rm bh}$}  &  \colhead{$\sigma$}  &  \colhead{$r_{\rm soi}$} \\
\colhead{${\rm M}_{\odot}$}  &  \colhead{km s$^{-1}$}  &  \colhead{pc ($\arcsec$)}
}
\startdata
$10^9$      &  293         &  50 (0.6)   \\
$10^8$      &  195         &  11.3 (0.14)   \\
$10^7$      &  130         &  2.5 (0.03)   \\
$10^6$      &   86         &  0.6 (0.007)   \\
$10^5$      &   57         &  0.13 (1.6E-3)   \\
$10^4$      &   38         &  0.03 (3.6E-4)   \\
$10^3$      &   25         &  0.007 (8.3E-5)   \\
$10^2$      &   17         &  0.001 (1.8E-5)   \\
\enddata
\tablecomments{Reversing the spiral galaxy $M_{\rm bh}$--$\sigma$ bisector relation
\citep[][their Table~4 entry ``All'']{2017MNRAS.471.2187D} we provide both the
stellar velocity dispersion that corresponds to the black hole masses listed
in column~1 and the expected sphere-of-influence (soi) of the black hole if
at a typical Virgo cluster distance of 17~Mpc.}
\end{deluxetable}

It is pertinent to ask, and interesting to know, what prospects there are for
high(er) spatial resolution observations than used thus far. 
In space, the 6.5m James Webb Space Telescope ({\it JWST}) will hopefully soon
accompany the 2.4m {\it HST}, with NIRCam \citep{2004SPIE.5487..628H} aboard
{\it JWST} providing a diffraction-limited spatial resolution of $\approx$70
milliarcseconds (mas), as defined by the PSF's FWHM at 2 microns.  This is comparable to the angular
resolution achieved at UV wavelengths with {\it HST's} long-slit Space
Telescope Imaging Spectrograph \citep[STIS:][]{1998PASP..110.1183W}.
The upcoming 24.5m Giant Magellan Telescope ({\it GMT}) is expected to
have a diffraction-limited resolution of $\sim$13 mas in the $J$-band
($\sim$22 mas in the $K$-band) feeding the {\it GMT} integral field
spectrograph \citep[GMTIFS:][]{2012SPIE.8446E..1IM}, while the Thirty Meter
Telescope ({\it TMT}) boasts 4 mas spaxels and 8 mas resolution from its
Infrared Imaging Spectrograph \citep[IRIS:][]{2016SPIE.9908E..1WL}.  The 40m
Extremely Large Telescope ({\it ELT}) will be equipped with the
High Angular Resolution Monolithic Optical and Near-Infrared
\citep[HARMONI:][]{2016SPIE.9908E..1XT} integral field spectrograph, also with
$4\times4$ mas spaxels.  This represents roughly an order of magnitude
improvement, and will enable one to resolve the sphere-of-influence around
Virgo cluster BHs of mass down to $10^6\,{\rm M}_{\odot}$ (see
Table~\ref{Tab-Calib}).  
For the 100$+$ galaxies within the Local Group that are more than ten times
closer than the Virgo cluster's mean distance, i.e., within 1.7~Mpc, the {\it
  TMT} and {\it ELT} will be able to probe BHs that are some ten times
smaller, encompassing the galaxies M33, NGC~185, NGC~205, NGC~300,
NGC~147, NGC~3109, NGC~6822, IC~10, IC~1613, IC~5152, UGC~4879, DDO~216,
DDO~210, DDO~221, Leo I, II \& III, Sextans A \& B, Antlia, etc.  One will
even be able to probe down to black hole masses of $10^4\,{\rm M}_{\odot}$ if
they are within 170 kpc, encompassing the many satellites of the Milky Way,
such as the Magellanic Clouds located 50 and 63 kpc away, Sextans, Ursa Minor, Draco,
Fornax, Sculptor, Carina, Pisces I, Crater II, Antlia 2, etc.  Although, for
BHs with a mass of $10^4$ M$_{\odot}$ and a soi equal to 0.03~pc (based on a host
system stellar has a velocity dispersion of 38 km s$^{-1}$), the number of stars within
the soi may be limited.  Excitingly, the soi around possible IMBHs within the Galaxy
\citep[see ][]{2017NatAs...1..709O, 2018MNRAS.478L..72R}, out to distances of
17 kpc, could be resolvable for masses down to $10^3\,{\rm M}_{\odot}$.

Impressively, the 
GRAVITY near-infrared ($K$-band) interferometric instrument involving all four of the 8~m
Very Large Telescopes (VLT) already provides 4~mas resolution, or 2~mas
resolution if using the four 1.8~m Auxiliary Telescopes 
\citep[e.g.,][]{gravitycollaboration2020detection}.   The planned optical interferometer
for the VLT, MAVIS, will have milliarcsecond spatial resolution at 550~nm
\citep{2020arXiv200909242M, 2021MNRAS.507.2192M}. 

Considering facilities with longer baselines, ALMA, with its 16~km baseline,
already provides 20 mas resolution at 230 GHz (1.3 mm), several times better
than most current optical/near-IR telescopes.
Radio observations by \citet{1995Natur.373..127M}
of maser emission from a circumnuclear disk in NGC~4258 (M106) were made using
a synthesized beam size of just $0.6\times 0.3$ mas, obtained using 22~GHz
(1.3 cm) interferometry on the Very Long Baseline Array (VLBA).\footnote{The
  VLBA can now achieve 0.12 mas (120 $\mu$as) resolution at a wavelength of
  3~mm, using the MK-NL baseline, and the ngVLA may spatially resolve SMBH binaries and triples
  \citep{2018ASPC..517..677B}.}  Indeed, this enabled confirmation that BHs
are real, as opposed to say a swarm of compact stellar-mass remnants.  Not
surprisingly, this result led to searches for more such maser detections
around BHs, and quite a few discoveries were made
\citep[e.g.,][]{2003ApJ...582L..11G, 2011ApJ...727...20K, 2016A&A...592L..13H}.
Most dramatically, recent
observations at 1.3~mm wavelengths taken with the {\it Event Horizon Telescope}
({\it EHT}) have
provided the highest resolution images to date.  With 20~$\mu$as resolution,
\citet{2019ApJ...875L...4E} probed not just within the soi, but were able to
see the silhouette of the event horizon around the SMBH in M87, located
$\sim$17 Mpc away in the Virgo cluster.  Such spatial resolution matches
the soi that a 100 solar mass black hole would have 17~Mpc away (see
Table~\ref{Tab-Calib}), although the radio flux from such a source may not be high
enough for the {\it EHT} \citep{2016Galax...4...54F}. 
The Spektr-M mission ({\it aka} the Millimetron Space Observatory, launch date
$\sim$2030) will place a 10-m dish 1.5 million kilometers from Earth.
It will operate at wavelengths from 0.07 to 10~mm. 
When joining the Earth-based interferometers, it will provide a 150-fold increase in
spatial resolution, heralding in a new era of astronomy, with 
nanoarcsecond spatial resolution around 130 nas.

\subsection{Potential future observations}

There are several follow-up observations and investigations which would yield
greater information and insight. 

The large collecting area of the upcoming ngVLA and SKA, plus 
the Five-hundred-meter Aperture Spherical radio Telescope \citep[{\it FAST}, aka
  Tianyan:][]{2006ScChG..49..129N, 2013IAUS..291..325L}, 
and the current SKA pathfinder 
MeerKAT \citep[originally the Karoo Array Telescope:][]{2009arXiv0910.2935B, 2009IEEEP..97.1522J}, 
and the Low-Frequency Array \citep[LOFAR:][]{2013A&A...556A...2V}
should prove valuable for detecting faint radio sources. 
Coupled with the improved spatial resolution from long-baseline
interferometry, one could search for masers around IMBHs 
\citep[e.g.,][]{2015aska.confE.119G} and probe the immediate vicinity of the
AGN and the base of their jets 
\citep{2000AJ....119.1695T, 2002AJ....123.1258H,  2015aska.confE.143P,
  2013ApJ...765...69D, 2021NatAs.tmp..133J}. 
As noted before, the radio luminosities can also be combined with the X-ray
luminosity for use in the `fundamental plane of black hole activity' \citep{
  2003MNRAS.345.1057M, 2004A&A...414..895F, 2015MNRAS.453.3447D,
  2016Ap&SS.361....9L, 2016MNRAS.455.2551N} to estimate the mass of the black
hole (section~\ref{subsec-radio}).
 
Reverberation mapping
of AGN can probe the gas clouds within the BH's sphere-of-influence, although the
assumptions about the orbital stability (virialized nature) and geometry of
these clouds, coupled with the use of a mean virial $f$-factor to convert
virial products, $r\Delta V^2/G$, into virial masses
\citep[e.g.,][]{1972ApJ...171..467B, 2000ApJ...540L..13P}, can hinder
confidence in the estimated black hole mass.  In application, the virial
factor is currently assumed to be constant for all AGN, and for IMBHs it is
further based upon the assumption that this constant value can be extrapolated
to masses less than $\sim$$10^6\,{\rm M}_{\odot}$, i.e., below masses used to establish
its value \citep[e.g.,][]{2004ApJ...613..682P, 2011MNRAS.412.2211G}.  
The characteristic time scale at which the power-spectrum of a galaxy's
central optical continuum 
variability flattens has also been shown to scale with black hole mass
\citep{2021Sci...373..789B}.  When the Eddington ratio is sufficiently high, so
that the variable optical continuum emission from the accretion disk
\citep[e.g.,][]{2003ApJ...599..173S, 2006ApJS..166...69S, 2018ApJ...868..152B} is not swamped by 
the photon shot noise of the  starlight in one's aperture, this will provide
yet another window of investigation.  The upcoming, large-scale time-domain
surveys such as LSST will be very useful in this regard
\citep[e.g.,][]{2014ApJ...782...37C, 2014IAUS..304...11I}. 

Thanks to facilities like the  Wide-field Infrared Survey Explorer \citep[WISE:][]{2010AJ....140.1868W} 
and the Spitzer Space Telescope \citep{2004ApJS..154....1W}, one can use mid-infrared diagnostics
to separate AGN from star formation.  The presence of 
mid-IR high-ionization lines, combined with the strength of polycyclic
aromatic hydrocarbon (PAH) emission features, can  provide a means to
discriminate between the dominant energy source within one's aperture 
\citep[e.g.,][]{2006ApJ...646..161D, 2009ApJ...704..439S, 2012ApJ...753...30S,
  2020ApJ...903...91Y}. 
The presence of high-ionization optical
emission lines can also support the presence of a black hole
\citep{1981PASP...93....5B, 1987ApJS...63..295V, 2006MNRAS.372..961K,
  2020ApJ...889..113M, 2020ApJ...898L..30M}. Furthermore, the existence of Doppler-broadened
emission lines --- used in calculating the virial product under the
assumption that the broadening traces the velocity dispersion due to
virialized motions around the black hole, as opposed to non virialized motion
or outflows \citep[e.g.,][see their Appendix~A]{2019ApJ...884...54M} --- 
can be used to infer the presence of an IMBH rather than a stellar-mass black
hole \citep{2015ApJ...809L..14B, 2018ApJ...863....1C}. 
We therefore intend to pursue the acquisition of 
Keck spectra to search for such optical emission lines in the Virgo cluster galaxies, 
enabling us to potentially derive a virial mass for the X-ray detected black
holes. 

The Kamioka Gravitational Wave Detector \citep[{\it
    KAGRA}:][]{2013PhRvD..88d3007A}, with its 3 km baseline, and the famous
{\it LIGO/VIRGO} facilities \citep{1992Sci...256..325A, 1997CQGra..14.1461C,
  2010CQGra..27h4006H, 2015CQGra..32b4001A}, are constrained to detect the
collision of BHs less massive than $\sim$200 M$_{\odot}$.  Thus far, {\it
  LIGO/VIRGO} have reported a bounty of BHs with masses tens of times the mass
of our Sun, along with the collisional-creation of a black hole with mass
equal to 98$^{+17}_{-11}$ M$_{\odot}$ \citep{2019arXiv191009528Z} and
142$^{+28}_{-16}$ M$_{\odot}$ \citep{2020arXiv200901075T}.  The proposed
underground Einstein Telescope \citep[ET:][]{2010CQGra..27s4002P,
  2011GReGr..43..485G, 2011PhRvD..83d4020H} is planning to have a 10~km
baseline with detector sensitivities that should enable it to detect IMBHs
across the Universe, as will the planned Cosmic Explorer
\citep[CE:][]{2019BAAS...51g..35R} with its 40~km baseline.  It is anticipated
that the planned space-based Deci-Hertz Interferometer Gravitational wave Observatory
\citep[{\it DECIGO}:][]{2011CQGra..28i4011K, 2020arXiv201211859I}, and the European Laser
Interferometer Space Antenna ({\it LISA}) 
\citep{1997CQGra..14.1399D} will also help to fill the
relative void between $\sim$$10^2$ and $\sim$$10^5$ M$_{\odot}$.  They will be
capable of capturing oscillations in the fabric of spacetime due to extreme-
and intermediate-mass ratio inspiral (EMRI and IMRI) events around IMBHs
\citep{2004CQGra..21S1595G, 2012A&A...542A.102M, 2015ApJ...814...57M,
  2017PhRvD..95j3012B, 2020arXiv200714403B}, and IMBH-IMBH mergers from dwarf
galaxy collisions \citep{2008ApJ...679L..89B, 2012ApJ...756L..18Y,
  2012ApJ...750..121G, 2014MNRAS.442.2909C, 2018ApJS..237...36P,
  2020ApJ...890....8C, 2020ApJ...891L..23Z, barausse2020massive}.

As seen here, it is known to be common for the centers of galaxies to house a
nuclear star cluster \citep{1983ApJS...53..375R, 1985AJ.....90.1681B,
  1985AJ.....90.1759S, 1997AJ....114.2366C, 1999AJ....118..208M,
  2002AJ....123.1389B}, including late-type galaxies and the dwarf early-type
galaxies of the Virgo cluster \citep{2006ApJS..164..334F,
  2006ApJS..165...57C}.  Nuclear star clusters can reside in a more gas-rich
environment than globular clusters because the gas escape speed, due to the
surrounding galaxy, is higher than in globular clusters.
Not surprisingly, such galaxy centers are suspected to be ripe fields for
cataclysmic disruptions and mergers of stars, neutron stars and black holes
\citep[e.g.,][]{1981SvAL....7..158D, 1988AZh....65..682I, 1990ApJ...356..483Q,
  2020MNRAS.497.2276P}.  They may also be the sites for some of the
hard to spatially constrain gravitational waves arising from the collision of
compact massive objects \citep{2016LRR....19....1A, 2016ApJ...826L..13A,
  2017PASA...34...69A, 2018LRR....21....3A, 2019ApJ...885L..19C,
  2020arXiv200901075T}, and the ejection site of high-velocity stars
\citep[e.g.,][]{2004ApJ...613.1133B, 2006ApJ...653.1203L, 2006ApJ...651..392S,
  2020MNRAS.491.2465K}. 
As done here, the stellar mass of the nuclear star cluster can be used to
predict the resident black hole mass.  This can be combined with multiple
black hole mass estimates from a wide array of independent black hole mass
scaling relations.  Such an approach was used on the spiral galaxy NGC~3319,
which has a centrally-located X-ray point-source, and which initially had a
black hole mass estimate of 3$\times10^2$--$3\times10^5$ M$_{\odot}$ based on
an assumed Eddington ratio of 0.001--1 \citep{2018ApJ...869...49J}, but for
whom an error-weighted meta-analysis of nine independent estimates of the
black hole mass yielded $M_{\rm bh}=(2.3^{+5.3}_{-1.6})\times10^4$ M$_{\odot}$
\citep{2021PASA...38...30D}.

\section{Potential dual and off-center AGN}\label{Sec_dual}

Normally, off-centered X-ray point-sources are associated with XRBs. 
Indeed,  the 
majority of off-centered ULXs have been found to be consistent with
super-Eddington, stellar-mass accretors \citep{2011NewAR..55..166F,
  2017ARA&A..55..303K}.  However, the current investigation suggests that some
IMBHs may have X-ray luminosities of $10^{38}$--$10^{40}$ erg s$^{-1}$, and 
thus comparable Eddington ratios 
($L_X/L_{\rm Edd}$, with $L_{\rm Edd} =                                 
1.3\times10^{38} M_{\rm bh}/M_{\odot}$ erg s$^{-1}$) to many SMBHs. 
It is therefore conceivable, if not inevitable, that some off-centered X-ray
point-sources with $L_X = 10^{38}$--$10^{40}$ erg s$^{-1}$ will be IMBHs. 

The destructive creative process of galaxy evolution can see galaxies stripped
to their nuclear core \citep[e.g.,][and references
  therein]{2003MNRAS.344..399B,2020MNRAS.492.3263G}, only to then merge and
become part of a larger galaxy than they ever were \citep[e.g.,][and
  references therein]{1975ApJ...196..407T, 2021Graham}.  This path, and no
doubt others, is expected to yield galaxies with off-centered massive black
holes, an informative clue to the growth of galaxies.  In
Section~\ref{Sec-Data}, we identified four galaxies with an additional, 
near-central X-ray point-source.  While these may well be XRBs, it is 
possible that they may be IMBHs. 

Pairs of active BHs with large, kpc-sized offsets have regularly been reported
in the literature.  They are 
typically associated with close pairs of galaxies, perhaps a parent plus
a satellite galaxy, and with the early stages of mergers
\citep{2008MNRAS.386..105B, 2009ApJ...702L..82C, 2010ApJ...717..209C,
  2011ApJ...740L..44F, 2015ApJ...806..219C, 2016ApJ...829...37B,
  2017ApJ...850..168K, 
  2019ApJ...882..181B, Rubinur_2020, li2021detectability, tubin2021complex}.  
After a significant galaxy 
accretion or merger event, the inspiral of the two massive black holes can
result in dual AGN --- if both black holes are active --- with at least one
black hole that is offset from the merged-galaxy center
\citep{2004ApJ...600..634B, 2005A&A...429L...9G, 2018ApJ...857L..22T,
  2020ApJ...896..113L}.  

NGC~6240 is one such galaxy with a dual X-ray AGN
having a separation of $\sim$1~kpc \citep{2003ApJ...582L..15K,
  2020arXiv200901824F}.  CXO J101527.2+625911, with its spatial offset of
1.26$\pm$0.05 kpc from its host galaxy's center, has additionally been shown
to have a velocity offset of 175$\pm$25 km s$^{-1}$ from the host galaxy
\citep{2017ApJ...840...71K}.  Dual black holes with sub-kpc separation are
also known \citep[e.g.,][]{2015ApJ...811...14M}.  In the radio galaxy 0402+379,
\citet{2006ApJ...646...49R}, see also \citet{2011MNRAS.410.2113B}, found a
binary SMBH --- in the form of a spatially resolved, double radio-emitting
nucleus --- with just a 7~pc separation.  Furthermore,  
many spectroscopic binaries, or at least systems with double-peaked emission
lines, are known \citep{1987ApJ...312L...1P, 2004ApJ...604L..33Z,
  2009ApJ...698..956C, komossa2021supermassive}, and the cyclical brightnesses
of quasar PSO J185 is suggestive of a binary SMBH at sub-pc separation
\citep{2019ApJ...884...36L}.

The dual X-ray point-source in NGC~4212, separated by $\sim$240~pc
(Figure~\ref{Fig-4212}), may be signaling the presence of a dual AGN,
although unlike with the post-merger, late-type Virgo cluster 
galaxy NGC~4424 \citep{2021Graham} --- which appears to show an off-centered
infalling nuclear star cluster with an X-ray point-source ---, there 
is no obvious evidence in the optical image for a recent merger
\citep{2018A&A...614A.143M}.  While the X-ray emission from the {\it center} of
NGC~4212 would tend to favor an association with the expected massive black
hole, the nature of the nearby slightly off-center companion X-ray
point-source is less certain.  Rather than being accreted, 
it may have formed {\it in situ} within, say, a 
star-forming clump \citep{pestoni2020generation}.  The same can be said about
NGC~4313, whose two inner X-ray point-sources are roughly twice as far apart
as the pair in NGC~4212.  While multiple black hole
scaling relations predict the existence of a central $10^5$--$10^6$
M$_{\odot}$ black hole in NGC~4212, and we have found an X-ray point-source
coincident with the galaxy center, in regard to the off-center X-ray
point-source in this galaxy, there is the likely possibility of an XRB
associated with a compact stellar-mass object
\citep[e.g.,][]{2011NewAR..55..166F, 2017ARA&A..55..303K}, or a distant AGN
\citep[e.g.,][]{2015MNRAS.450..787S}.

The eventual coalescence of binary black holes may result in a
gravitational wave recoil of the merged black holes, again resulting in an
off-center position for the new larger black hole \citep{2004ApJ...607L...9M,
  2005LRR.....8....8M, 2006ApJ...653L..93B, 2007ApJ...659L...5C,
  2007ApJ...661..430H, 2008ApJ...689L..89K, 2008MNRAS.390.1311B,
  2010PhRvD..81j4009S, 2016MNRAS.456..961B, 2019ApJ...885L...4S}.
Such recoils might take the surrounding star cluster with them
\citep[e.g.,][]{2009ApJ...699.1690M}. 

Offsets of 10--10$^2$ parsecs from the optical centers of galaxies, as defined
by their outer isophotes, have been observed among the dense
$\approx$10$^{6\pm1}\,{\rm M}_{\odot}$ nuclear star clusters in low-mass Virgo
cluster galaxies \citep{2000A&A...359..447B, 2003A&A...407..121B}.  For
reference, 1$\arcsec$ at 17 Mpc is 82 parsecs.  This offset is 
likely due, in part, to the shallow gravitational potential well of these
low-S\'ersic index galaxies \citep{1988MNRAS.232..239D, 1994MNRAS.268L..11Y,
  2000AJ....119..593J, 2005MNRAS.362..197T}, which can result in oscillations
or wandering of the cluster about the galaxy center.  The association of
massive black holes with these nuclear star clusters has been known for over a
decade \citep{2007ApJ...655...77G, 2008AJ....135..747G, 2008ApJ...678..116S,
  2009MNRAS.397.2148G}, and it has long been known that the two entities 
coexisted in the Milky Way \citep{1969Natur.223..690L, 1972AJ.....77..292S,
  1974ngbh.procR....R} and M32 \citep{1935ApJ....82..192S,
  1970ARA&A...8..369B, 1984ApJ...283L..27T}.  Therefore, similar offsets
between some AGN X-ray point-sources and the optical center of the parent
galaxy are to be expected, and have been found in low-mass galaxies \citep{2020ApJ...898L..30M}.  
Indeed, even in some large galaxies, massive AGN 
have been found slightly ($\sim$10~pc) off-center
\citep[e.g.,][]{2010ApJ...717L...6B, 2014ApJ...795..146L,
  2014ApJ...796L..13M}, but more so in dwarf galaxies
\citep{2020ApJ...898L..30M, 2020ApJ...888...36R} and simulations of dwarf
galaxies \citep[e.g.,][]{2019MNRAS.482.2913B, 2019MNRAS.486..101P}.  This is
also to be expected if the dynamical friction timescale correlates inversely
with the black hole mass.

\section{Summary}\label{Sec_summary}

As detailed above, there is a wealth of opportunity to further pursue the IMBH
candidates identified here and those expected to reside in other low-mass
galaxies.  This is possible through longer X-ray exposures, simultaneous and
deep radio observations, high signal-to-noise optical spectra, new mid-IR
diagnostic tools, plus a wealth of emerging 
facilities with higher spatial resolution, and gravitational wave detectors.

We have discovered central, or close to central, 
X-ray point-sources in eleven Virgo cluster spiral galaxies, ten of which are expected 
to harbor an IMBH.\footnote{NGC~4212 is predicted to have a black hole mass of 
$10^5$--($2\times10^6)\,{\rm M}_{\odot}$ (GSD19).}
This adds to the three already known in the literature: 
NGC~4178 \citet{2012ApJ...753...38S}, 
NGC~4713 \citet{2015ApJ...814...11T},
and NGC~4470 (GSD19). 
Collectively, this represents nearly half of our sample of 33+1 spiral galaxies
expected to possess an IMBH.  This contrasts notably with the 10\% (central
X-ray point-source) detection rate in a sample of 30 Virgo cluster early-type
galaxies expected to possess an IMBH \citep{2019MNRAS.484..794G}, even though
both samples had comparable exposure times of typically more than a couple of
hours per galaxy.  We suggest that this outcome may not necessarily reflect
the occupation fraction of IMBHs, but rather the Eddington ratios in these two
samples.  The amount of cool gas available is likely to be a valuable clue
for future observing campaigns pursuing AGN in low-mass and dwarf
galaxies. This notion also meshes with the findings from
\citet{2009MNRAS.397..135K}, which reports higher Eddington ratios in
star-forming galaxies than in quiescent, i.e., non-star-forming, galaxies.

Scouring the literature, we found that four of these 14 galaxies (NGC: 4212;
4313; 4607; and 4713) have been identified as LINERs or LINER/H\,{\footnotesize II}
composites, plus NGC~4178 has a 
high-ionization [Ne\,{\footnotesize V}]  14.32 $\mu$m emission line suggestive of an AGN. 
An additional two galaxies (NGC~4197 and NGC~4330) have 
$L_{0.5-10\,{\rm keV}} \approx 10^{40}$ erg s$^{-1}$, making these {\it
  nuclear} X-ray sources likely AGN according to \citet{2010ApJ...724..559L}. 
That is, half of our 14 late-type galaxies with a nuclear X-ray point-source 
have evidence for hosting a massive black hole rather than a stellar-mass XRB. 
Furthermore, of the four 
(from the 11 new) galaxies for which the central X-ray emission was
sufficiently strong enough to
measure its spectrum (NGC: 4197; 4330; 4492; 4607), the data favored a
power-law over a blackbody curve.  In the case of NGC~4197, which has the
brightest central X-ray point-source, and for which we predict M$_{\rm
  bh}=6\times10^4$ M$_{\odot}$ (see Table~\ref{Tab-IMBH}), we found its 
spectrum to be consistent with the low/hard state of a
greater-than-(stellar-mass) black hole.

We have also detected a clear, dual X-ray point-source in NGC~4212, with the
off-center point-source located 2.9 arcsec (240~pc) away from the
centrally-located source.  Further observation is required to establish if it
is the first dual IMBH, a notion which we consider speculative put
conceivable.  We note that NGC~4470, NGC~4492 and NGC~4313 are also
new targets of interest due to their (weaker) dual X-ray point-sources, with
one of each pair of point-sources residing at the center of each of these
galaxies, and the partner 170, 550 and 590~pc distant.

\begin{acknowledgments}

This research was supported under the Australian Research
  Council's funding scheme DP17012923 and 
  the French Center National d'Etudes Spatiales (CNES), and 
is based upon work supported by Tamkeen under the NYU Abu Dhabi
Research Institute grant CAP$^3$. 
RS warmly thanks Curtin University for their hospitality during the planning
stage of this project. 
Support for this work was provided by the National Aeronautics and Space
  Administration through Chandra Award Number LP18620568 issued
  by the Chandra X-ray  
  Center, which is operated by the Smithsonian Astrophysical Observatory for and
  on behalf of the National Aeronautics Space Administration under contract NAS8-03060.
Data underlying this article are available in the Chandra Data Archive
 (CDA: \url{https://cxc.harvard.edu/cda/}),
 and the Next Generation Virgo Cluster Survey website
 (\url{https://www.cfht.hawaii.edu/Science/NGVS/}).
Based on observations made with the NASA/ESA Hubble Space Telescope, and
 obtained from the Hubble Legacy Archive, which is a collaboration between the
 Space Telescope Science Institute (STScI/NASA), the Space Telescope European
 Coordinating Facility (ST-ECF/ESA) and the Canadian Astronomy Data Centre
 (CADC/NRC/CSA).
The HST imaging data used in this paper may be obtained from the Barbara
    A.\ Mikulski Archive for Space Telescopes (MAST).
This research has made use of NASA’s Astrophysics Data System (ADS)
 Bibliographic Services and of the NASA/IPAC Extragalactic Database (NED),
 which is operated by the Jet Propulsion Laboratory, California Institute
 of Technology, under contract with NASA.

\end{acknowledgments}

\software{
{\sc IRAF} \citep{1986SPIE..627..733T,1993ASPC...52..173T},
{\it Isofit/Cmodel} \citep{2015ApJ...810..120C},
{\sc Profiler} \citep{2016PASA...33...62C},
{\sc xspec} \citep[v12.9.1:][]{1996ASPC..101...17A},
{\sc ciao} \citep[v4.12:][]{2006SPIE.6270E..1VF},
{\sc ftools} \citep{1995ASPC...77..367B}. 
}

\newpage

\bibliography{Paper-newX-IMBHs}{}
\bibliographystyle{aasjournal}

\end{document}